# Calibrated bootstrap for uncertainty quantification in regression models

Glenn Palmer[1], Siqi Du[2], Alexander Politowicz[2], Joshua Paul Emory[2], Xiyu Yang[2], Anupraas Gautam[1], Grishma Gupta[1], Zhelong Li[2], Ryan Jacobs[2], Dane Morgan[2*]

[1]Department of Computer Sciences, University of Wisconsin-Madison, Madison, WI, USA.

[2]Department of Materials Science and Engineering, University of Wisconsin-Madison, Madison, WI, USA.

*Corresponding author. E-mail: ddmorgan@wisc.edu

**Abstract**

Obtaining accurate estimates of machine learning model uncertainties on newly predicted data is essential for understanding the accuracy of the model and whether its predictions can be trusted. A common approach to such uncertainty quantification is to estimate the variance from an ensemble of models, which are often generated by the generally applicable bootstrap method. In this work, we demonstrate that the direct bootstrap ensemble standard deviation is not an accurate estimate of uncertainty and propose a calibration method to dramatically improve its accuracy. We demonstrate the effectiveness of this calibration method for both synthetic data and physical datasets from the field of Materials Science and Engineering. The approach is motivated by applications in physical and biological science but is quite general and should be applicable for uncertainty quantification in a wide range of machine learning regression models.

## Introduction

Machine learning is seeing an explosion of application in physical and biological science for predicting properties of chemical and materials systems, expanding the widely used tools of Quantitative Structure Property/Activity Relationship (QSPR/QSAR) modeling. Machine learning applications frequently make use of supervised regression to predict a desired target property as a function of easily accessible features. However, uncertainty quantification (UQ) remains a major challenge for these models. UQ approaches for regression models generally fall into 4 categories that include methods based on: (1) feature data, (2) ensembles of models, (3) a statistical estimator to approximate the variance of predictions, and (4) Bayes' theorem. Recent work on UQ for QSPR/QSAR has investigated all of these approaches.[1–7] Despite this work, methods appear to vary in effectiveness between datasets[1], and there is no obvious choice for a UQ method that is simultaneously easy to implement, generalizable, and empirically validated.

A widely used and extremely general approach to UQ in machine learning is based on the ensemble method, which is a technique that makes predictions using multiple machine learning models, and outputs a weighted average as the final prediction. A common technique to construct an ensemble of predictors is through bootstrap aggregating, or "bagging." In the bagging approach, multiple models are each trained on a subset of data from the training set, where each subset is constructed by choosing points uniformly at random with replacement.[8] A well-established machine learning approach that uses bagging is random forests, which are made up of a bootstrap ensemble of decision trees.[9]

Because ensemble methods work by computing a distribution of predicted values, they naturally lend themselves to computing estimates of uncertainty. There has been substantial interest in developing theoretically sound and computationally efficient methods for UQ in ensemble methods in the past few decades.[10–18] Particular attention has been given to the use of bootstrap and jackknife resampling methods in classification,[11–14] including methods of rescaling estimated error rates to make them more accurate.[12] Work has also been done on the use of ensemble methods for variance estimation in regression. One thread of research has focused on applying a secondary resampling technique such as bagging or the related jackknife method to bagged learners to estimate variance.[10,11,16] This resampling on top of the original bootstrapping can be very computationally intensive, so a major focus has been on making these methods more efficient. However, efforts toward making these resampling techniques more computationally efficient can lead to biased estimates of variance, and efforts to correct these biases appear to have mixed results empirically.[16] In a second thread of related work, for the specific context of random forests and other tree-based models, Lu and Hardin[18] provided a promising framework for estimating the distribution of prediction errors for a test-set observation $x$ using a weighted sum of prediction errors for out-of-bag cohabitants of $x$ in the trees comprising the model. However, there is no way at present to apply this approach outside of tree-based models.

One general and common UQ technique is to use the standard deviation of the predicted values from the ensemble as an uncertainty estimate.[1,2,5,6] While this technique has in some cases been shown to correlate well with the uncertainty of predictions,[19] it is not clear that this standard deviation should accurately predict the standard error of predictions in general. In fact, if the test set includes points very different from training data, the ensemble standard deviation has been shown to substantially underestimate uncertainty.[2,3]

Therefore, despite the above previous work, development of general, flexible, and accurate uncertainty estimation techniques for regression predictions of bagged learners is still an open challenge.[10] We will show in this work that while standard deviation of the predicted values from the ensemble generally gives a poor estimate of uncertainty, the estimate can be greatly improved through calibration.

In the context of this work, calibration refers to any transformation of an uncertainty estimate to make it more accurate that can then be applied to new predictions, and it is a natural approach to take when uncertainty estimates can be assessed in some manner. In particular, once an uncertainty estimate has been computed, it is almost always possible to use some form of cross-validation data to judge its accuracy and, if necessary, modify the estimate to be more accurate. For example, Kuleshov et al.[20] proposed a calibration method for deep learning uncertainty estimates by generalizing the method of Platt scaling[21] used in classification, and demonstrated that it could be used to produce accurate confidence intervals. Beyond improving accuracy, one benefit of using some sort of calibration post-processing step is that it can allow for more interpretable values from UQ methods which initially only provide estimates of relative uncertainty. For example, when using uncertainty estimates based on distances in feature space, Hirschfeld et al.[1] proposed calibrating the estimates so that they can be interpreted as variances. In their calibration method, the distance-based uncertainty estimate $U(x)$ is assumed to be linearly related to the prediction variance $\hat{\sigma}^2(x)$ such that $\hat{\sigma}^2(x) = a\,U(x) + b$. The values of $a$ and $b$ can be found by minimizing the sum of the negative log likelihoods that each predicted value in the validation set was drawn from a normal distribution with mean equal to the true value of the point, and variance equal to $a\,U(x) + b$. (See equations (11) and (12) in Hirschfeld et al.[1]) Janet et al.[2] showed that for neural network predictions, the distance to available data in the latent space of the neural network, when calibrated by a linear rescaling similar to Hirschfeld et al., provided a more accurate error estimate than using the standard deviation of predictions by an ensemble of neural networks. Both Hirschfeld et al.[1] and Janet et al.[2] provide calibrations for specific cases to improve uncertainty quantification from feature distances. However, neither demonstrate that the approaches can be used generally and neither explore the use of calibration for correcting ensemble uncertainties. We do both these things in the present work.

In this paper, we develop a new method of machine learning model UQ based on calibration of bootstrapped errors from ensemble models. We focus on predicting standard deviation as our UQ metric. We begin by investigating the accuracy of the standard deviation of bootstrap ensemble predictions in estimating the true standard error of predicted values. We refer to this bootstrap estimate as $\hat{\sigma}_{uc}$ (where the subscript *uc* refers to "uncalibrated" and we use the "$\widehat{\ }$" notation to refer to predicted values from the machine leaning model(s)). Then, we implement a method of calibrating $\hat{\sigma}_{uc}$ to yield a calibrated estimate $\hat{\sigma}_{cal}$ (where the subscript *cal* refers to "calibrated"), and demonstrate the effectiveness of this calibration in producing highly accurate uncertainty estimates across multiple sets of both synthetic and physical data. We will use the general symbol $\hat{\sigma}$ when referring nonspecifically to one or both of $\hat{\sigma}_{uc}$ and $\hat{\sigma}_{cal}$. We evaluate $\hat{\sigma}$ for random forests, bootstrap ensembles of linear ridge regression models, bootstrap ensembles of Gaussian process regression (GPR) models, and bootstrap ensembles of neural networks. We perform these evaluations at some level for a total of 10 data sets.

We use two methods of evaluating the accuracy of $\hat{\sigma}$. First, we examine the distribution of the ratios of residuals to $\hat{\sigma}$, which we call the "r-statistic" distribution:

$$r = \frac{residual}{\hat{\sigma}}$$

as proposed by Ling et al.[22] and further applied by Lu et al.[23] If $\hat{\sigma}$ accurately represents the standard deviation of residuals and the model has no bias, the r-statistic distribution should be normally distributed with a mean of zero and a standard deviation of one. Thus, the closeness of the r-statistic distribution to a standard normal distribution is a method of assessing the accuracy of $\hat{\sigma}$. One weakness of assessment with the r-statistic distribution is that it does not provide a way to directly evaluate whether larger (smaller) values of $\hat{\sigma}$ correspond to larger (smaller) residuals. To address this weakness, we also plot binned values of $\hat{\sigma}$ (x-axis) against the root mean square of the residuals (y-axis) in each bin (see Methods for details). We call this an RMS residual vs. $\hat{\sigma}$ plot and it was first introduced for this type of analysis by Morgan and Jacobs.[19] For perfectly accurate $\hat{\sigma}$ and infinite sampling, we would expect this plot to have a slope of one and an intercept of zero, i.e., the magnitude of $\hat{\sigma}$ should correlate perfectly with the root mean square of the model residuals. By comparing the points on this plot to the identity function, one can understand whether $\hat{\sigma}$ is underestimating or overestimating the standard deviation of residuals at various levels of uncertainty. We show that $\hat{\sigma}_{uc}$ correlates with the actual standard deviation of residuals (i.e., larger ensemble standard deviations correspond to larger standard deviations of residuals), but that for different models and data sets, they systematically underestimate or overestimate the true values.

After evaluating $\hat{\sigma}_{uc}$ using the above methods, we implement a log-likelihood optimization calibration scheme similar to the one from Hirschfeld et al.[1] described above, and evaluate the quality of $\hat{\sigma}_{cal}$ after calibration. The only difference between our log-likelihood optimization method and the log-likelihood optimization method described above from Hirschfeld et al. is that we assume the estimated standard deviation is linearly related to the true standard deviation, rather than the variances being linearly related. While this type of method has been used to calibrate distance-based uncertainty estimates,[1,2] to our knowledge it has not been applied to ensemble-based approaches. We show that this calibration method is highly effective at calibrating the ensemble standard deviation as an estimate of uncertainty. Finally, we compare our values of $\hat{\sigma}_{cal}$ for GPR to the Bayesian uncertainty estimates typically used for a single GPR model, which we refer to as $\hat{\sigma}_{GPR}$ and show that $\hat{\sigma}_{cal}$ is more accurate than either uncalibrated or calibrated values of $\hat{\sigma}_{GPR}$.

Overall, our results suggest that the calibrated ensemble standard deviation as an uncertainty estimate can be used across many different machine learning models and data sets. The method can be applied to any model where the bootstrap and CV ensemble generation is not computationally prohibitive. The approach is therefore likely to be of significant utility for many machine learning applications.

## Results

### Recalibration performance of random forest models on synthetic and physical (diffusion and perovskite) datasets

In this section, we show the effects of calibration on select combinations of models and datasets to illustrate a number of key points. The complete set of all model and dataset combinations we considered are included in the Supplementary Information and show results completely consistent with those shown in the main text. First, we used the r-statistic distribution and RMS residual vs. $\hat{\sigma}$ plots to evaluate our calibration method for random forest models using the synthetic,[24] diffusion,[23] and perovskite[25] datasets (see Datasets sub-section in the Materials and methods section for more information). These results are shown in Figure 1. Note that before calibration, there is a consistent positive slope in the RMS residual vs. $\hat{\sigma}$ plots as $\hat{\sigma}_{uc}$ becomes larger – that is, larger uncertainty estimates correspond to larger residual values. However, for all three data sets, the uncalibrated points on the RMS residual vs. $\hat{\sigma}$ plots fall below the identity function, meaning they are overestimating the uncertainty of predictions. After calibration, all three r-statistic distributions appear to be reasonably close to standard normal distributions, and the binned uncertainty estimates lie very closely on the identity function line. Note that points in Figure 1 which are not filled in represent bins that have fewer than 30 points and therefore may deviate from the identity function just due to poor sampling. In general, we find that calibrated points with good sampling are quite close to the identity function line. Overall, the results in Figure 1 demonstrate that our calibration method performs very well for ensembles of random forest models. To further illustrate the power and general applicability of our method, we have performed our recalibration method using random forests on seven additional datasets from the materials science community (see Datasets sub-section in the Methods for more information). In all cases, the $\hat{\sigma}_{uc}$ values are greatly improved with our calibration approach. See Figures 55-68 in the Supplementary Information for r-statistic and RMS residual vs. $\hat{\sigma}$ plots for these additional datasets.

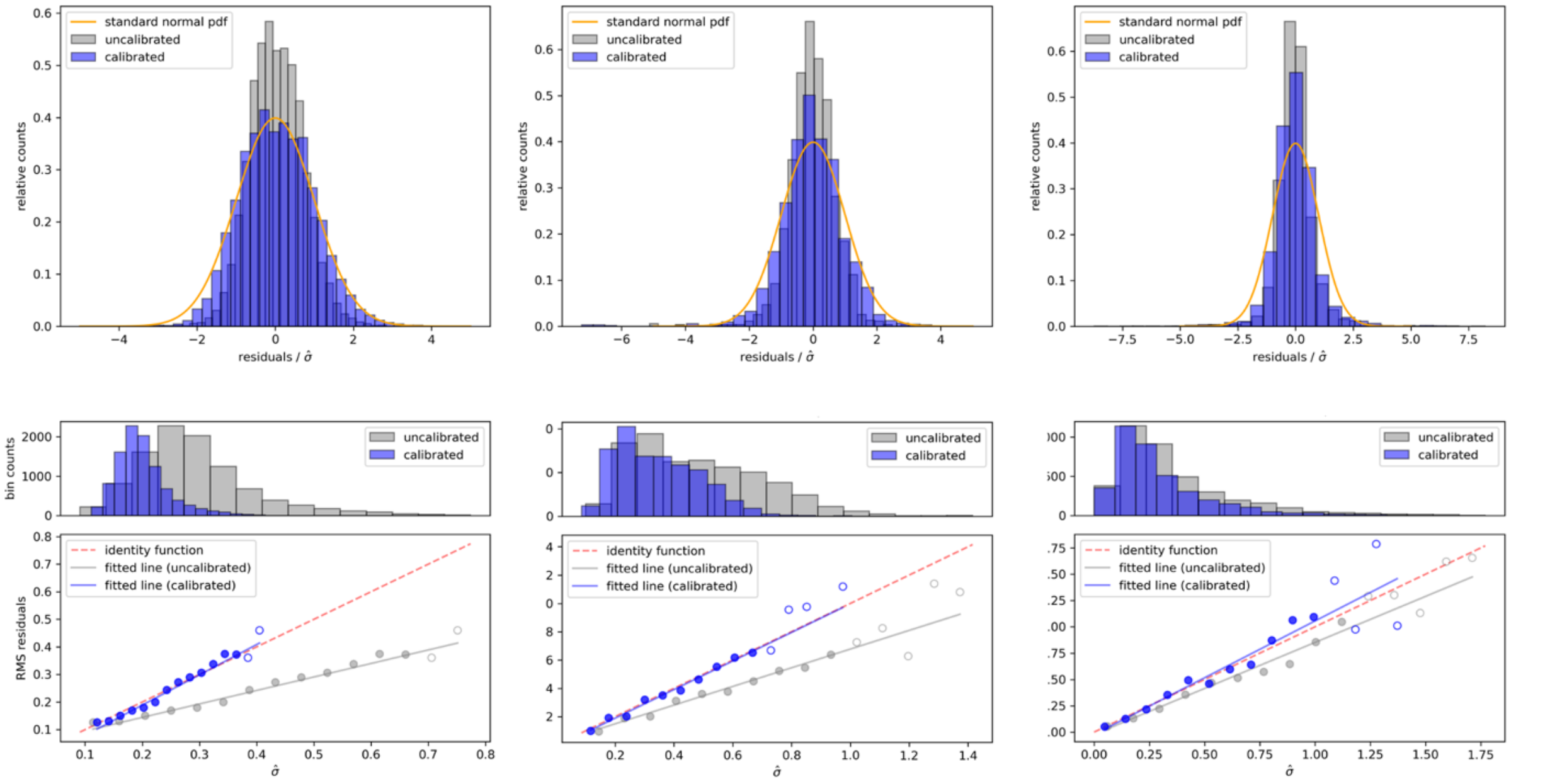

**Fig. 1. Recalibration performance of random forest models on synthetic and physical (diffusion and perovskite) datasets.** Distributions of r-statistic values (top row) and RMS residual vs. $\hat{\sigma}$ plots (bottom row) for random forest, using the synthetic (left column), diffusion (middle column), and perovskite (right column) data sets, shown with both uncalibrated and calibrated $\hat{\sigma}$

values. Markers which are not filled in represent bins with fewer than 30 points. Statistics for each plot are summarized in Table 1.

**Recalibration performance of ensembles of Gaussian process regression and linear ridge regression models on the diffusion dataset**

In Figure 2, we show the r-statistic and RMS residual vs. $\hat{\sigma}$ plots for the diffusion dataset using a bootstrap ensemble of 200 GPR models and a bootstrap ensemble of 500 linear ridge regression models. Before calibration, the GPR values of $\hat{\sigma}_{uc}$ appear to underestimate the true uncertainty, while after calibration, the r-statistic distribution appears very close to standard normal, and the well-sampled points lie closely on the identity function line. We did not expect our method to work as well for linear ridge regression, since the uncertainty in linear regression models is typically dominated by bias, while the bootstrap standard deviation attempts to capture variance (see further discussion of this issue below). As shown in Figure 2, the r-statistic distribution for the calibrated linear ridge regression uncertainty estimates appears to be close to a standard normal distribution. However, the fitted line to the calibrated points on the RMS residual vs. uncertainty estimate plot has a slope of 0.597, substantially less than 1. We also ran a test using neural network models, using a smaller ensemble and just one data set as the fitting is much more computationally demanding. Specifically, we used an ensemble of 25 neural networks on the diffusion data set and found our calibration scheme performed well for well-sampled cases, with calibrated errors close to the identity line (See Supplementary Information Figures 53-54). Additional study is needed to more thoroughly explore the UQ behavior of neural network ensembles, specifically, to assess how well our approach performs for different types of network architectures and non-bootstrap methods of generating ensembles (e.g., selection of different models during training, restarting from different initial weights, and dropout).

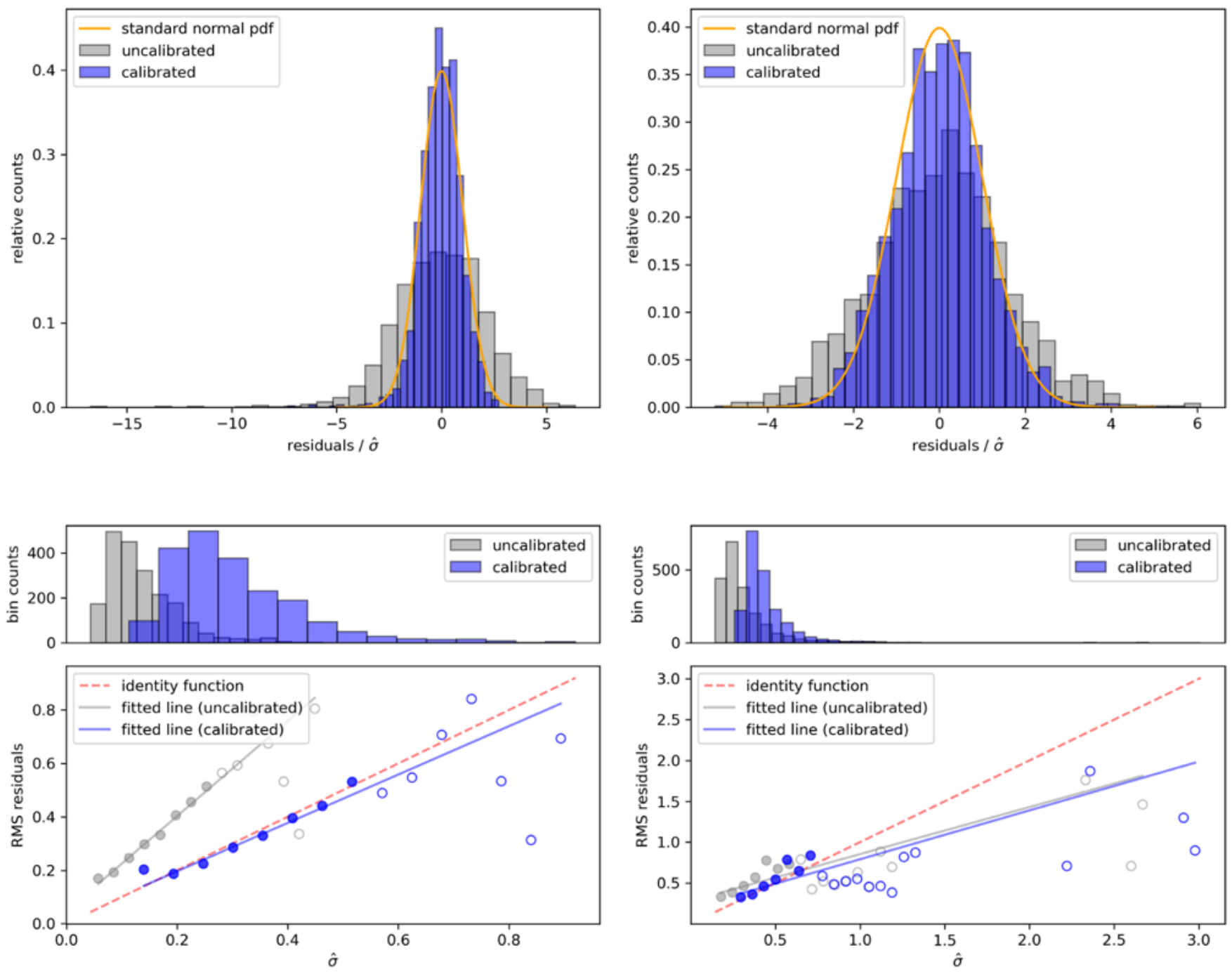

**Fig. 2. Recalibration performance of ensembles of Gaussian process regression and linear ridge regression models on the diffusion dataset.** Distributions of r-statistic values (top row) and RMS residual vs. $\widehat{\sigma}$ plots (bottom row) for Gaussian process regression (left column) and linear ridge regression (right column), using the diffusion dataset, shown with both uncalibrated and calibrated $\widehat{\sigma}$ values. Markers which are not filled in represent bins with fewer than 30 points. Statistics for each plot are summarized in Table 1.

## Comparison of ensemble and Bayesian error estimates of Gaussian Process Regression

To provide a comparison to our bootstrap ensemble method of uncertainty estimation for GPR, in Figure 3 we show the results of our calibration method applied to the Bayesian standard deviation estimates $\hat{\sigma}_{GPR}$, which are often used for UQ for GPR. In the RMS residual vs. $\hat{\sigma}_{GPR}$ plot, the uncalibrated $\hat{\sigma}_{GPR}$ values do not seem to be strongly correlated with the residuals. Because of this, after calibration, several well-sampled points deviate substantially from the identity function line. Note that as shown in Table 1, the uncalibrated and calibrated fitted lines have slopes of 0.109 and 0.275, respectively. Furthermore, both the uncalibrated and calibrated r-statistic distributions have a sharp peak close to 0, suggesting that the $\hat{\sigma}_{GPR}$ values may be overestimating the true uncertainty.

When interpreting this comparison of $\hat{\sigma}_{GPR}$ to our method based on the bootstrap ensemble (see Figure 2), it is important to note that the superiority of our method is confined to the setting in which we hope to obtain accurate uncertainty estimates for predictions on a test set we know is similar to our training set. When test set values are substantially different from a training set, the $\hat{\sigma}_{cal}$ will likely underestimate the true uncertainty, perhaps severely. In that setting, $\hat{\sigma}_{GPR}$ may do a better job of indicating a higher degree of uncertainty, although $\hat{\sigma}_{GPR}$ converges asymptotically to zero (or a constant) for features far from the training data due to the kernel becoming zero. However, for our present goal of obtaining accurate uncertainty estimates for predictions of reasonably similar data, this test demonstrates that our approach is significantly more accurate than the widely used $\hat{\sigma}_{GPR}$ values, even when they are calibrated.

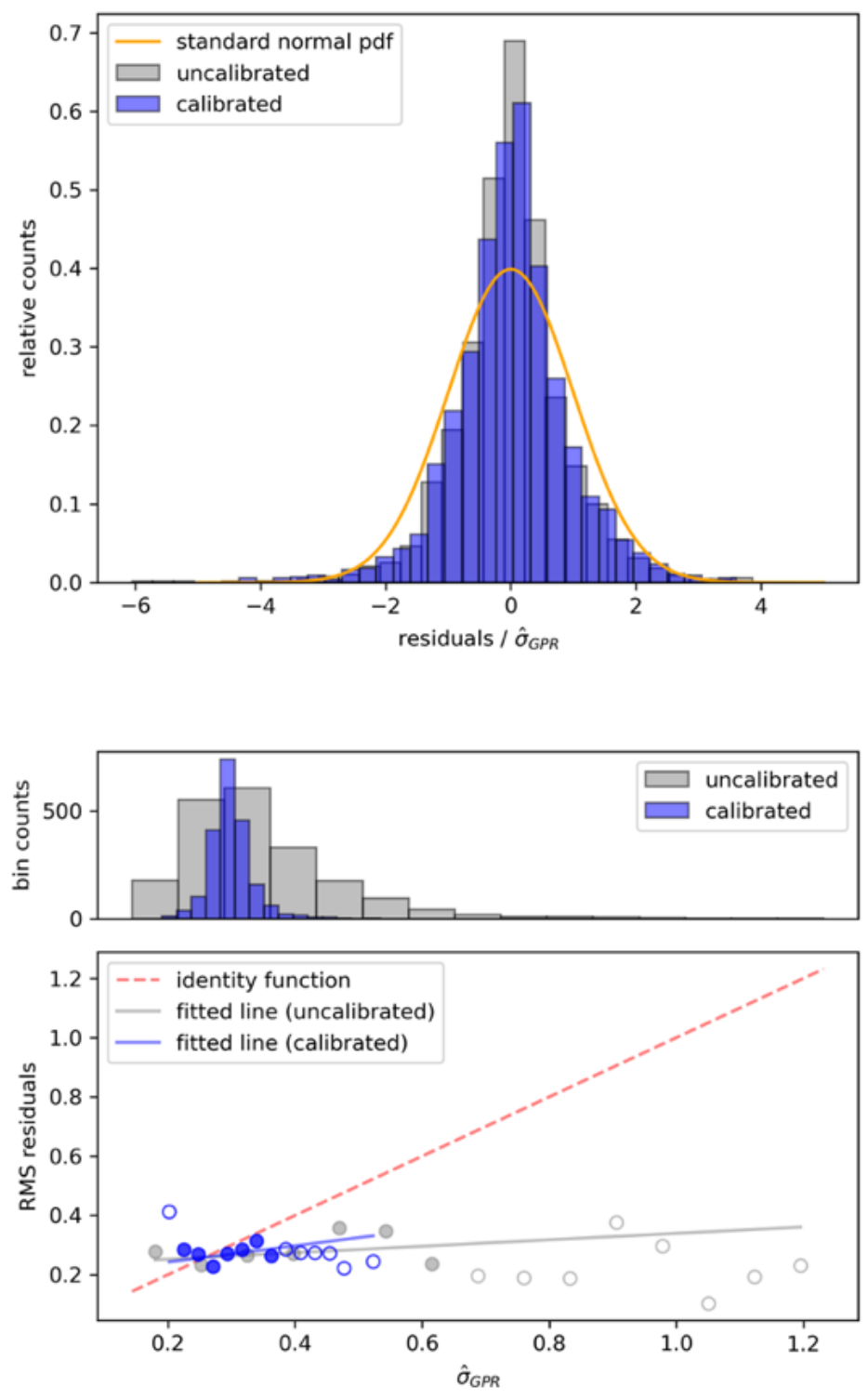

**Fig. 3. Recalibration performance of Gaussian process regression Bayesian uncertainty estimates on the diffusion dataset.** Distributions of r-statistic values (top) and RMS residual vs. $\hat{\sigma}_{GPR}$ plot (bottom) for Bayesian uncertainty estimates obtained from a single Gaussian process regression model, using the diffusion dataset. Both uncalibrated and calibrated $\hat{\sigma}_{GPR}$ values are shown. Markers which are not filled in represent bins with fewer than 30 points. Statistics for each plot are summarized in Table 1.

**Recalibration performance of random forest models on synthetic data with varying amounts of noise**

In Figure 4, we show the results of adding varying amounts of noise to the synthetic data set, and then using our calibration method with random forest. We generated seven noisy training and test sets by adding Gaussian noise with mean 0 and standard deviation equal to 0.1, 0.2, 0.3, 0.4, 0.5, 1.0, and 2.0 times the standard deviation of the original training set (cases 0.1, 0.2, 0.3, and 0.5 are shown in Figure 1 and all the cases are in the Supplementary Information in Figures 23-36). For the 0.1 and 0.2 noise cases, our method appears to work similarly well to the no-noise case. For the 0.3 noise case, the calibrated points on the RMS residual vs. $\hat{\sigma}$ plot appear to diverge somewhat from the identity function line. Then, for the 0.5 noise case, the values of $\hat{\sigma}_{cal}$ have started collapsing to a constant value. For the 1.0 and 2.0 noise cases, the $\hat{\sigma}_{cal}$ values approximately converge to a constant of 1 standard deviation of the noisy data set. The trend of convergence toward a constant $\hat{\sigma}_{cal}$ is expected, since, as the noise increasingly dominates the underlying true values, this trend in the calibration is the only option to obtain accurate uncertainty estimates. In the limiting case, when the training and test values are normally distributed and totally independent of the features, the best a model can do is to predict the training set mean and estimate the training set standard deviation as the uncertainty. Overall, these results indicate that our calibration method is robust to small and even quite significant (0.2 times the standard deviation of the original training set) amounts of noise and handles very large amounts of noise in the way we would expect. It is somewhat unclear how well the approach works with intermediate amounts of noise of about 0.3 times the standard deviation of the original training set, and additional study is needed for this case.

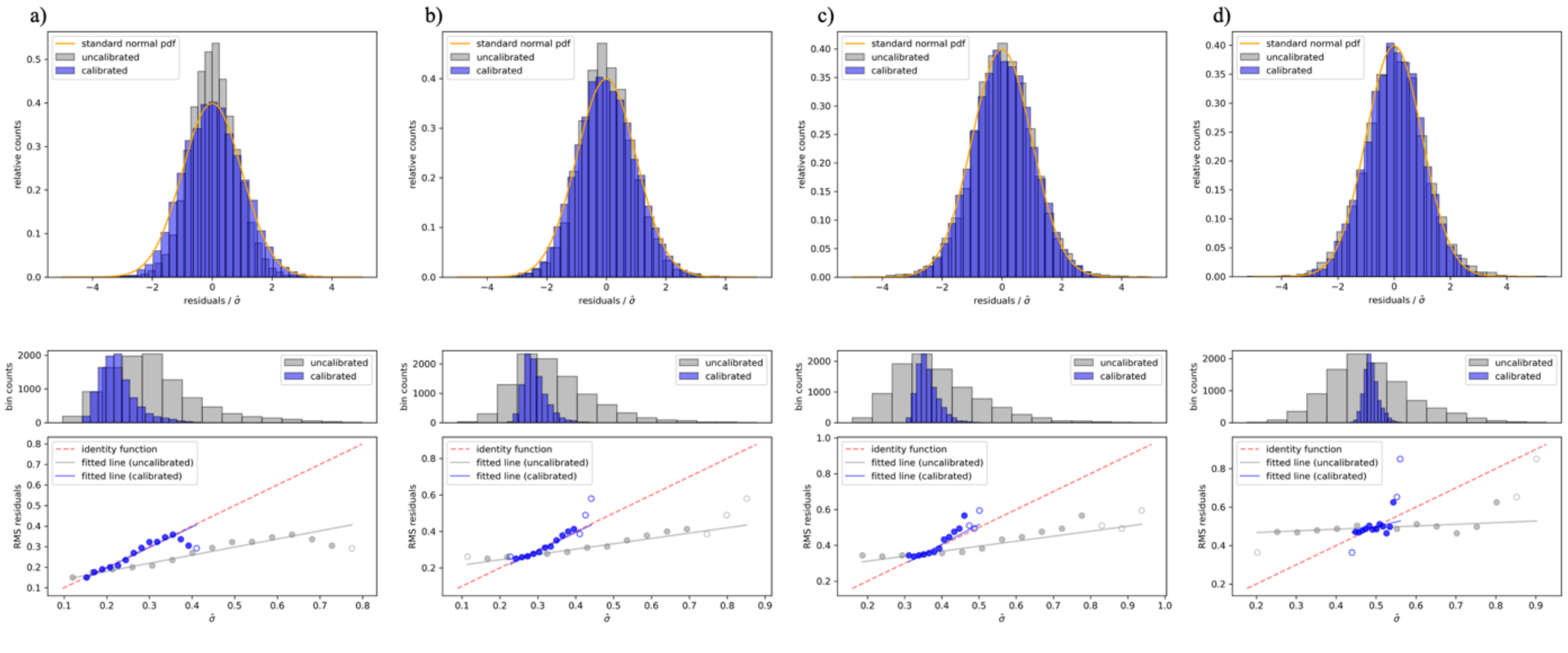

**Fig. 4. Recalibration performance of random forest models on synthetic data with varying amounts of noise.** Distributions of r-statistic values and RMS residual vs. $\hat{\sigma}$ plots for random

forest, using the synthetic dataset with varying amounts of noise added. Gaussian noise with mean zero and standard deviation equal to (a) 0.1, (b) 0.2, (c) 0.3, and (d) 0.5 times the standard deviation of the training set with no noise was added to the four plots, respectively. Both uncalibrated and calibrated $\hat{\sigma}$ are shown. Markers which are not filled in represent bins with fewer than 30 points. Statistics for each plot are summarized in Table 1.

In Table 1, we present the r-statistic mean and standard deviation, the fitted slopes, y-intercepts, and $R^2$ values, and the calibration factors *a* and *b* for all the plots in Figures 1-4. For additional plots and a complete table of these values for all tests we ran, please see Table 1 and plots in the Supplementary Information.

**Table 1. Summary of values from Figs. 1-4.** Given are the mean and standard deviation of the r-statistic distribution for $\hat{\sigma}_{cal}$ and $\hat{\sigma}_{uc}$, the slope, intercept, and $R^2$ values for the linear fits on the RMS residual vs. $\hat{\sigma}$ plots (referred to here as "RvE" for brevity), and the calibration factors a and b, which scale the uncalibrated uncertainty estimates U(x) as aU(x)+b.

| | | | r-stat mean | r-stat stdev | RvE slope | RvE intercept | RvE $R^2$ | a | b |
|---|---|---|---|---|---|---|---|---|---|
| Figure 1 | RF, Synthetic | uncalibrated | 0.043 | 0.675 | 0.489 | 0.046 | 0.964 | 0.445 | 0.070 |
| | | calibrated | 0.060 | 0.938 | 1.099 | -0.031 | 0.964 | | |
| | RF, Diffusion | uncalibrated | -0.022 | 0.707 | 0.660 | 0.017 | 0.971 | 0.647 ± 0.047 | 0.033 ± 0.025 |
| | | calibrated | -0.027 | 0.979 | 1.010 | -0.013 | 0.970 | | |
| | RF, Perovskite | uncalibrated | -0.014 | 0.766 | 0.875 | -0.023 | 0.972 | 0.807 ± 0.038 | 0.002 ± 0.003 |
| | | calibrated | -0.019 | 0.941 | 1.080 | -0.024 | 0.953 | | |
| Figure 2 | LR, Diffusion | uncalibrated | 0.001 | 1.574 | 0.575 | 0.279 | 0.561 | 0.927 ± 0.115 | 0.154 ± 0.041 |
| | | calibrated | 0.002 | 1.058 | 0.597 | 0.193 | 0.521 | | |
| | GPR, Diffusion | uncalibrated | -0.094 | 2.182 | 1.779 | 0.047 | 0.963 | 1.867 ± 0.207 | 0.048 ± 0.029 |
| | | calibrated | -0.039 | 0.963 | 0.905 | 0.015 | 0.929 | | |
| Figure 3 | GPR Bayesian, Diffusion | uncalibrated | 0.007 | 0.884 | 0.109 | 0.231 | 0.127 | 0.233 ± 0.114 | 0.219 ± 0.042 |
| | | calibrated | 0.006 | 0.920 | 0.275 | 0.188 | 0.128 | | |
| Figure 4 | RF, Synthetic, 0.1 noise | uncalibrated | 0.037 | 0.775 | 0.394 | 0.101 | 0.926 | 0.394 | 0.105 |
| | | calibrated | 0.047 | 0.979 | 1.001 | -0.005 | 0926 | | |
| | RF, Synthetic, 0.2 noise | uncalibrated | 0.020 | 0.914 | 0.292 | 0.187 | 0.914 | 0.290 | 0.194 |
| | | calibrated | 0.008 | 0.977 | 1.008 | -0.009 | 0.914 | | |
| | RF, Synthetic, 0.3 noise | uncalibrated | 0.038 | 1.017 | 0.279 | 0.256 | 0.804 | 0.253 | 0.264 |
| | | calibrated | 0.036 | 1.004 | 1.103 | -0.035 | 0.804 | | |
| | | uncalibrated | 0.039 | 1.081 | 0.086 | 0.451 | 0.166 | 0.173 | 0.405 |
| | RF, Synthetic, 0.5 noise | calibrated | 0.026 | 1.009 | 0.495 | 0.250 | 0.166 | | |

**Discussion**

It is useful to consider the nature of machine learning model errors to better understand why our calibrated bootstrap approach is so effective and some of the methods limitations. The total expected squared error in model can be represented as:[26]

$$E\left[\left(F(X)+\epsilon-\hat{F}(X)\right)^2\right]=\left(E[\hat{F}(X)]-F(X)\right)^2+E\left[\left(\hat{F}(X)-E[\hat{F}(X)]\right)^2\right]+\sigma^2.$$

Here, the expectation is the average over all possible training data sets of size $n$, which we can imagine to be randomly sampled from the total possible space of data. The three right-hand side terms from left to right are the model bias squared, model variance, and noise variance, respectively. The model bias (or just bias) is the difference between the expected value of our model averaged over all training set samplings $E[\hat{F}(X)]$ and the underlying true function $F(X)$. The model variance (or just variance) is the squared spread in $\hat{F}(X)$ relative to its average, again taken over all training set samplings. In the absence of noise, the bootstrap approach is an estimate of the variance term in this expression. However, it is an imperfect estimate as the sampling is not all possible training data sets of size $n$ but instead the finite number of bootstrap resampling data sets. We have attempted to use enough bootstrap samples to remove the finite sampling error from being significant, but the nature of the bootstrap data sets is still a source of error. For models with little bias, as is the case for almost all of the models studied here, our approach can be understood to be calibrating the variance estimate. It is perhaps not totally surprising that the approximate sampling from bootstrap gives a variance estimate that is off but strongly correlated with the correct answer, given that it is an approximation to the proper sampling to estimate this correct answer. These considerations also suggest that the present method may be less accurate or fail entirely when bias is a dominant source of error. Such situations would include cases where the predicted test data is far outside the domain of the model. As an example of our model failing for out-of-domain test data, in the Supplementary Information Figures 69-72 we show our recalibration method and associated parity plots on the diffusion dataset for different test sets that are outside the training data, which show, as expected, that the model error cannot be accurately predicted.

Overall, we have demonstrated that across multiple models and datasets, our calibrated bootstrap method of estimating uncertainty is highly accurate. Particularly noteworthy are its exceedingly accurate estimates for random forest across all observed datasets, as demonstrated by the r-statistic and RMS residual vs. $\hat{\sigma}$ plots, and its superior performance for GPR predictions when compared to the Bayesian UQ method typically used in that setting. When our results are taken together, they suggest that the calibrated bootstrap method could be of significant utility for applied scientists in need of better UQ. However, further work is needed to establish its limitations, particularly for predictions on data that is far from the domain of the training data. In addition, the specific approach here is just one of a large family of related methods that can be generated by considering other ensembles besides bootstrap, other uncertainty estimators besides standard deviation (e.g. using confidence intervals), and other calibration approaches, and further exploration of this family of methods could yield additional useful approaches.

**Materials and methods**

**Models.** We used four types of machine learning models to evaluate our method: random forest, Gaussian process regression (GPR), and linear ridge regression, and neural networks. Random forest, GPR, and linear ridge regression were implemented using scikit-learn.[27] The neural networks were implemented in Keras using the TensorFlow backend.[28,29] For the scikit-learn models, all values not specified and any explicitly referenced as default values are set to default values used as of January 2021 in scikit-learn version 0.23.2. For random forest, we used 500 decision trees, a max depth of 30, and default values for all other settings. For GPR, we used as the kernel the sum of the ConstantKernel, the Matern kernel with length scale set to 2.0 and the default value 1.5 of the smoothness parameter nu, and the WhiteKernel with the noise level set to 1. We set alpha, the value added to the diagonal of the kernel matrix to prevent numerical issues, to $10^{-4}$, and the n_restarts_optimizer parameter to 30. An ensemble of Keras neural networks was applied to the diffusion dataset only. For this case, the individual neural networks consisted of three layers: an input layer of 20 nodes, a hidden layer with 10 nodes, and an output layer with a single node. The network was fully connected, used rectified linear activation for each layer, and was optimized using the Adam optimizer using mean squared error as the loss function. An ensemble of these neural networks was built using the BaggingRegressor model contained in scikit-learn, where each estimator in the BaggingRegressor was made using the KerasRegressor method in Keras, which provides a means to have scikit-learn-like functionality applied to Keras models. When constructing the bootstrap ensemble of GPR models, we used 200 models to keep the total computation time tractable. For linear ridge regression, we used scikit-learn default values, and used 500 models to construct the bootstrap ensembles. For neural networks, only 25 models were used to keep the computation time tractable.

**Datasets**. We used ten total datasets to evaluate our methods, but the bulk of our analysis was concentrated on just three of these datasets. Here, we provide an overview of the three datasets which we evaluated in the most detail. First, we used a set of synthetic data based on a method proposed by Friedman.[24] There were five features $x_0$ through $x_4$ for each point, each drawn uniformly at random from the interval $[0, 0.5]$. Then, y-values were generated for each point using the function:

$$y = 30\, sin\,(4\pi\, x_0\, x_1) + 20(x_2 - 0.5)^2 + 10x_3 + 5x_4.$$

We used the above process to generate both a training set of 500 points and a test set of 10,000 points.

Next, we used two physical datasets from the materials science community. The first dataset (referred in this work as the "diffusion" dataset) is a computed database of impurity diffusion activation energies for 15 pure metal hosts and 408 host-impurity pairs.[23] The second dataset (referred in this work as the "perovskite" dataset) is a computed database of perovskite oxide thermodynamic phase stabilities.[25]

In addition to the above datasets, we performed additional select tests on seven more physical datasets from the materials science community. These datasets consist of experimental steel yield strengths,[30] experimental thermal conductivities,[31] calculated maximum piezoelectric displacements,[32] calculated saturation magnetization of Heusler compounds,[33,34] calculated bulk moduli of assorted materials,[35] calculated electronic bandgaps of double perovskite oxides,[36] and experimental superconducting critical temperatures of a large assortment of materials.[37]

**Train/test splitting.** For all data sets, we use a nested cross-validation approach to obtain the value of $\hat{\sigma}_{cal}$ and assess the performance of our calibrated UQ models. First, each dataset (synthetic or physical) was split into numerous train and test subsets (we call this the level 1 split). The test data is held out until the end and used to evaluate the performance of the calibrated UQ models. The training data is then subsequently split into training and validation subsets (we call this the level 2 split) and is used to assess the uncalibrated UQ models and obtain the calibration parameters for that training dataset. For the synthetic dataset, we trained models on the entire training set (i.e, there was no level 1 split as test data could be readily generated), and then made predictions, along with uncertainty estimates for those predictions on the entire test set. To obtain calibration scaling factors (see below), we used 5-fold cross-validation repeated four times on the training set, for a total of 20 level 2 train/validation splits. For both materials data sets, we generated multiple level 1 80%/20% train/test splits through 5-fold cross-validation, and used the predictions accumulated over those splits to evaluate our method. For the diffusion dataset, we used 25 level 1 train/test splits, while for the perovskite dataset, because of its larger size, we used 10 level 1 train/test splits. The different numbers of level 1 train/test splits were chosen to yield reasonable statistics in and reasonable time, which led to smaller numbers of splits for the larger data sets. Within each of the training sets, we again calculated the calibration scaling factors using 5-fold cross-validation repeated four times, for a total of 20 level 2 train/validation splits. For interpretability on the plots (described below), we scaled all $\hat{\sigma}$ values and residuals so that they were in units of one standard deviation of the entire training set.

**r-statistic plots.** To create the r-statistic plot for a given model, uncertainty-estimate method, and dataset, we calculate the ratio of the residual and $\hat{\sigma}$ value for each prediction. We plotted these ratios in a histogram with 30 bins, so that the width of the bins was 1/30 of the range from the smallest (most negative) ratio value to the largest (most positive) ratio value. For comparison we plotted the probability density function of a standard normal distribution and overlaid it on the histogram.

**RMS residual vs. $\hat{\sigma}$ plots.** To create the RMS residual vs. $\hat{\sigma}$ plots for a given model, uncertainty-estimate method, and dataset, we used the residuals and $\hat{\sigma}$ values for all predictions, scaled by the standard deviation of the dataset as described above. For most plots, we used 15 bins for the $\hat{\sigma}$ values, so that the width of each bin was 1/15 of the range from the smallest $\hat{\sigma}$ value to the largest $\hat{\sigma}$ value. However, to ensure that estimates were spread over multiple bins in the range where most uncertainty estimates fell, we decreased this bin size if necessary to ensure that the lowest 90% of the $\hat{\sigma}$ values were spread across at least five bins.

With the $\hat{\sigma}$ values organized into bins, we calculated the root mean square of the residuals corresponding to the $\hat{\sigma}$ values in each bin. We made a scatter plot with one point for each bin, where the horizontal axis represents the $\hat{\sigma}$ value and the vertical axis represents the root mean square of the corresponding residual values. We fit a line to this scatter plot using least-squares linear regression with scikit-learn,[27] with the fit weighted by the number of points in each bin. For comparison we also overlaid a line with a slope of one and a y-intercept of zero. We also included above each of these plots a histogram of the number of data points in each bin to enable assessment of sampling quality.

**Calculating calibration factors**. We calculated calibration factors to correct $\hat{\sigma}_{uc}$ using a method similar to one applied by Hirschfeld et al.,[1] and described briefly in the introduction. Given the set of $\hat{\sigma}$ values and residuals for the cross-validation predictions, we labeled each uncalibrated $\hat{\sigma}$ value as $\hat{\sigma}_{uc}(x)$, and sought to find the corresponding calibrated uncertainty estimates $\hat{\sigma}_{cal}(x)$ that accurately predict the standard error of predictions. To do so, we assumed that the prediction standard error was linearly related to $\hat{\sigma}_{uc}(x)$, such that for some $a$ and $b$, we have $\hat{\sigma}_{cal}(x) = a\hat{\sigma}_{uc}(x) + b$. To find appropriate values of $a$ and $b$, we found the values that minimized the sum of the negative log-likelihoods that each residual $R(x)$ from cross-validation predictions was drawn from a normal distribution with a mean of 0 and a standard deviation of $a\hat{\sigma}_{uc} + b$. That is, letting $D_{cv}$ be the set of cross-validation data, we solved the optimization problem:

$$a, b = argmin_{a\prime,b\prime} \sum_{x,y \in D_{cv}} \ln 2\pi + \ln(a'\hat{\sigma}_{uc}(x) + b')^2 + \frac{R(x)^2}{(a\prime\hat{\sigma}_{uc}(x)+b\prime)^2}.$$

To solve the above problem, we used the Nelder-Mead optimization algorithm as implemented in scipy.[38] Once these optimal values of $a$ and $b$ were obtained in the above manner, we used them to calibrate the $\hat{\sigma}_{uc}$ values for test set predictions by scaling as $a\hat{\sigma}_{uc}(x) + b$.

**Convergence data.** For each model and data set, we made plots to determine the number of bootstrap models necessary for the calibration-factor calculations to converge to a consistent value (see Figures 37-52 in the Supplementary Information). To calculate calibration factors for this purpose, we did 5-fold cross-validation repeated four times with each entire data set, and calculated calibration factors through the log-likelihood optimization method described above. For random forest and ridge regression models with all three data sets, we calculated factors using 50, 100, 200, 500, and 1000 bootstrap models. We repeated this calculation 10 times and plotted the mean and standard deviation of these results on a scatter plot (see Figures 37-40, 43-46, and 49-52 in the Supplementary Information). Because of computational constraints, we did these calculations for the GPR model only with 50, 100, and 200 bootstrap models, and repeated each calculation five times. Based on the resulting convergence plots, random forest and ridge regression models appeared to converge by 500 models, and GPR appeared to approximately converge by 100 models. Therefore, we expect that our results from using 500 models for random forest and ridge regression and 200 models for GPR are reasonable. Our neural network tests used only 25 models which is likely too few for robust convergence but was enough to show qualitatively that our approach works well for this type of model.

**Adding noise.** To evaluate how our calibration method responds when dealing with noisy data, we added varying amounts of Gaussian noise to both the training and test sets of our synthetic data set. We used seven different amounts of noise, all drawn from a normal distribution with a mean of zero, but standard deviations of 0.1, 0.2, 0.3, 0.4, 0.5, 1.0, and 2.0 times the original standard deviation of the training set. We then used the calibration method described above and evaluated our results with the r-statistic and RMS residual vs. $\hat{\sigma}$ plots.

**Acknowledgements:** Financial support provided by the National science foundation provided for Glenn Palmer (Award # 1545481), Siqi Du, Alexander Politowicz, Joshua Paul Emory, and Xiyu Yang (Award # 1636950 and 1636910), and Ryan Jacobs and Dane Morgan (Award # 1931298). Financial support for Anupraas Gautam and Grishma Gupta provided by the University of Wisconsin Harvey D. Spangler Professorship. We thank Max Hutchinson, Matthias Rupp, and Logan Ward for many helpful discussions.

**Author contributions:** G.P., R.J., and D.M. developed the UQ method presented. G.P., S.D., A.P., J.P.E., X.Y, A.G., G.G., Z.L, and R.J. wrote code and analyzed output data. G.P. wrote the first draft of the manuscript. G.P, R.J., and D.M. edited and contributed to later drafts of the manuscript.

**Competing interests:** The authors declare no competing interests.

**Data and materials availability:** We have made the python code used to perform all the calculations and generate all figures publicly available on GitHub (https://github.com/uw-cmg/ML-error). We have also added the methods in this paper to the Materials Simulation Toolkit for Machine Learning (MAST-ML) toolkit for easy application by future users.[39] The MAST-ML toolkit can be found at https://github.com/uw-cmg/MAST-ML. Also in the GitHub repository above, we have included the synthetic, diffusion, and perovskite datasets we used in csv format in the "SI" folder. We have not included the additional data sets we studied as they were not studied as extensively here and are all readily available through the provided references. We have also included csv files with the data from all convergence plots, as well as all test-set residuals, $\hat{\sigma}_{uc}$ values, $\hat{\sigma}_{cal}$ values, and calibration factors we obtained and used to create the r-statistic and RMS residual vs. $\hat{\sigma}$ plots.

*Sci.* **176**, (2020).

**Supplementary Materials**

*Table S1: Summary of values from all r-statistic and RMS residual vs. uncertainty estimates in the SI. Given are the mean and standard deviation of the r-statistic distribution for uncalibrated and calibrated error estimates, the slope, intercept, and $R^2$ values for the linear fits on the RMS residual vs. uncertainty estimate plots, and the calibration factors a and b, which scale the uncalibrated uncertainty estimates U(x) as aU(x)+b.*

| | | r-stat mean | r-stat stdev | RvE slope | RvE intercept | RvE $R^2$ | a | b |
|---|---|---|---|---|---|---|---|---|
| RF, Friedman | uncalibrated | 0.043 | 0.675 | 0.489 | 0.046 | 0.964 | 0.445 | 0.070 |
| | calibrated | 0.060 | 0.938 | 1.099 | -0.031 | 0.964 | | |
| RF, Diffusion | uncalibrated | -0.022 | 0.707 | 0.660 | 0.017 | 0.971 | 0.647 ± 0.047 | 0.033 ± 0.025 |
| | calibrated | -0.027 | 0.979 | 1.010 | -0.013 | 0.970 | | |
| RF, Perovskite | uncalibrated | -0.014 | 0.766 | 0.875 | -0.023 | 0.972 | 0.807 ± 0.038 | 0.002 ± 0.003 |
| | calibrated | -0.019 | 0.941 | 1.080 | -0.024 | 0.953 | | |
| LR, Friedman | uncalibrated | 0.645 | 7.878 | 12.182 | -0.274 | 0.582 | 8.213 | -0.061 |
| | calibrated | 0.101 | 1.091 | 1.483 | -0.184 | 0.582 | | |
| LR, Diffusion | uncalibrated | 0.001 | 1.574 | 0.575 | 0.279 | 0.561 | 0.927 ± 0.115 | 0.154 ± 0.041 |
| | calibrated | 0.002 | 1.058 | 0.597 | 0.193 | 0.521 | | |
| LR, Perovskite | uncalibrated | -0.080 | 4.566 | 2.567 | 0.240 | 0.698 | 4.447 ± 0.154 | -0.034 ± 0.017 |
| | calibrated | -0.021 | 1.107 | 0.580 | 0.258 | 0.685 | | |
| GPR, Friedman | uncalibrated | 0.093 | 1.582 | 2.027 | -0.008 | 0.929 | 1.875 | -0.006 |
| | calibrated | 0.061 | 0.984 | 1.081 | -0.001 | 0.929 | | |
| GPR, Diffusion | uncalibrated | -0.094 | 2.182 | 1.779 | 0.047 | 0.963 | 1.867 ± 0.207 | 0.048 ± 0.029 |
| | calibrated | -0.039 | 0.963 | 0.905 | 0.015 | 0.929 | | |
| GPR Bayesian, Friedman | uncalibrated | 0.041 | 0.542 | 1.505 | -0.069 | 0.891 | 0.839 | -0.019 |
| | calibrated | 0.072 | 1.119 | 1.793 | -0.035 | 0.891 | | |
| GPR Bayesian, Diffusion | uncalibrated | 0.007 | 0.884 | 0.109 | 0.231 | 0.127 | 0.233 ± 0.114 | 0.219 ± 0.042 |
| | calibrated | 0.006 | 0.920 | 0.275 | 0.188 | 0.128 | | |
| GPR Bayesian, Perovskite | uncalibrated | -0.006 | 1.097 | 0.850 | 0.049 | 0.814 | 1.114 ± 0.036 | 0.000 ± 0.000 |
| | calibrated | -0.005 | 0.992 | 0.756 | 0.052 | 0.829 | | |
| RF, Friedman, 0.1 noise | uncalibrated | 0.037 | 0.775 | 0.394 | 0.101 | 0.926 | 0.394 | 0.105 |
| | calibrated | 0.047 | 0.979 | 1.001 | -0.005 | 0926 | | |
| RF, Friedman, 0.2 noise | uncalibrated | 0.020 | 0.914 | 0.292 | 0.187 | 0.914 | 0.290 | 0.194 |
| | calibrated | 0.008 | 0.977 | 1.008 | -0.009 | 0.914 | | |
| RF, Friedman, 0.3 noise | uncalibrated | 0.038 | 1.017 | 0.279 | 0.256 | 0.804 | 0.253 | 0.264 |
| | calibrated | 0.036 | 1.004 | 1.103 | -0.035 | 0.804 | | |
| RF, Friedman, 0.4 noise | uncalibrated | -0.072 | 0.978 | 0.188 | 0.347 | 0.750 | 0.202 | 0.369 |
| | calibrated | -0.068 | 0.936 | 0.928 | 0.005 | 0.750 | | |
| RF, Friedman, 0.5 noise | uncalibrated | 0.039 | 1.081 | 0.086 | 0.451 | 0.166 | 0.173 | 0.405 |
| | calibrated | 0.026 | 1.009 | 0.495 | 0.250 | 0.166 | | |
| RF, Friedman, 1.0 noise | uncalibrated | 0.046 | 1.137 | 0.056 | 0.702 | 0.115 | 0.017 | 0.744 |
| | calibrated | 0.036 | 0.980 | 3.316 | -1.764 | 0.115 | | |
| RF, Friedman, 2.0 noise | uncalibrated | -0.030 | 1.107 | 0.193 | 0.789 | 0.577 | -0.021 | 0.978 |
| | calibrated | -0.011 | 1.006 | -9.036 | 9.622 | 0.577 | | |
| NN, Diffusion | uncalibrated | -0.101 | 2.309 | 1.53 | 0.10 | 0.88 | 1.874 ± 0.216 | 0.020 ± 0.014 |
| | calibrated | -0.046 | 1.040 | 0.79 | 0.07 | 0.91 | | |
| RF, Bulk modulus | uncalibrated | 0.026 | 1.102 | 0.78 | 0.11 | 0.88 | 0.700 ± 0.084 | 0.041 ± 0.013 |
| | calibrated | 0.023 | 1.034 | 1.06 | -0.01 | 0.90 | | |
| RF, Double perov. gap | uncalibrated | 0.007 | 0.774 | 0.72 | 0.01 | 0.99 | 0.747 ± 0.039 | 0.003 ± 0.015 |
| | calibrated | 0.009 | 1.026 | 0.96 | 0.01 | 0.98 | | |
| RF, Heusler | uncalibrated | 0.008 | 1.040 | 1.36 | -0.14 | 0.93 | 1.073 ± 0.059 | -9.346 ± 7.802 |
| | calibrated | 0.008 | 1.042 | 1.25 | -0.09 | 0.93 | | |
| RF, Piezo | uncalibrated | 0.018 | 1.356 | 0.57 | 0.38 | 0.62 | 0.3640 ± 0.0604 | 0.3978 ± 0.0389 |
| | calibrated | -0.003 | 1.016 | 1.24 | -0.20 | 0.42 | | |
| RF, Steel strength | uncalibrated | 0.005 | 1.118 | 0.93 | 0.06 | 0.93 | 0.905 ± 0.068 | 13.007 ± 6.412 |
| | calibrated | 0.003 | 1.069 | 1.03 | 0.01 | 0.94 | | |
| RF, Super-conductor | uncalibrated | 0.010 | 1.051 | 0.85 | 0.04 | 1.00 | 0.852 ± 0.019 | 0.025 ± 0.004 |
| | calibrated | 0.007 | 1.005 | 1.00 | 0.01 | 1.00 | | |
| RF, Thermal conduct. | uncalibrated | -0.082 | 2.338 | 0.86 | 0.21 | 0.94 | 1.0419 ± 0.1251 | 0.0794 ± 0.0161 |
| | calibrated | -0.032 | 1.058 | 0.81 | 0.10 | 0.93 | | |

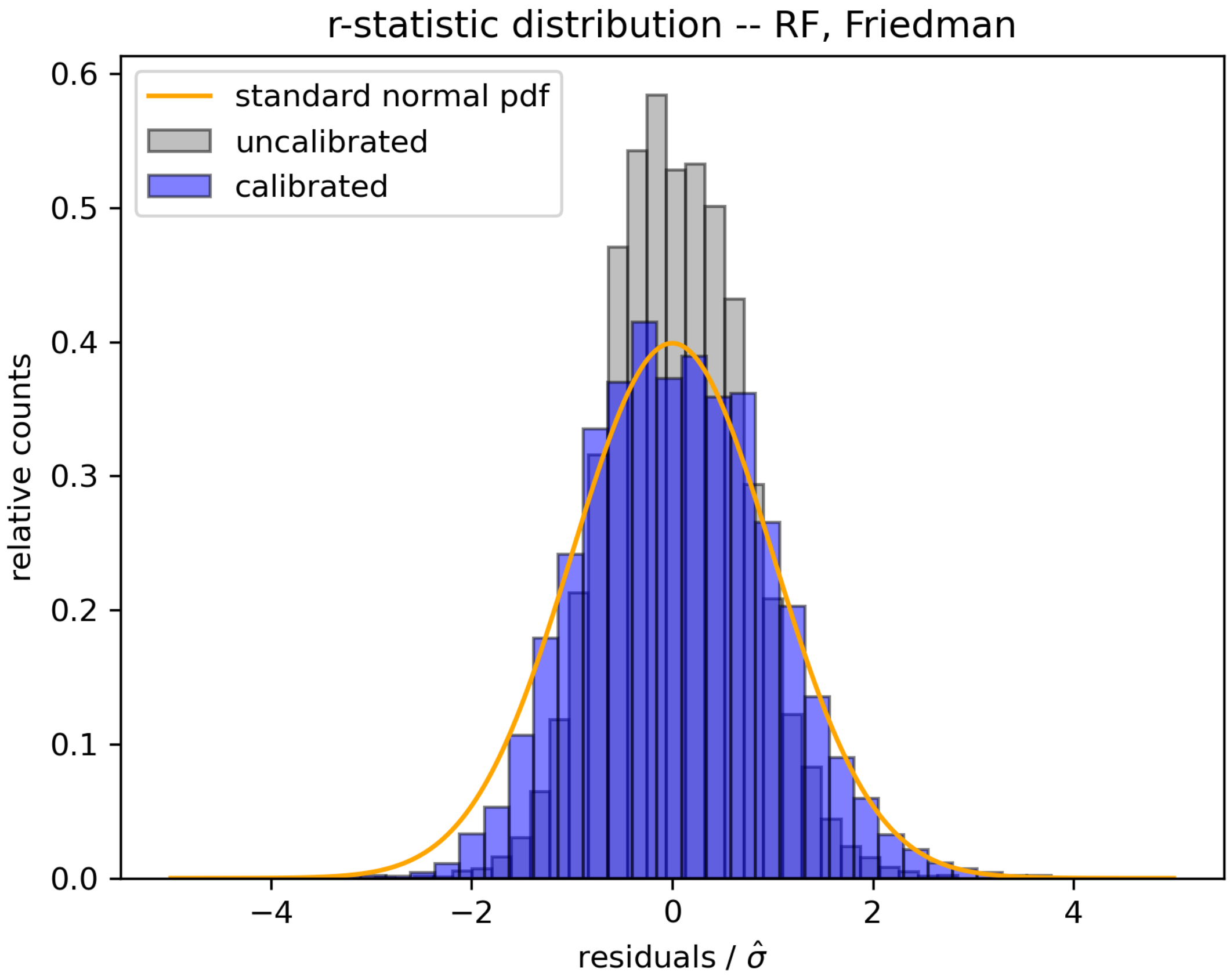

*Figure 1: Friedman random forest r-statistic*

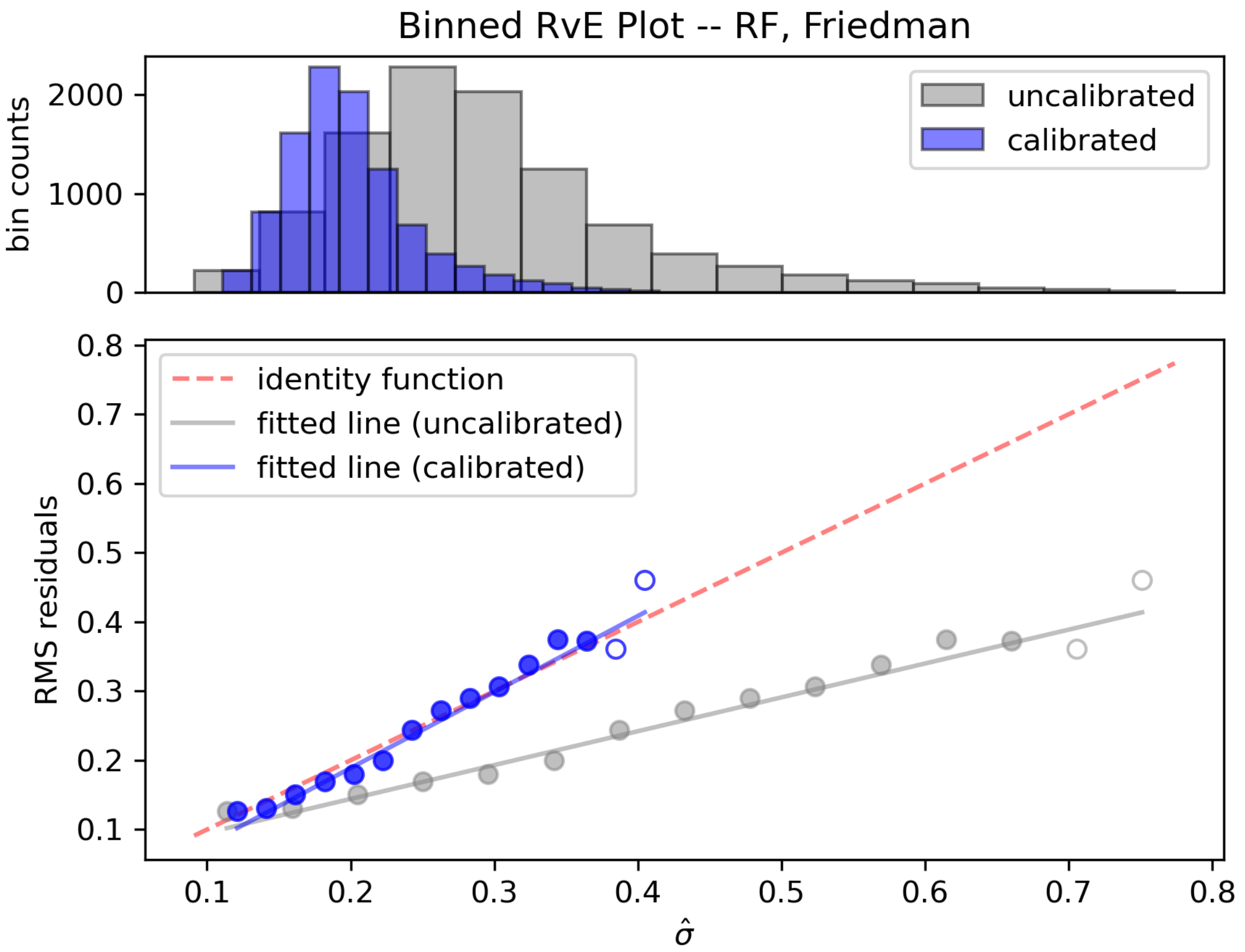

*Figure 2: Friedman random forest RvE*

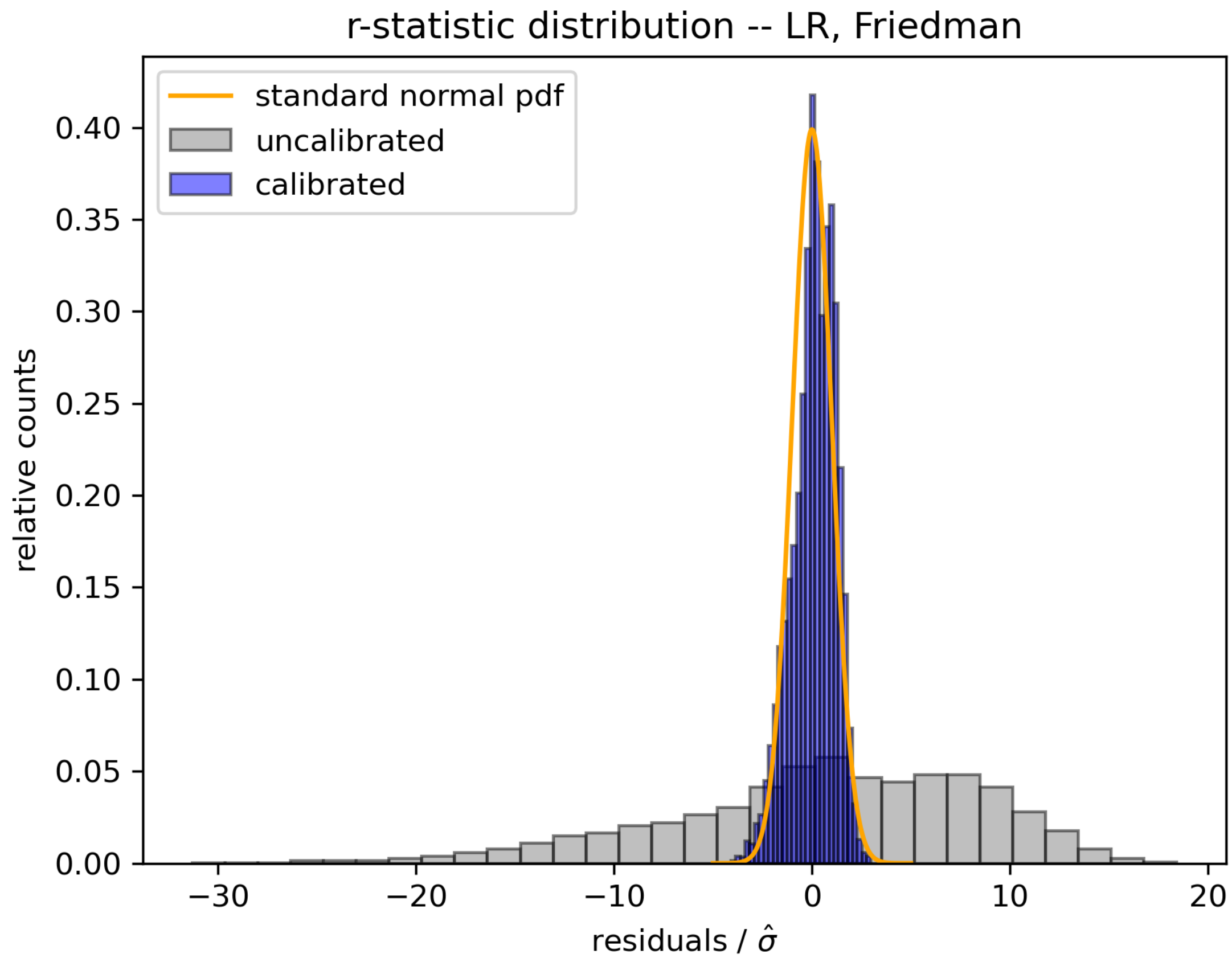

*Figure 3: Friedman linear ridge regression r-statistic*

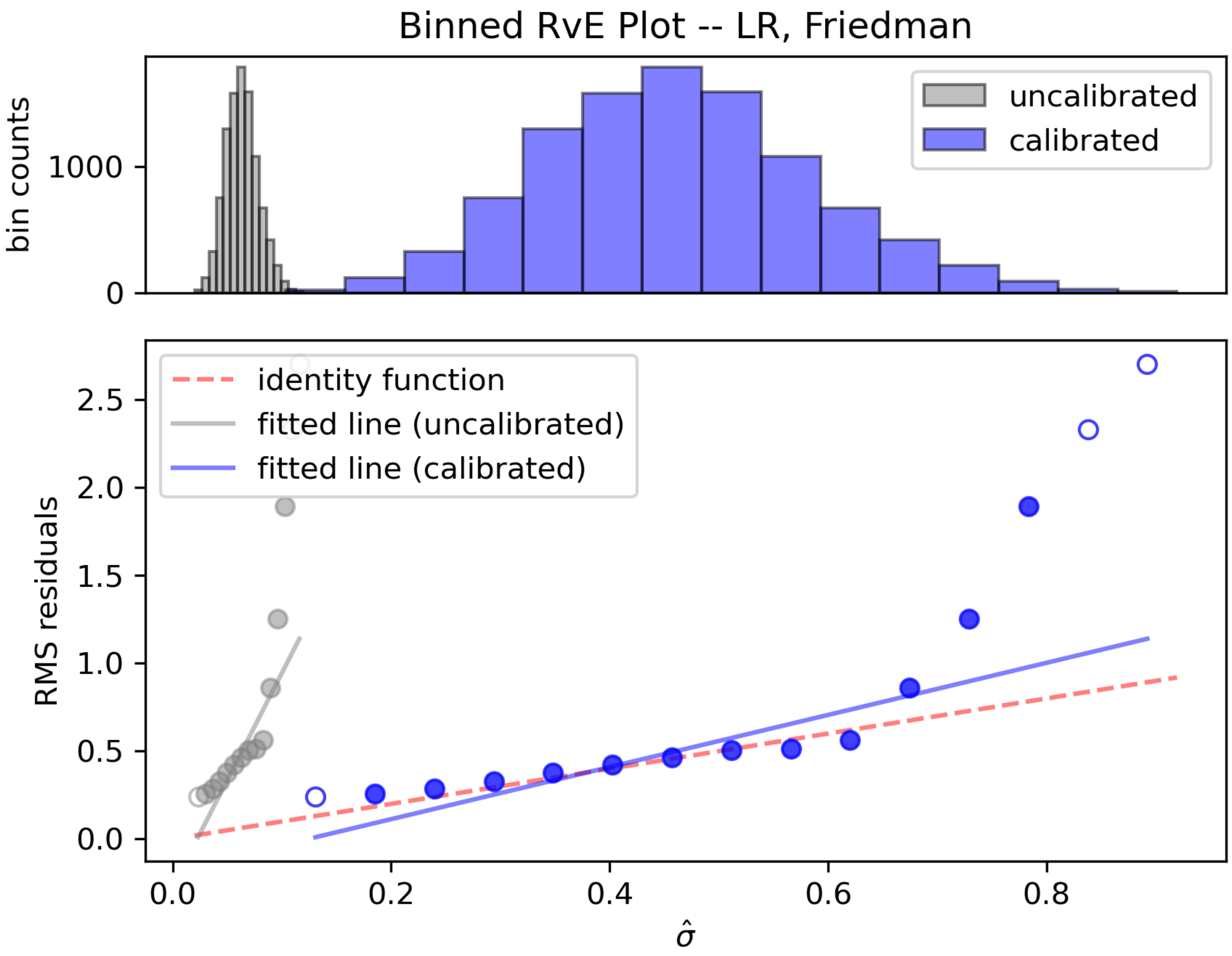

*Figure 4: Friedman linear ridge regression RvE*

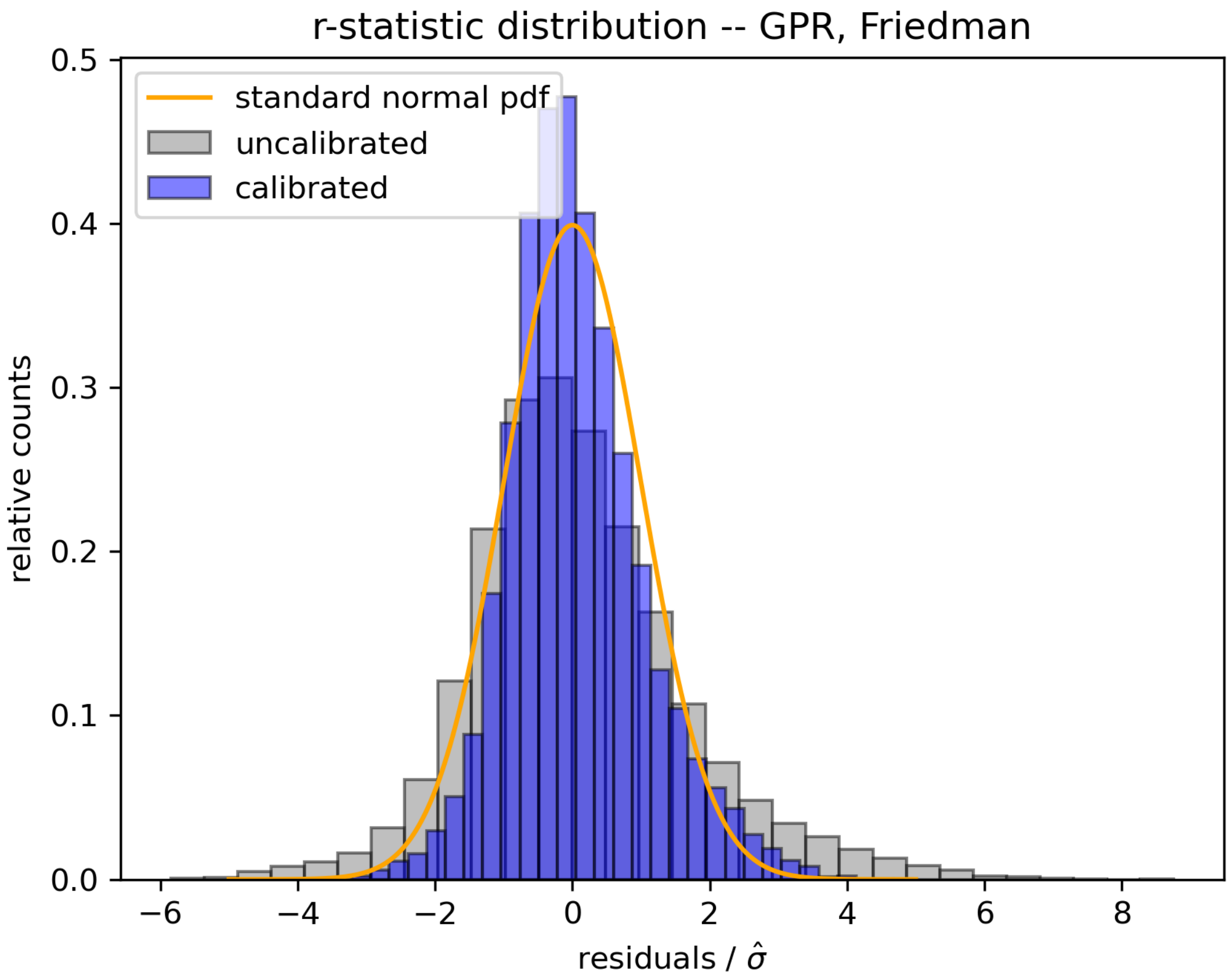

*Figure 5: Friedman GPR r-statistic*

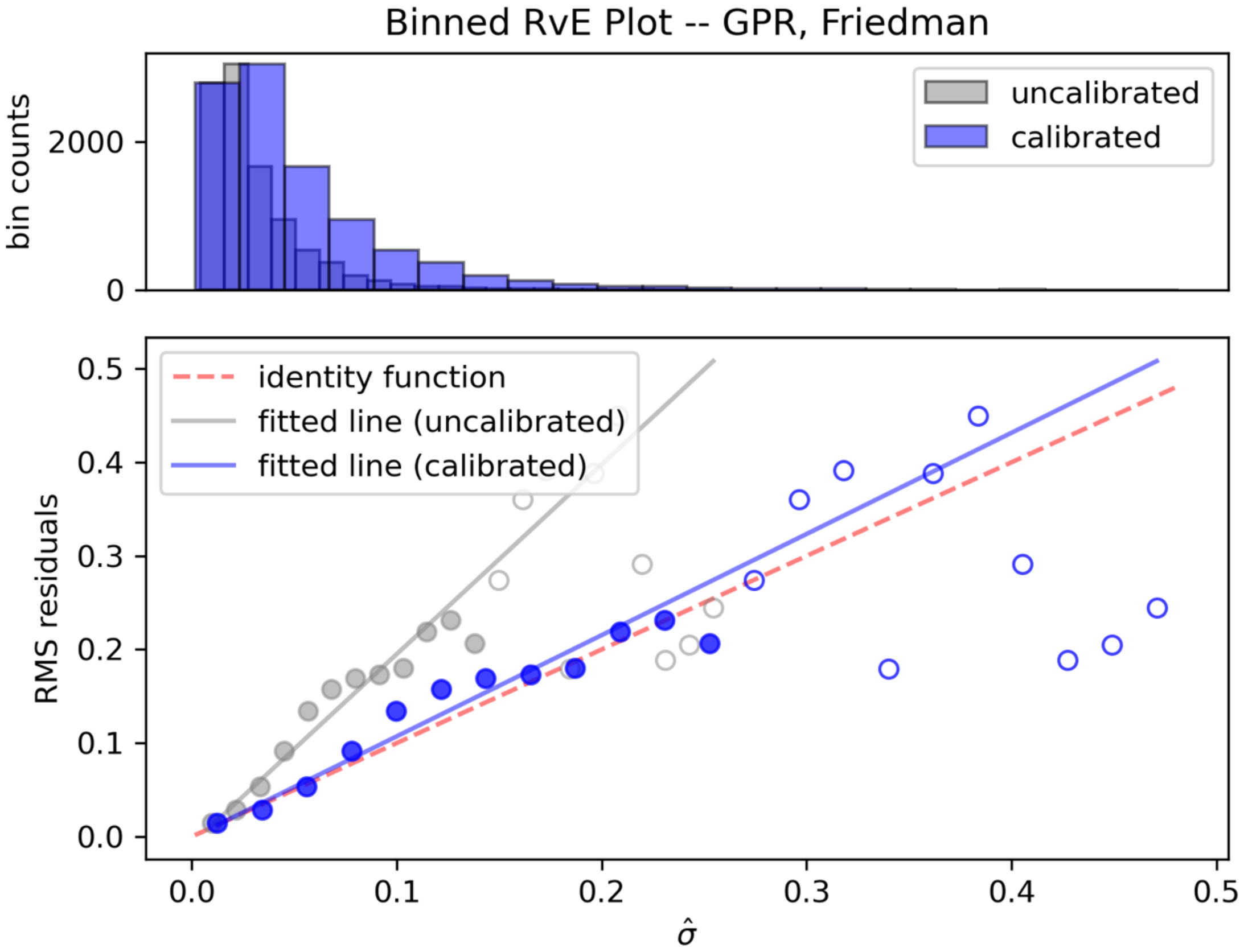

*Figure 6: Friedman GPR RvE*

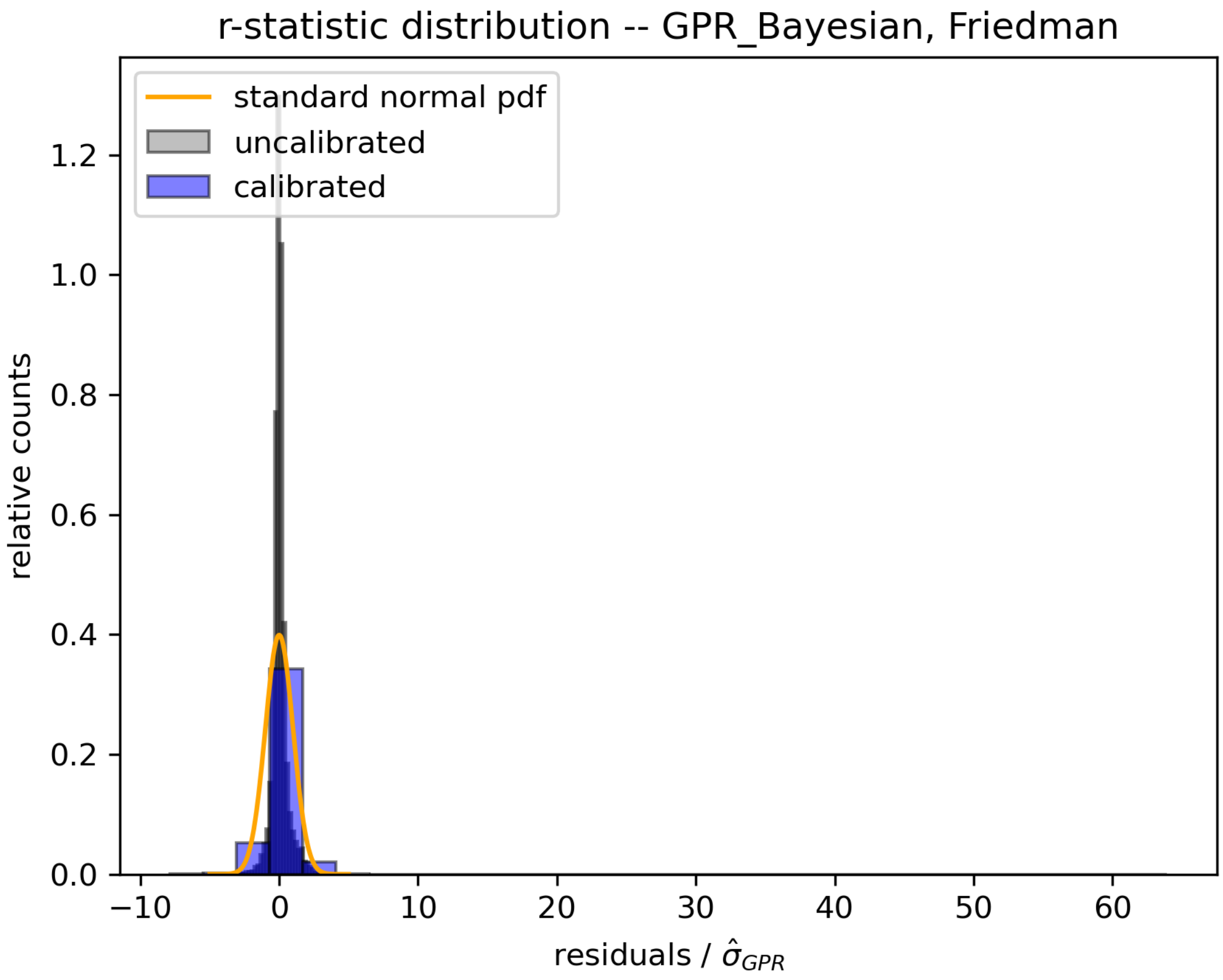

*Figure 7: Friedman Bayesian GPR r-statistic*

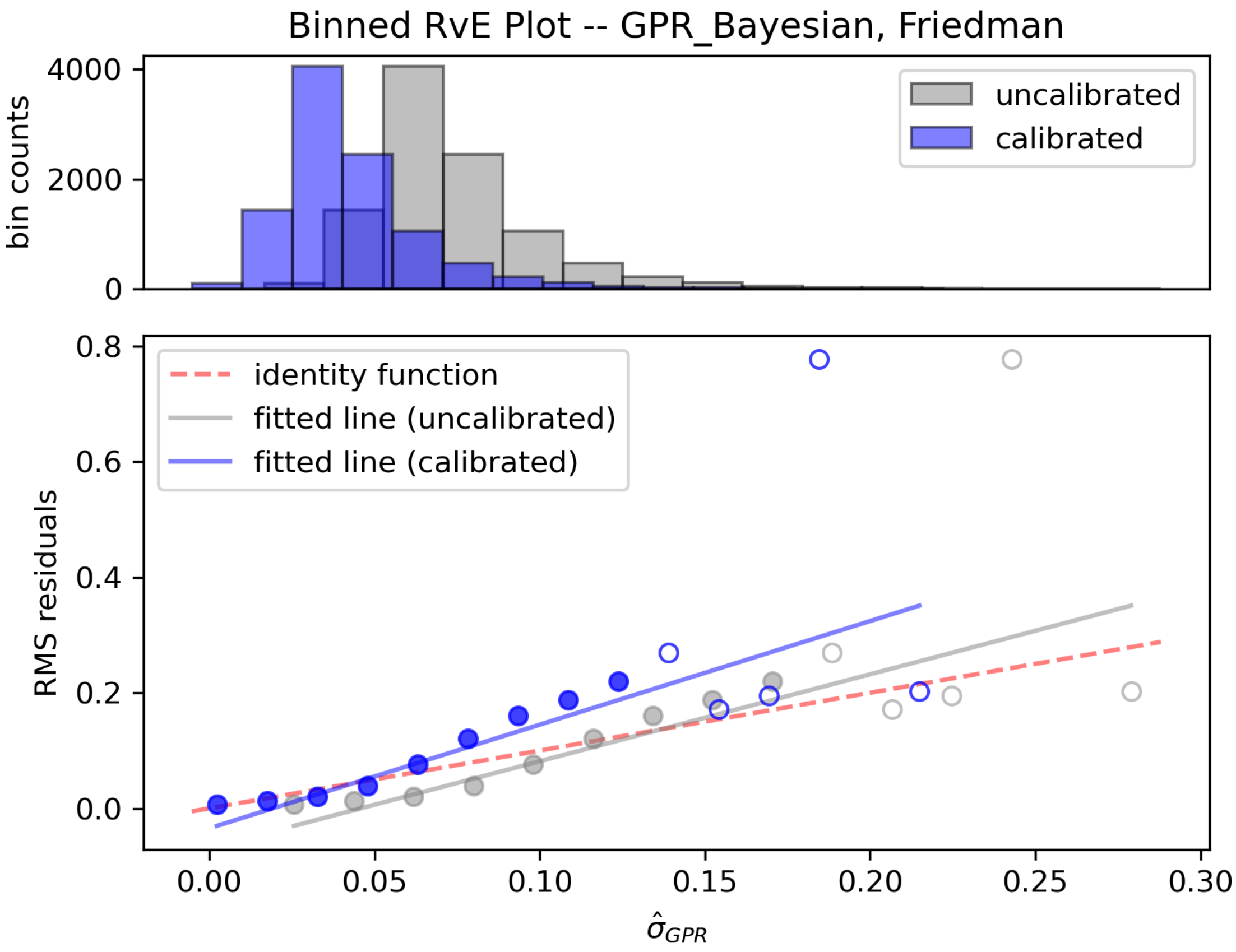

*Figure 8: Friedman Bayesian GPR RvE*

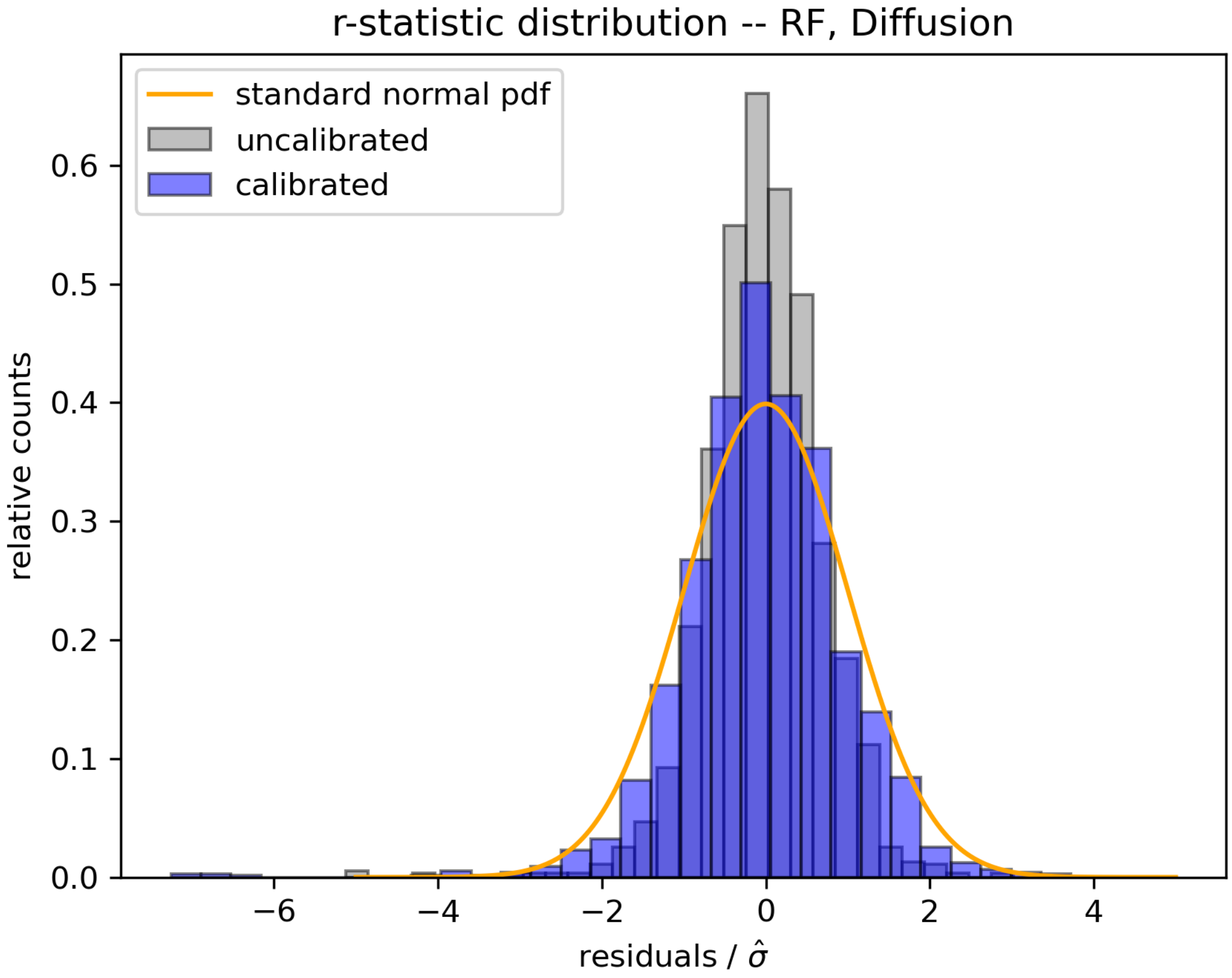

*Figure 9: Diffusion random forest r-statistic*

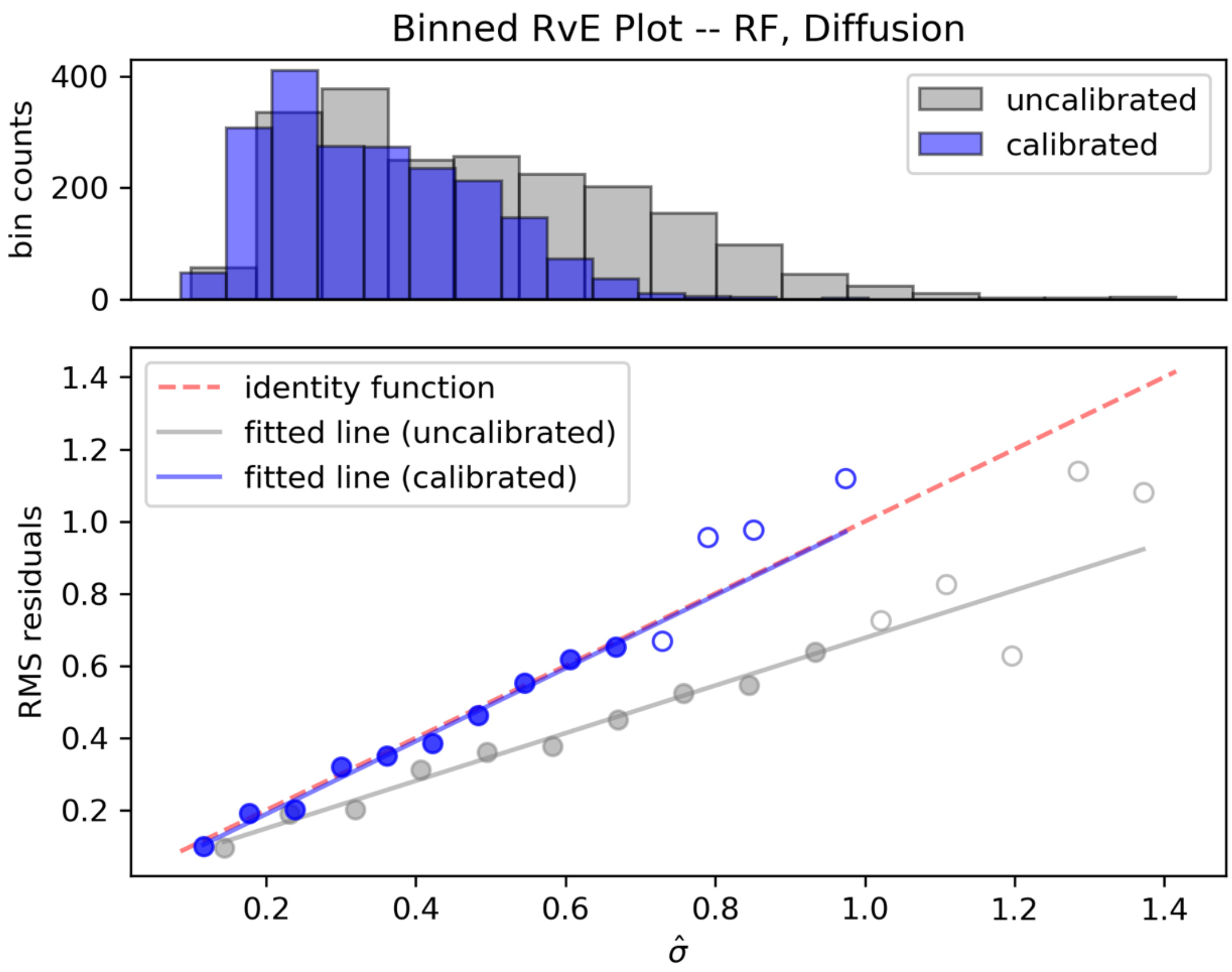

*Figure 10: Diffusion random forest RvE*

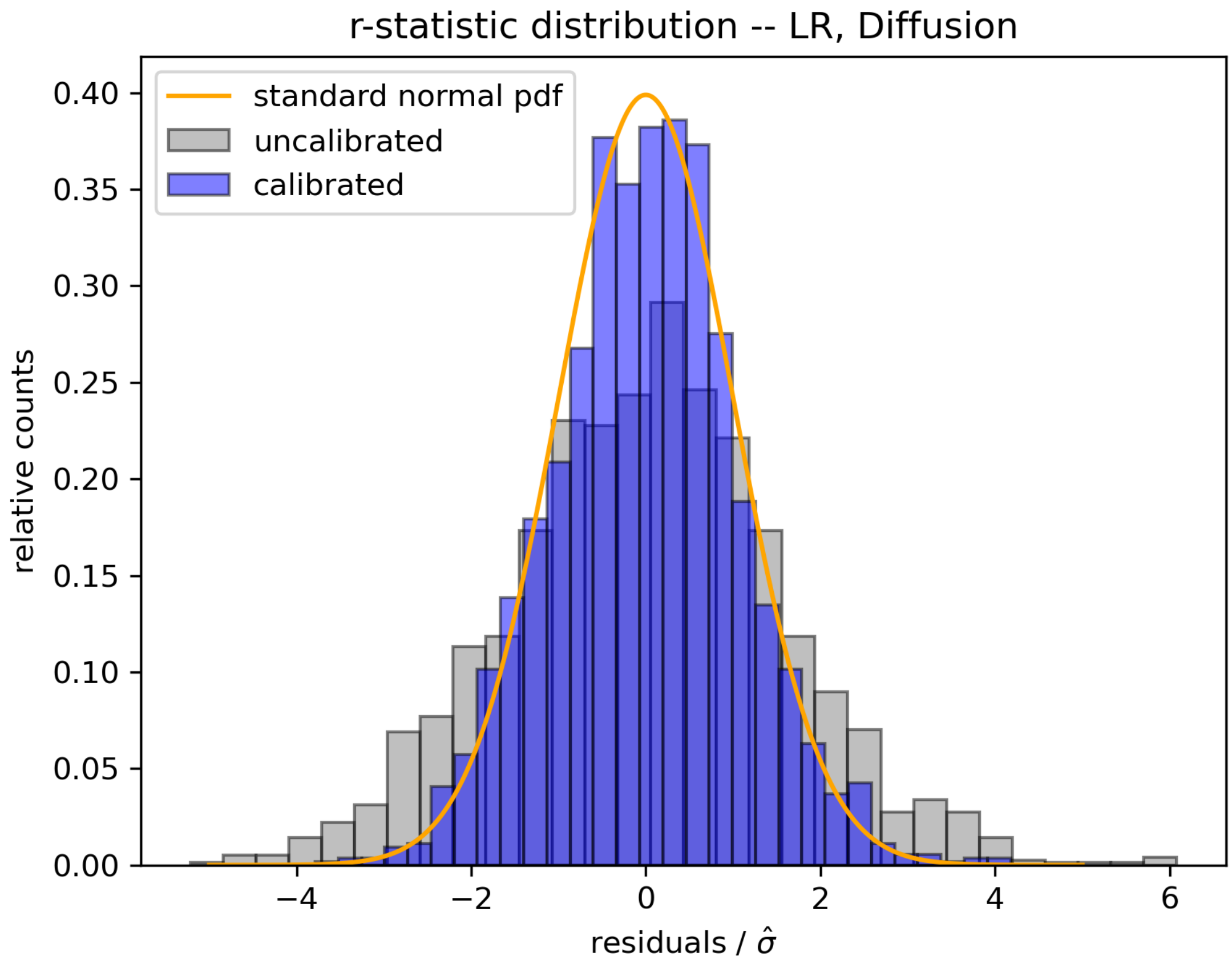

*Figure 11: Diffusion linear ridge regression r-statistic*

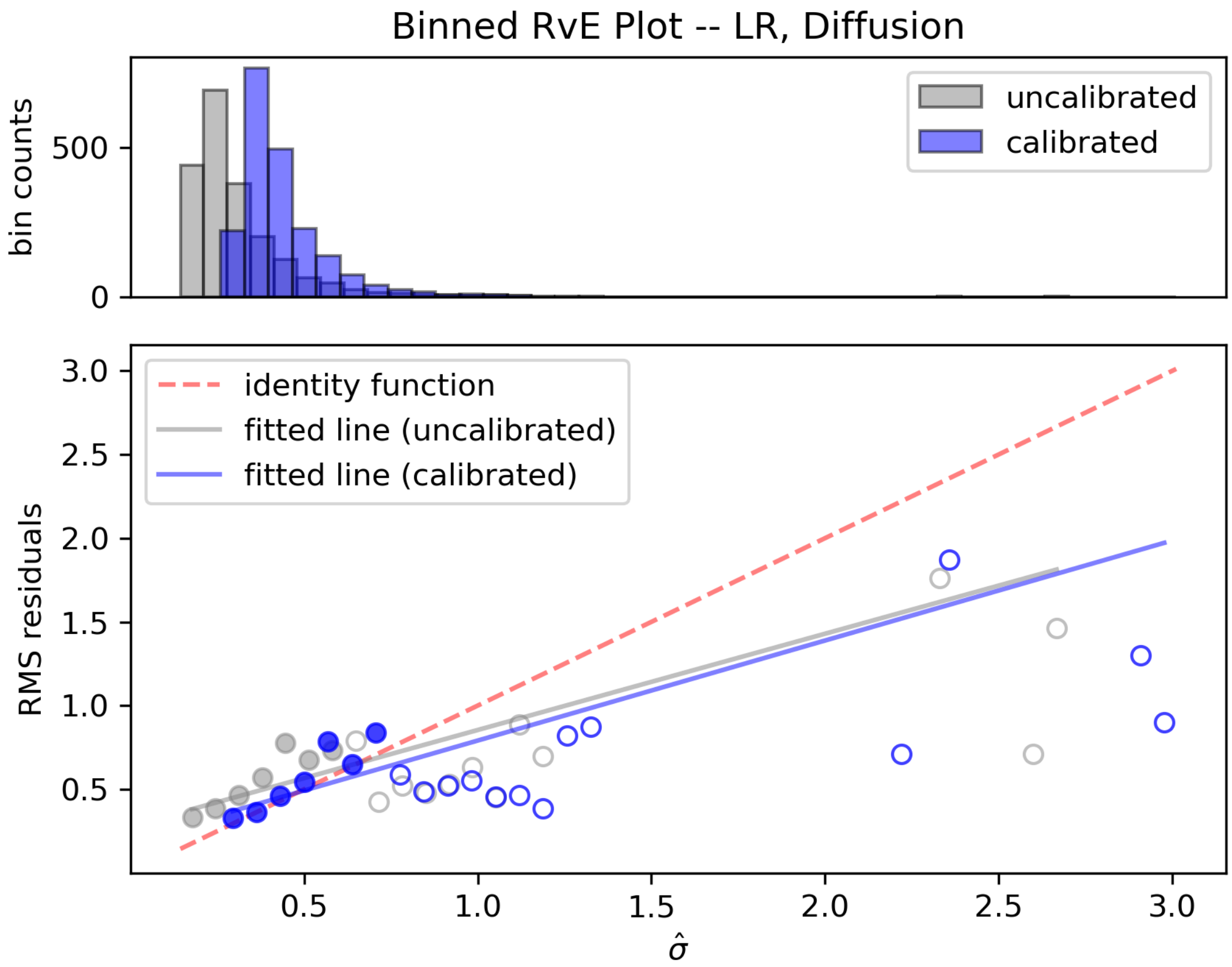

*Figure 12: Diffusion linear ridge regression RvE*

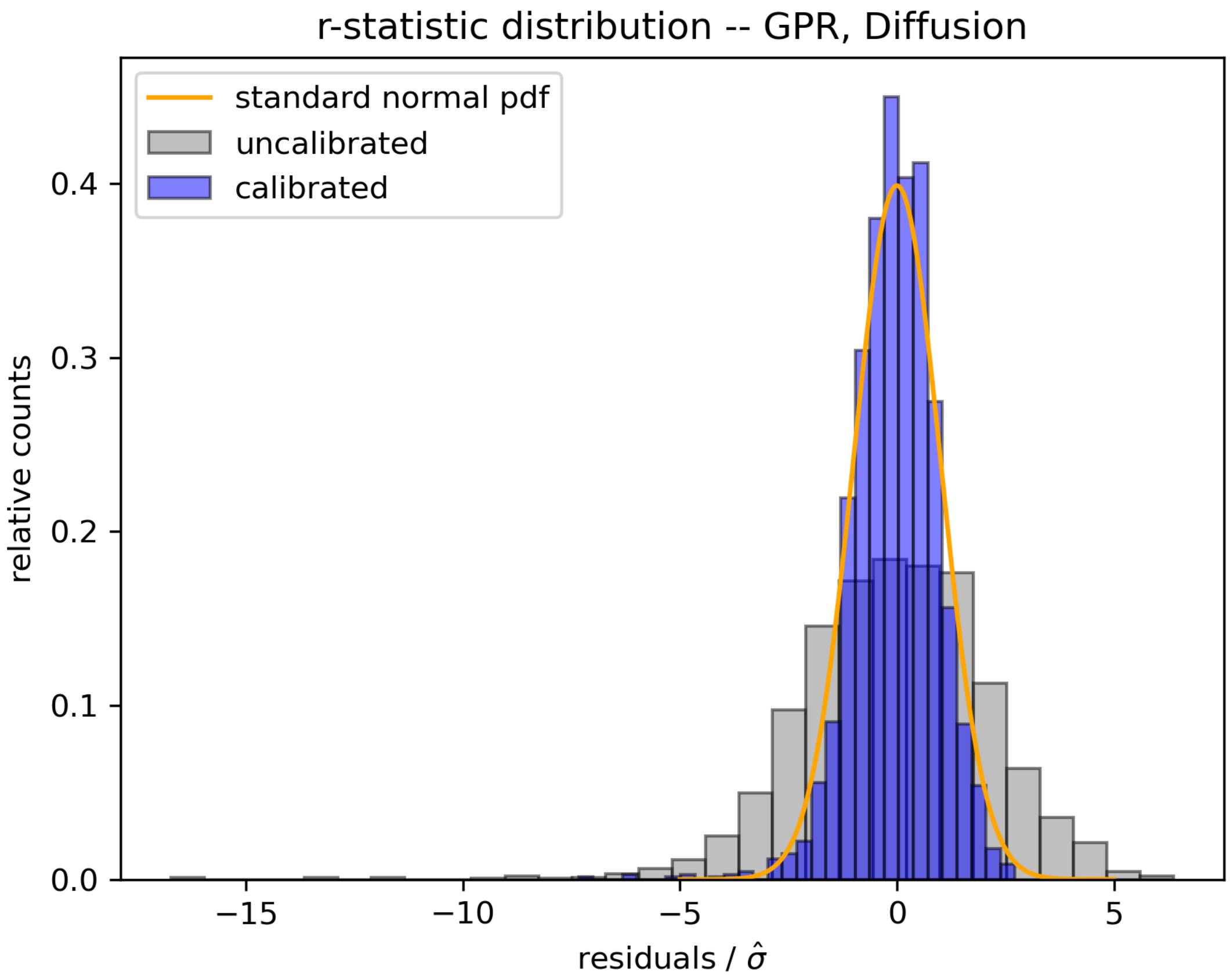

*Figure 13: Diffusion GPR r-statistic*

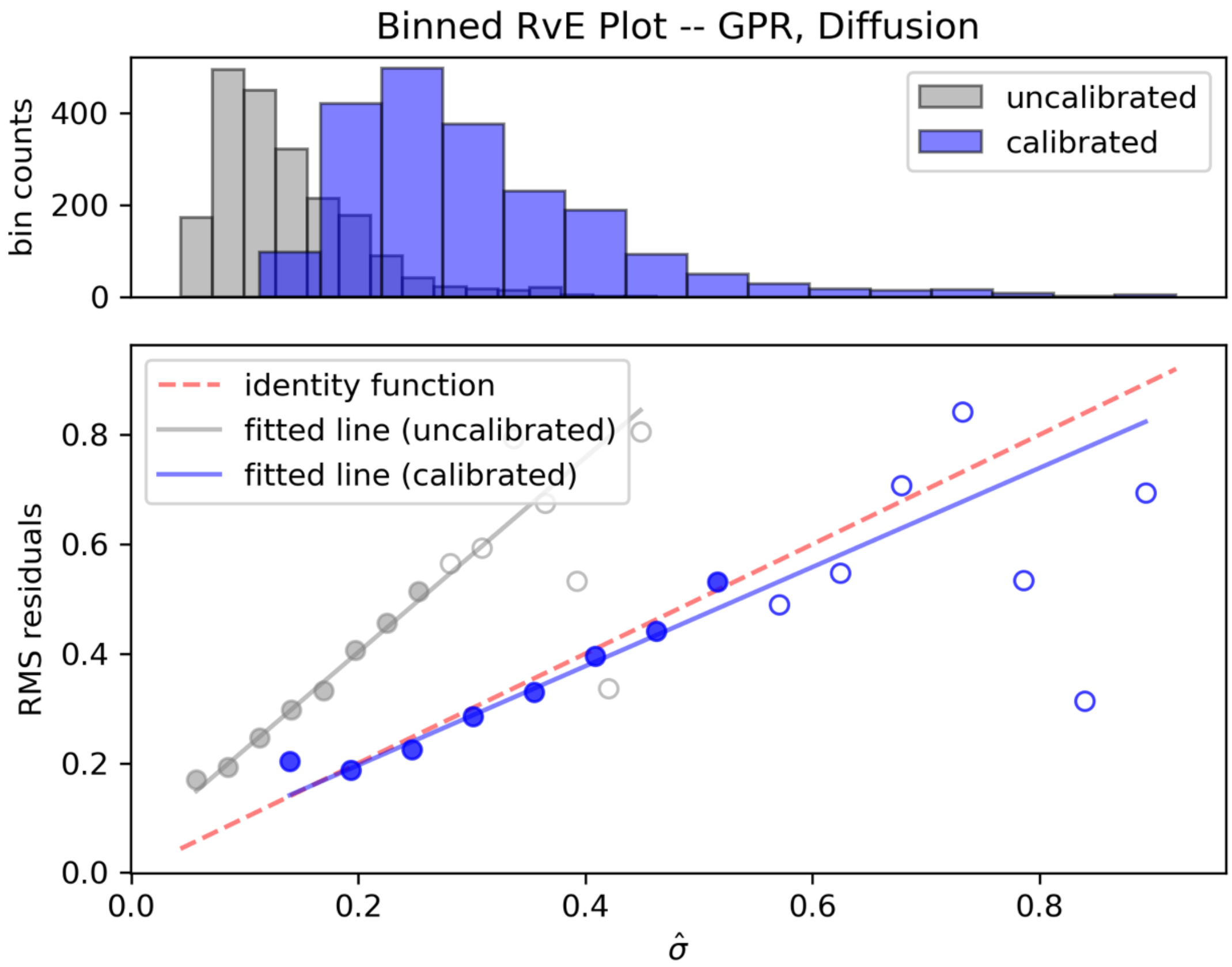

*Figure 14: Diffusion GPR RvE*

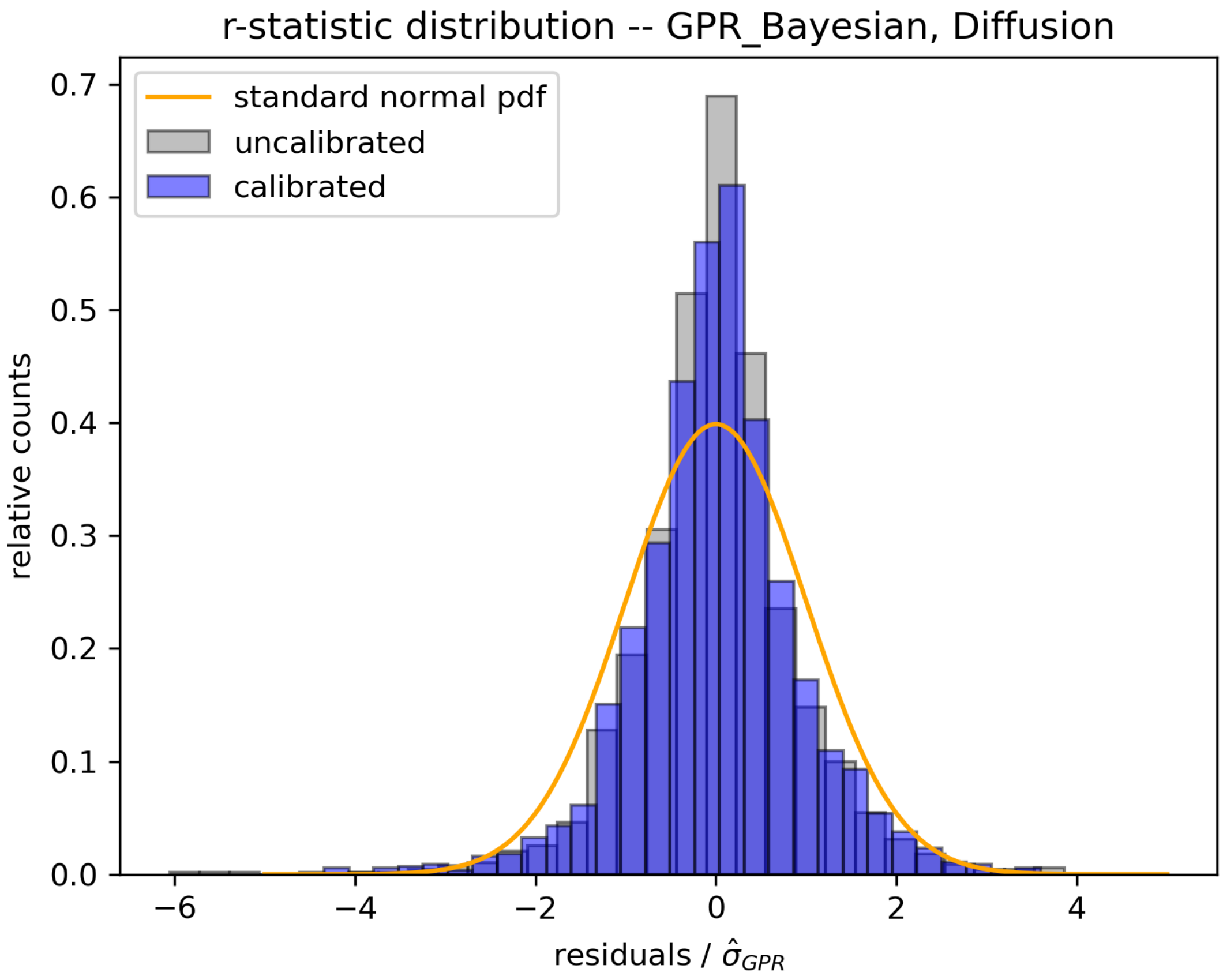

*Figure 15: Diffusion Bayesian GPR r-statistic*

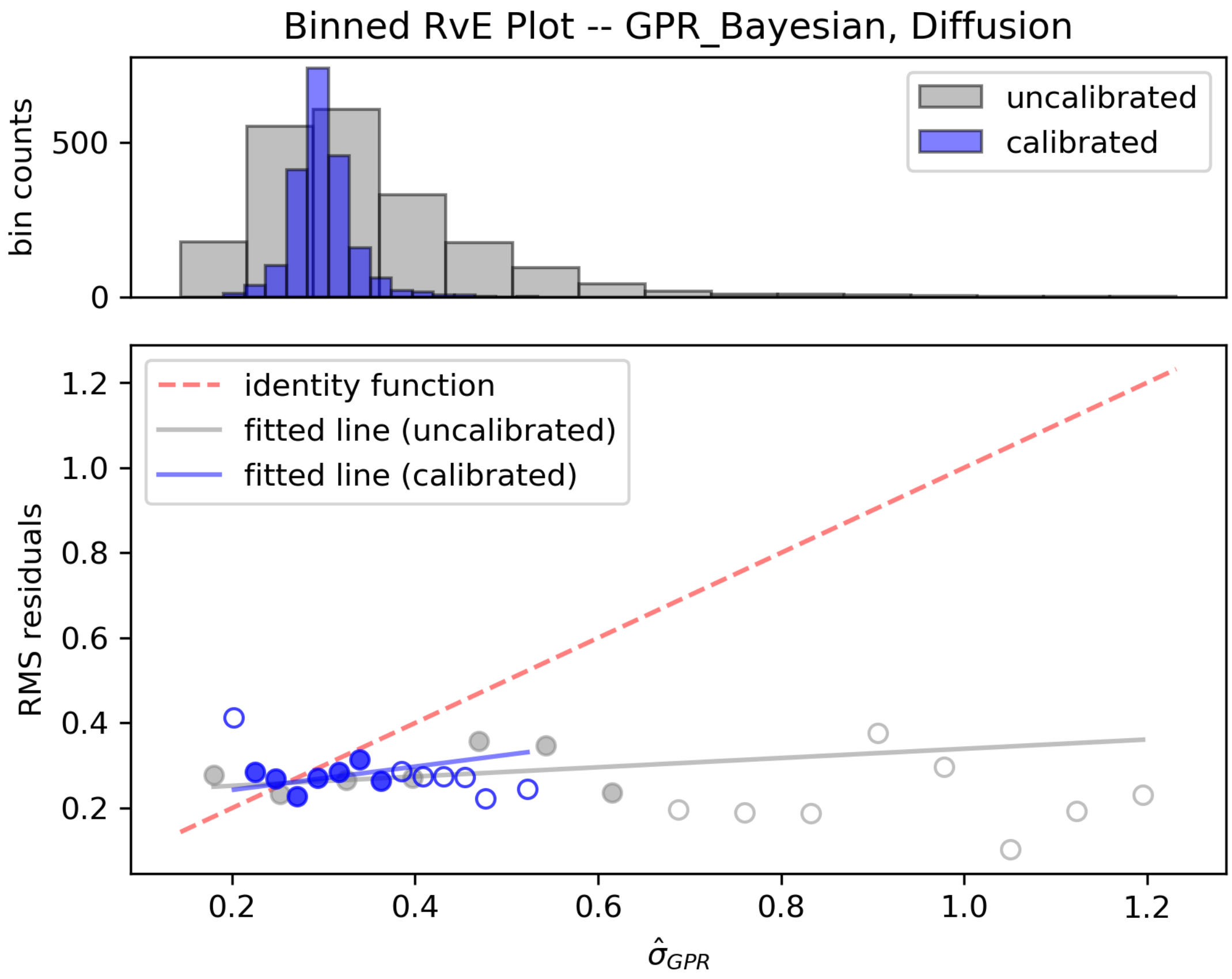

*Figure 16: Diffusion Bayesian GPR RvE*

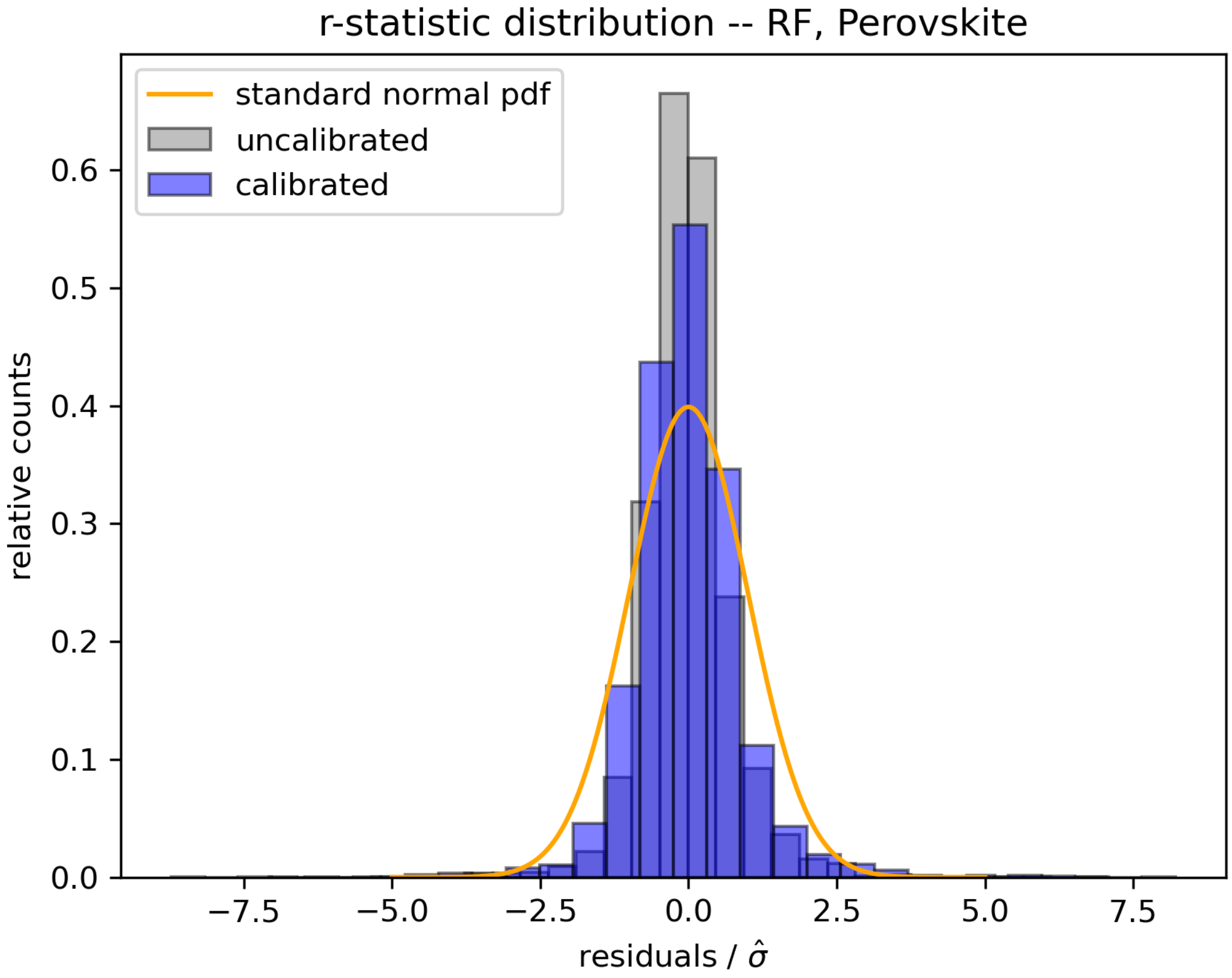

*Figure 17: Perovskite random forest r-statistic*

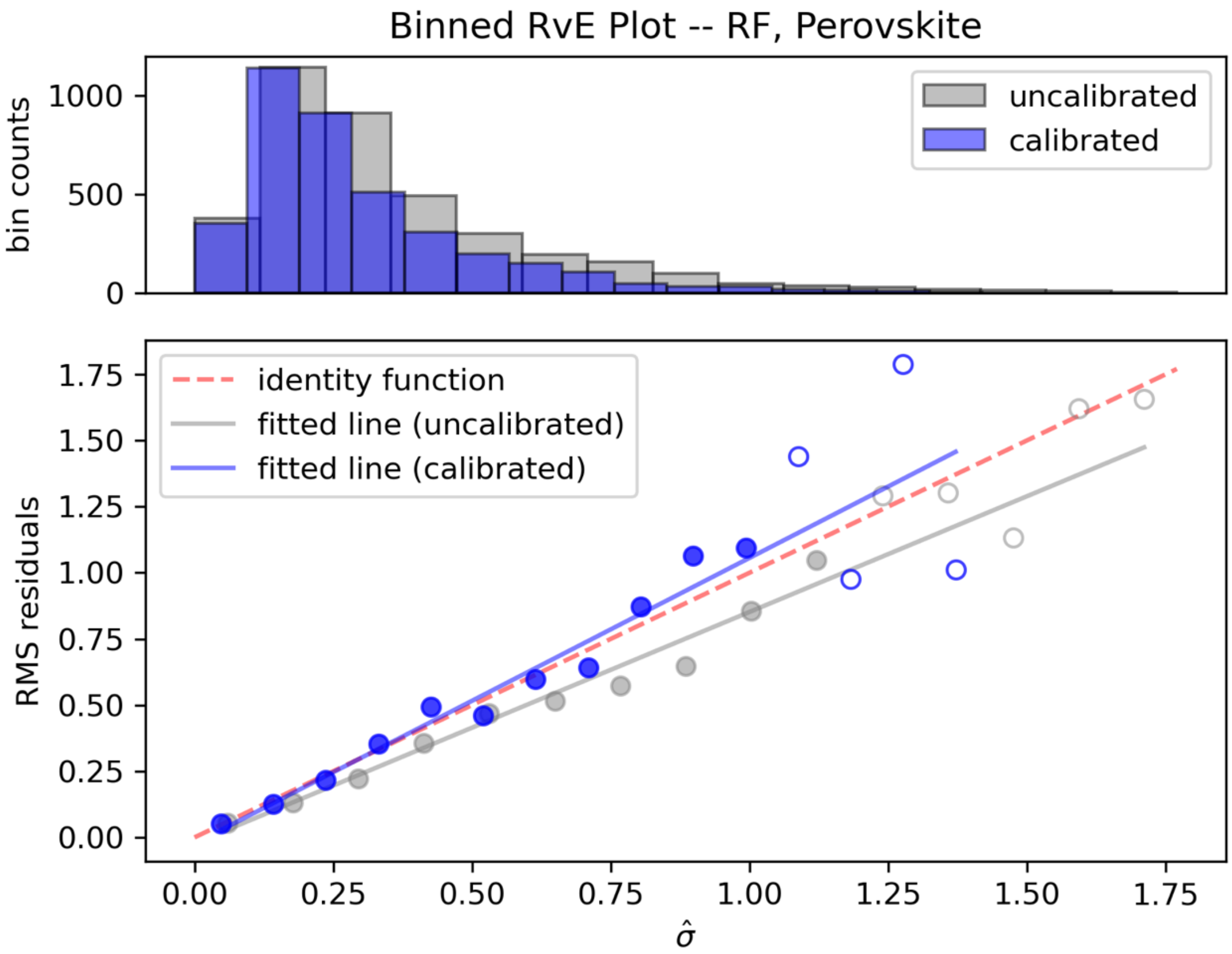

*Figure 18: Perovskite random forest RvE*

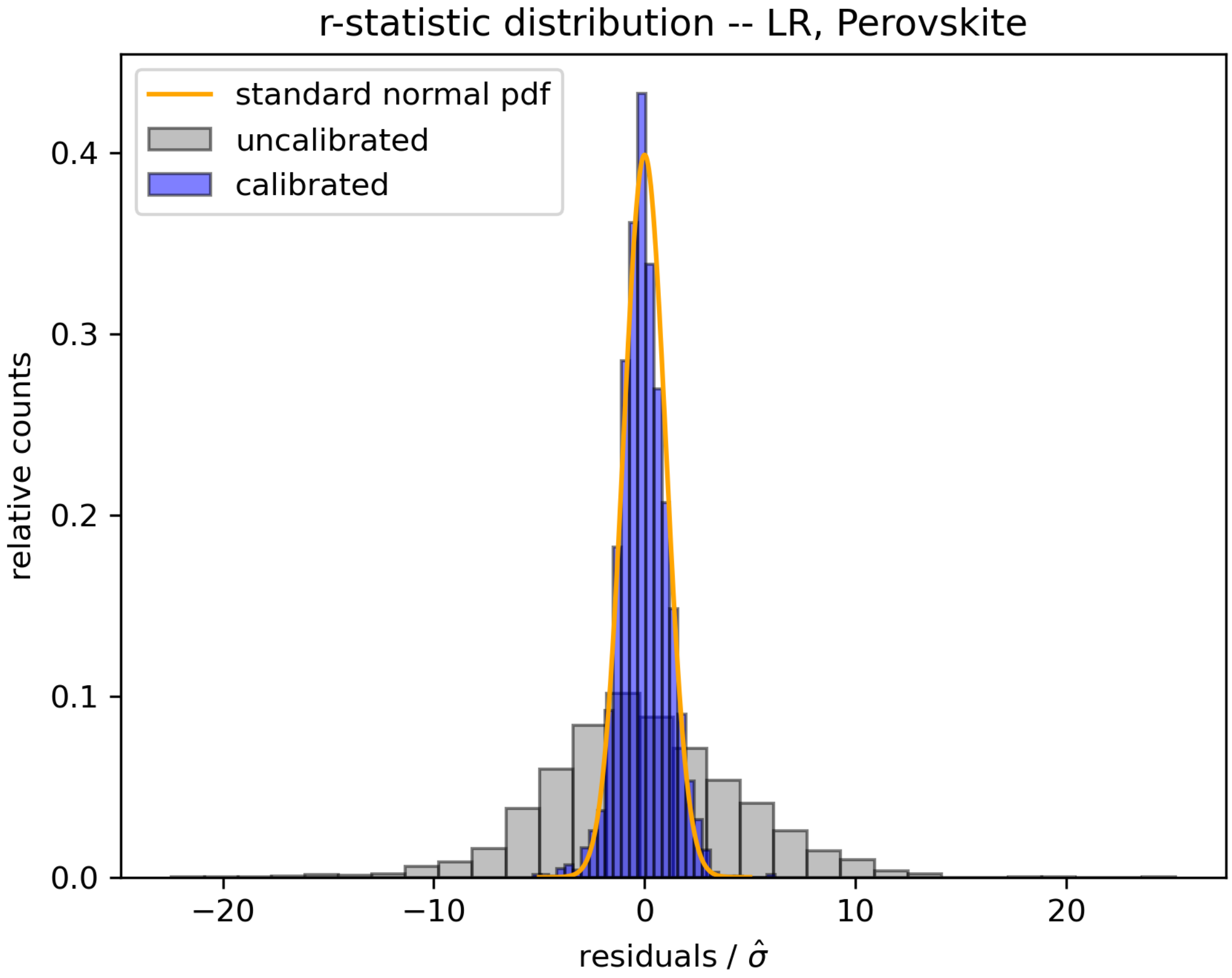

*Figure 19: Perovskite linear ridge regression r-statistic*

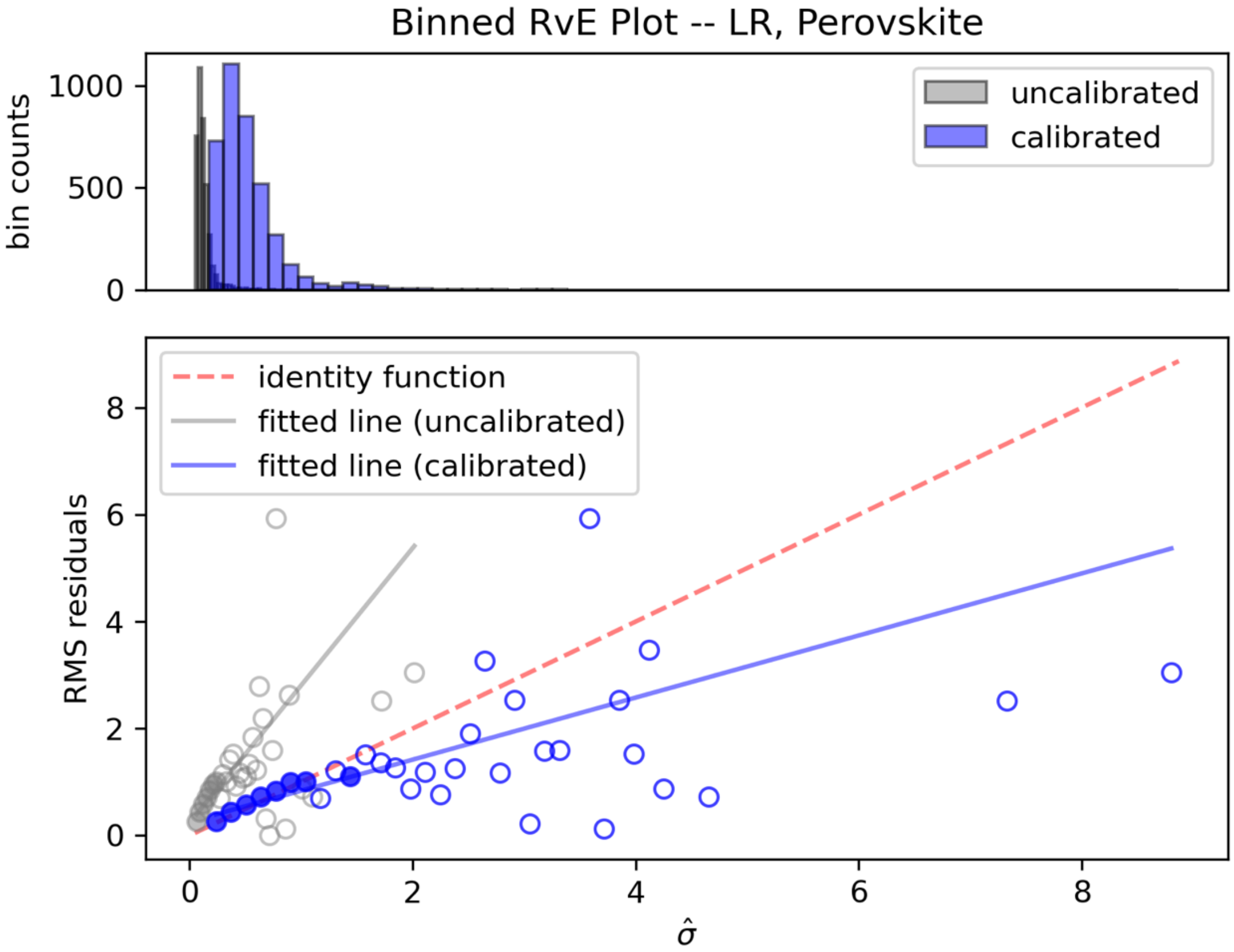

*Figure 20: Perovskite linear ridge regression RvE*

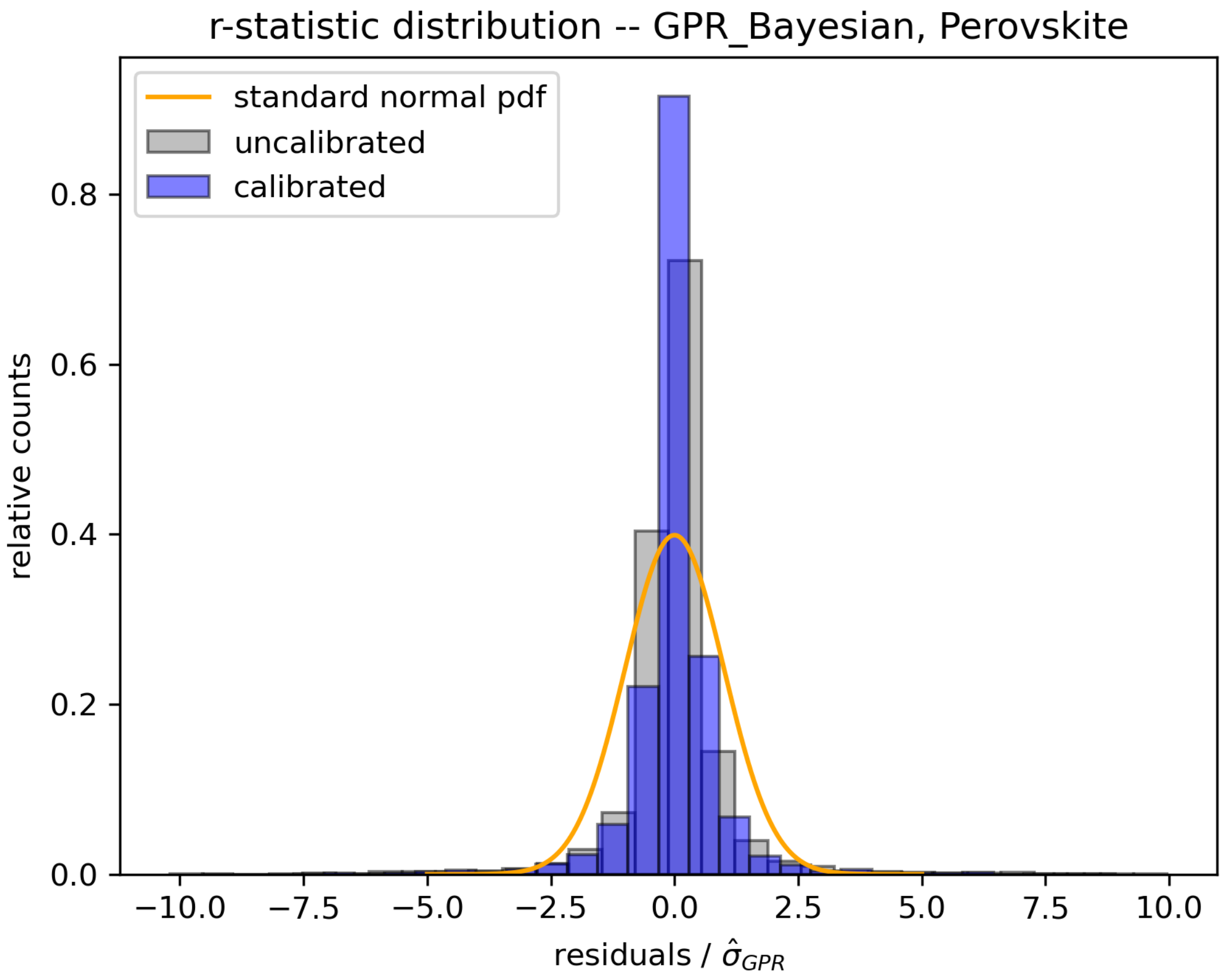

*Figure 21: Perovskite Bayesian GPR r-statistic*

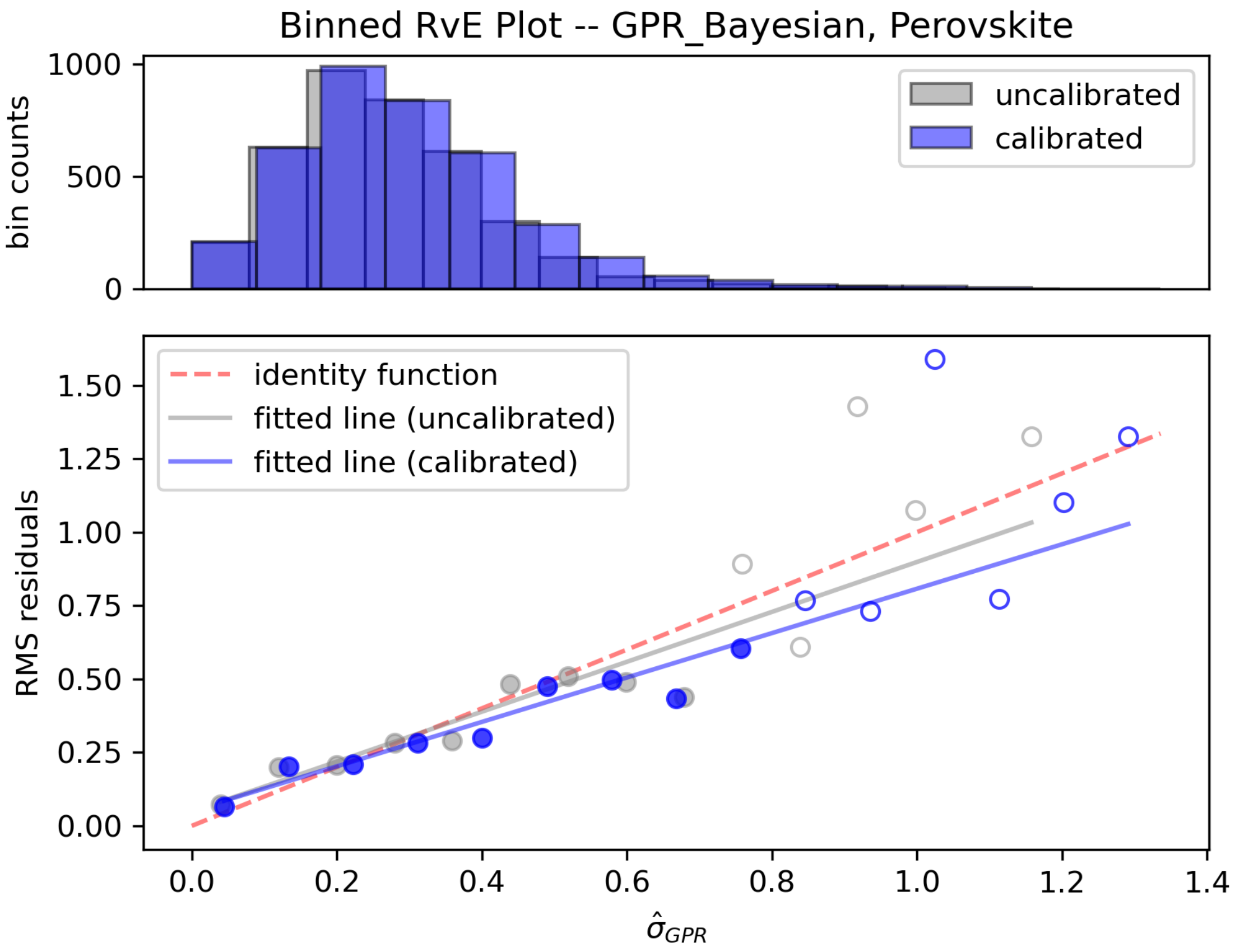

*Figure 22: Perovskite Bayesian GPR RvE*

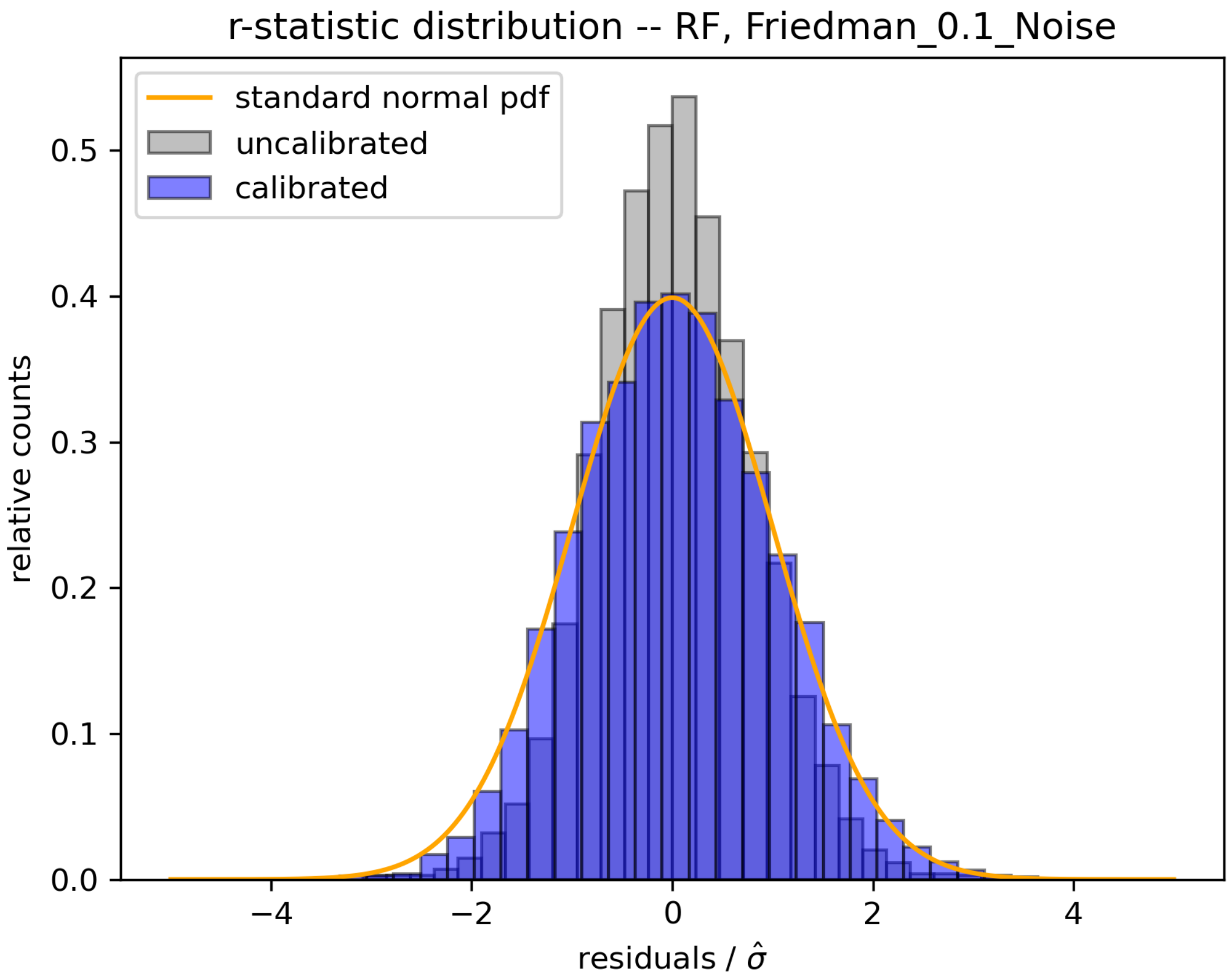

*Figure 23: Friedman with 0.1 sigma gaussian noise r-statistic*

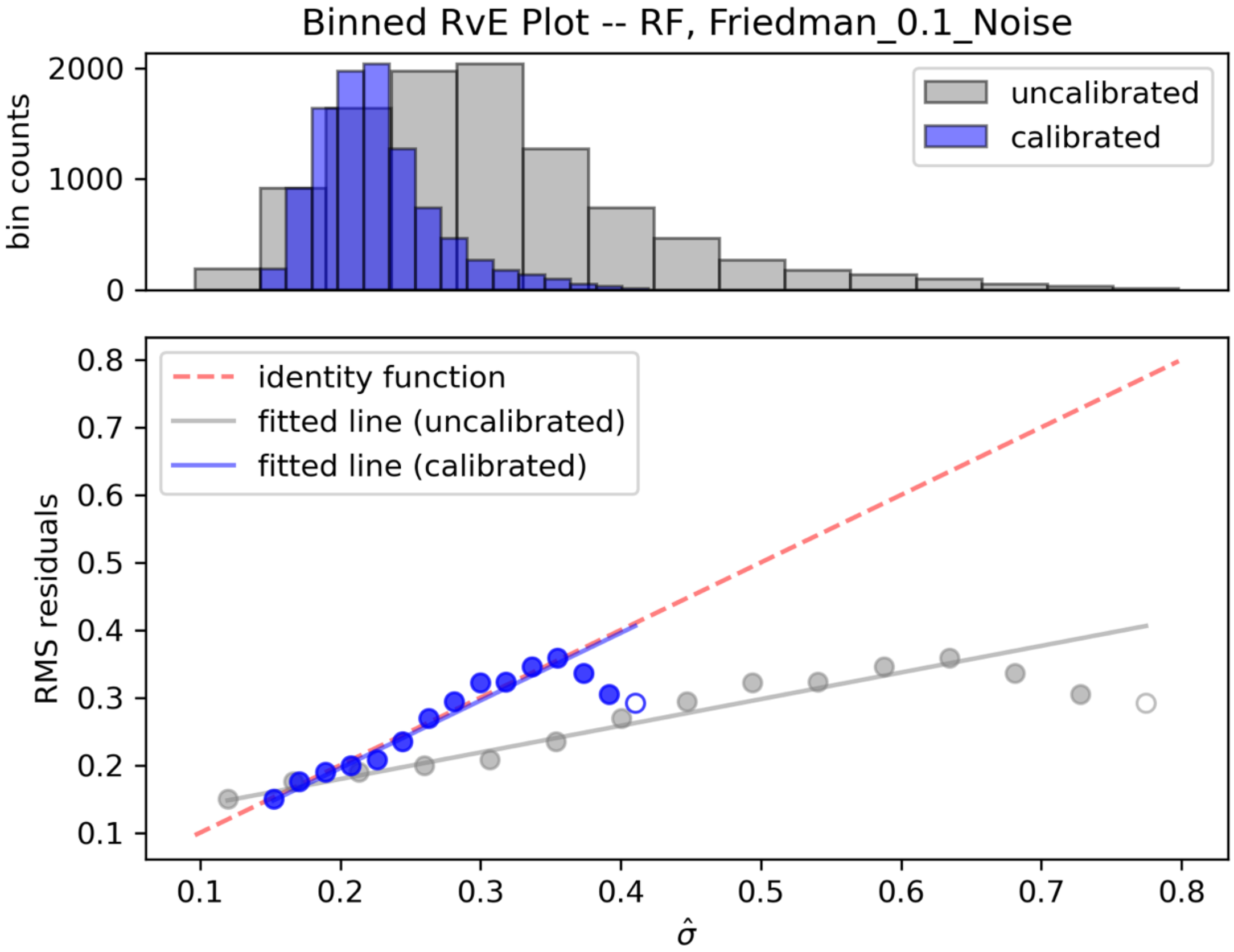

*Figure 24: Friedman with 0.1 sigma gaussian noise RvE*

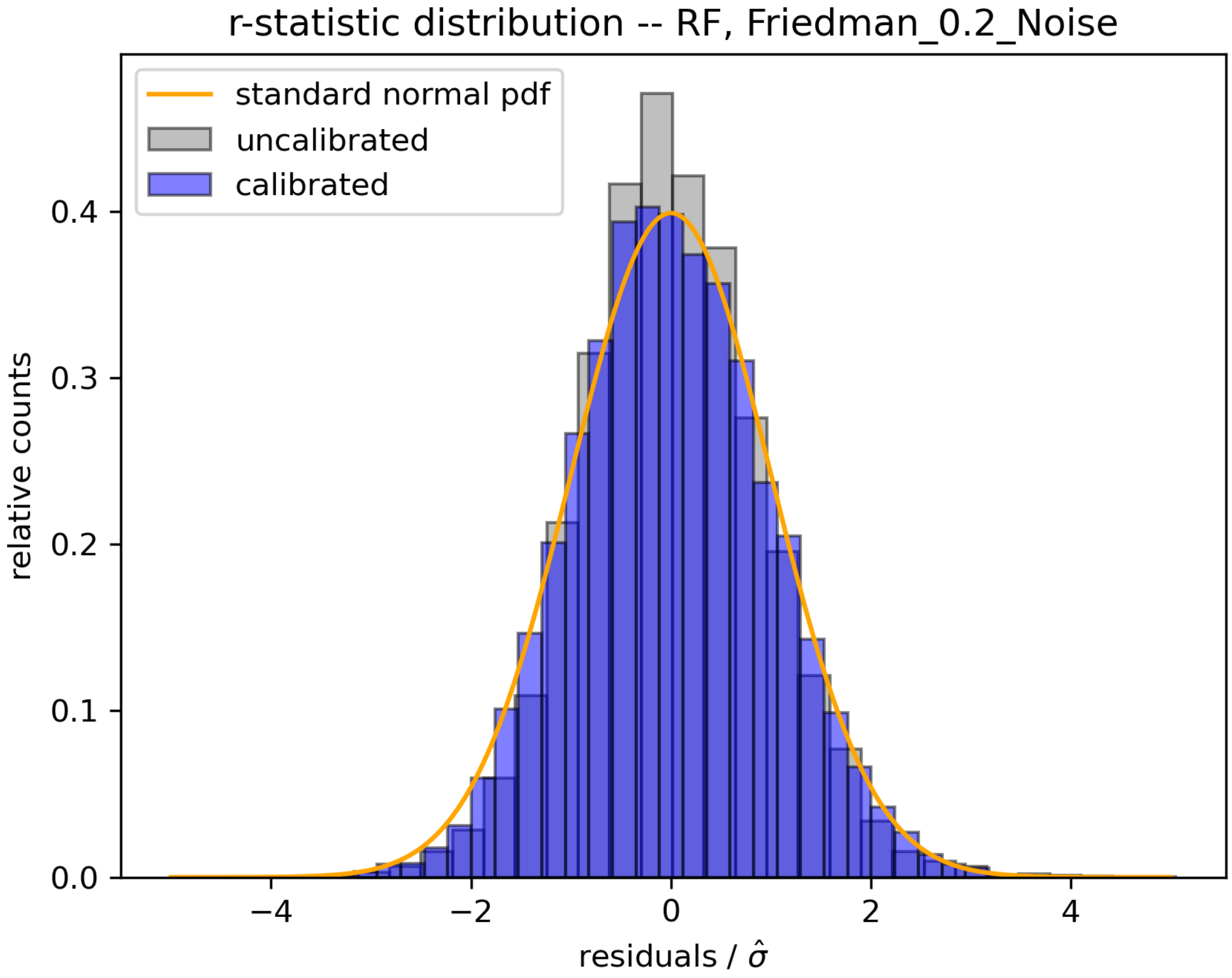

*Figure 25: Friedman with 0.2 sigma gaussian noise r-statistic*

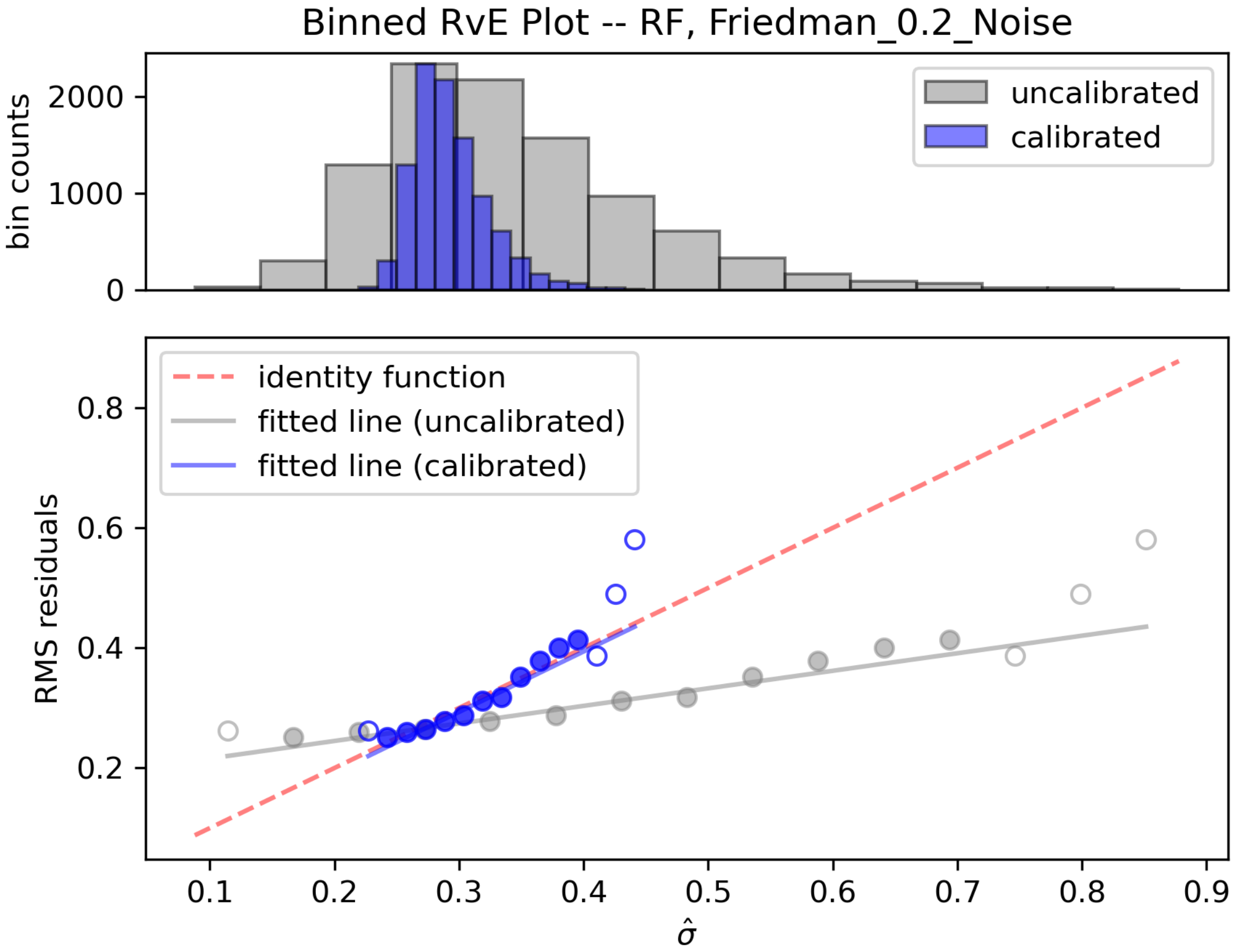

*Figure 26: Friedman with 0.2 sigma gaussian noise RvE*

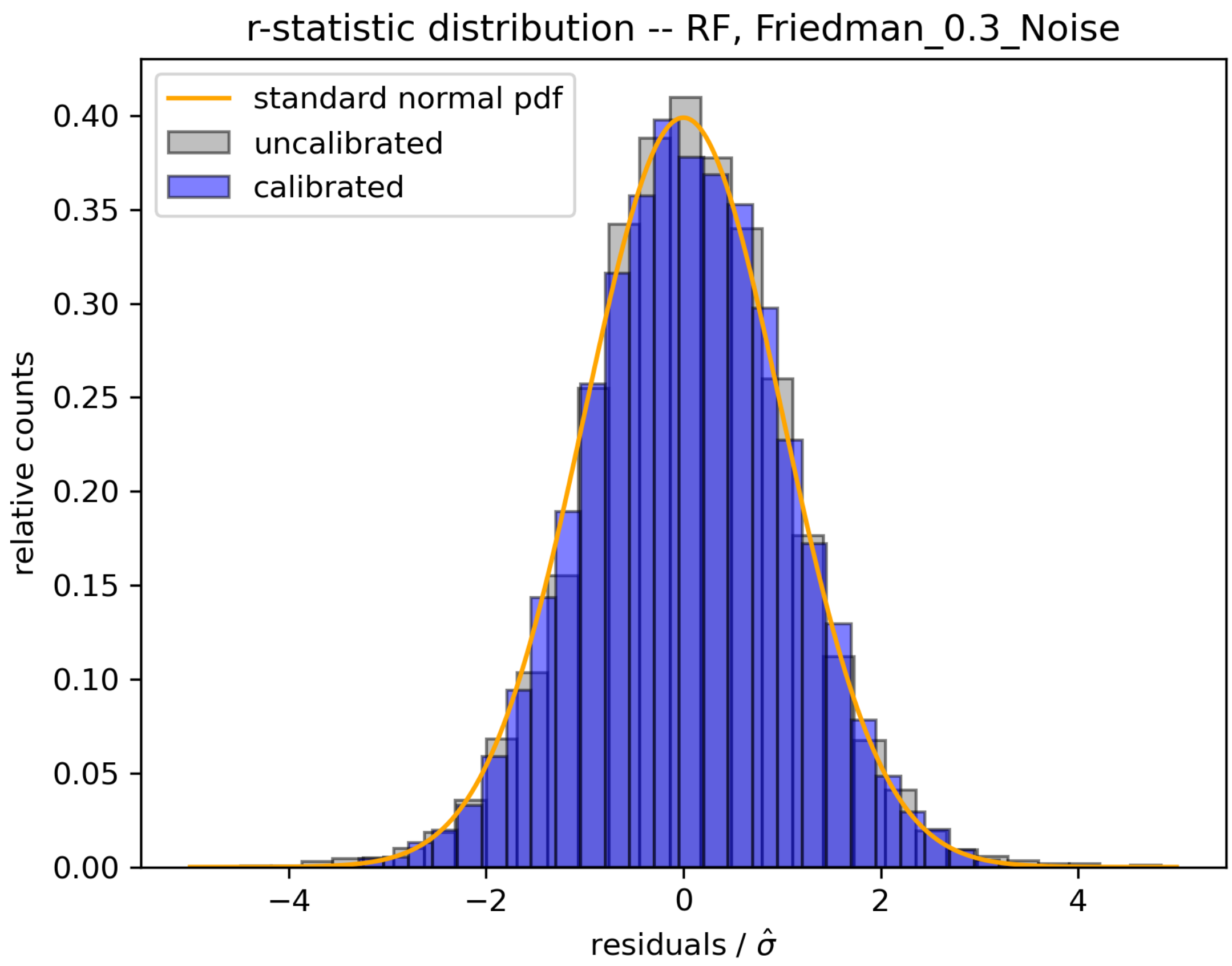

*Figure 27: Friedman with 0.3 sigma gaussian noise r-statistic*

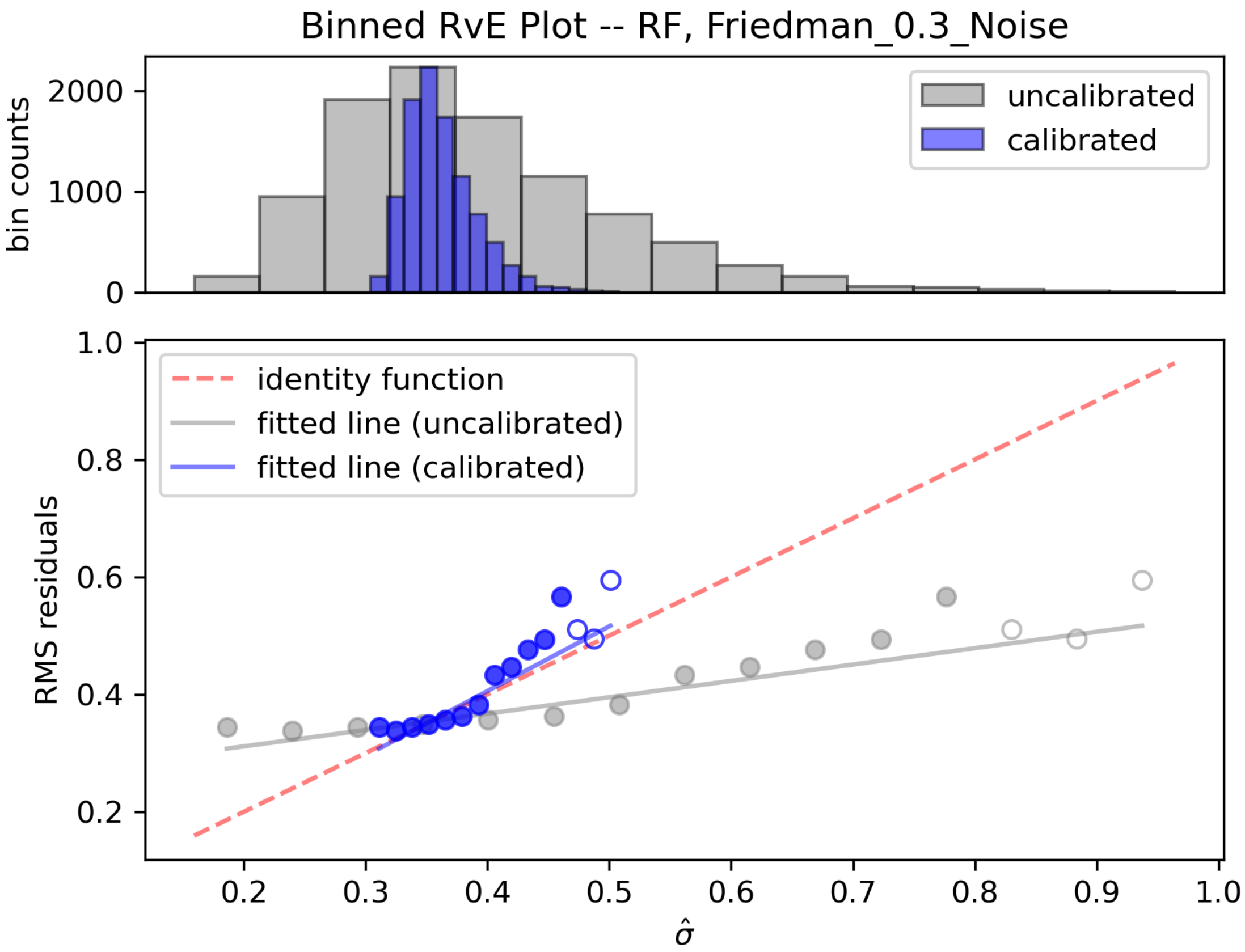

*Figure 28: Friedman with 0.3 sigma gaussian noise RvE*

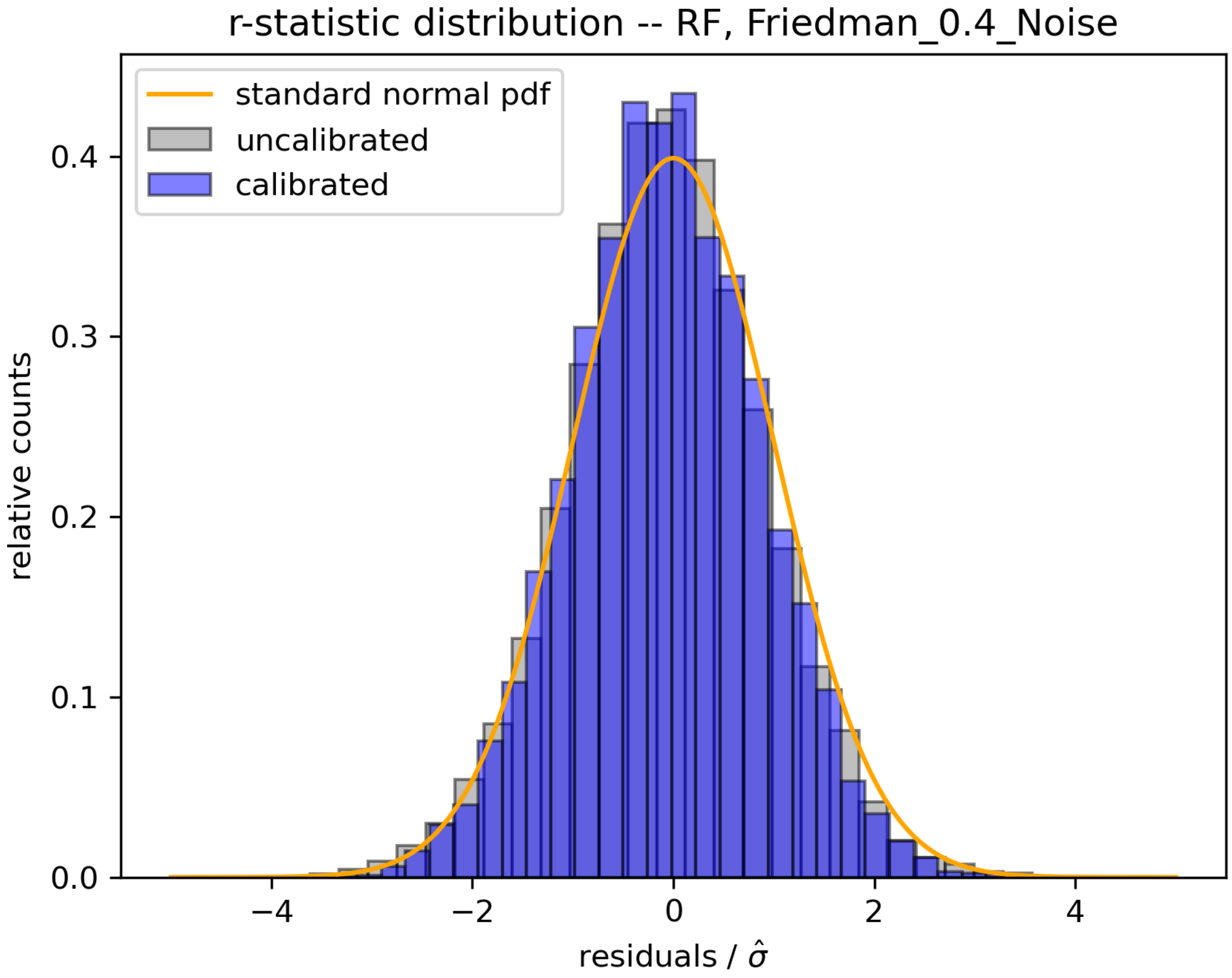

*Figure 29: Friedman with 0.4 sigma gaussian noise r-statistic*

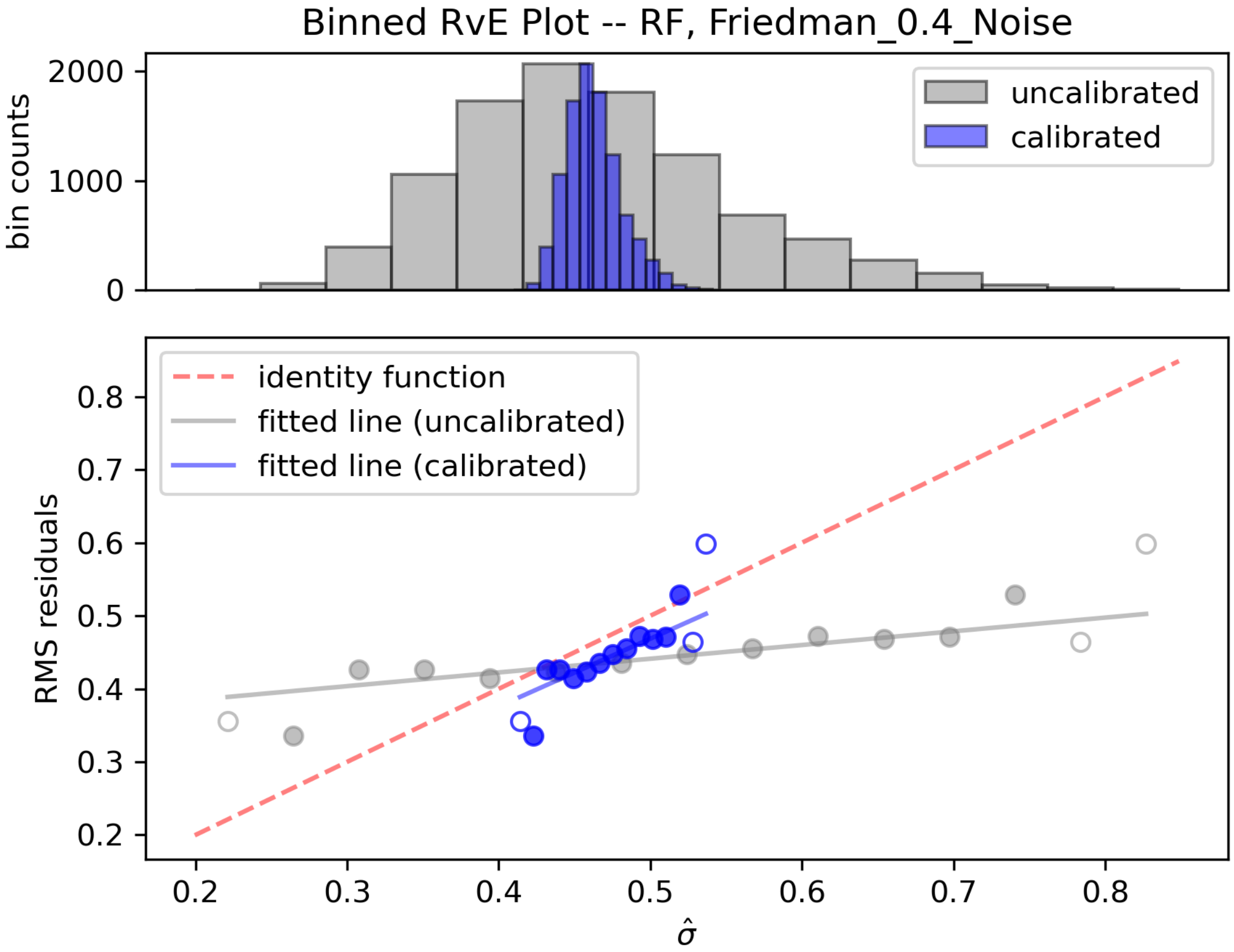

*Figure 30: Friedman with 0.4 sigma gaussian noise RvE*

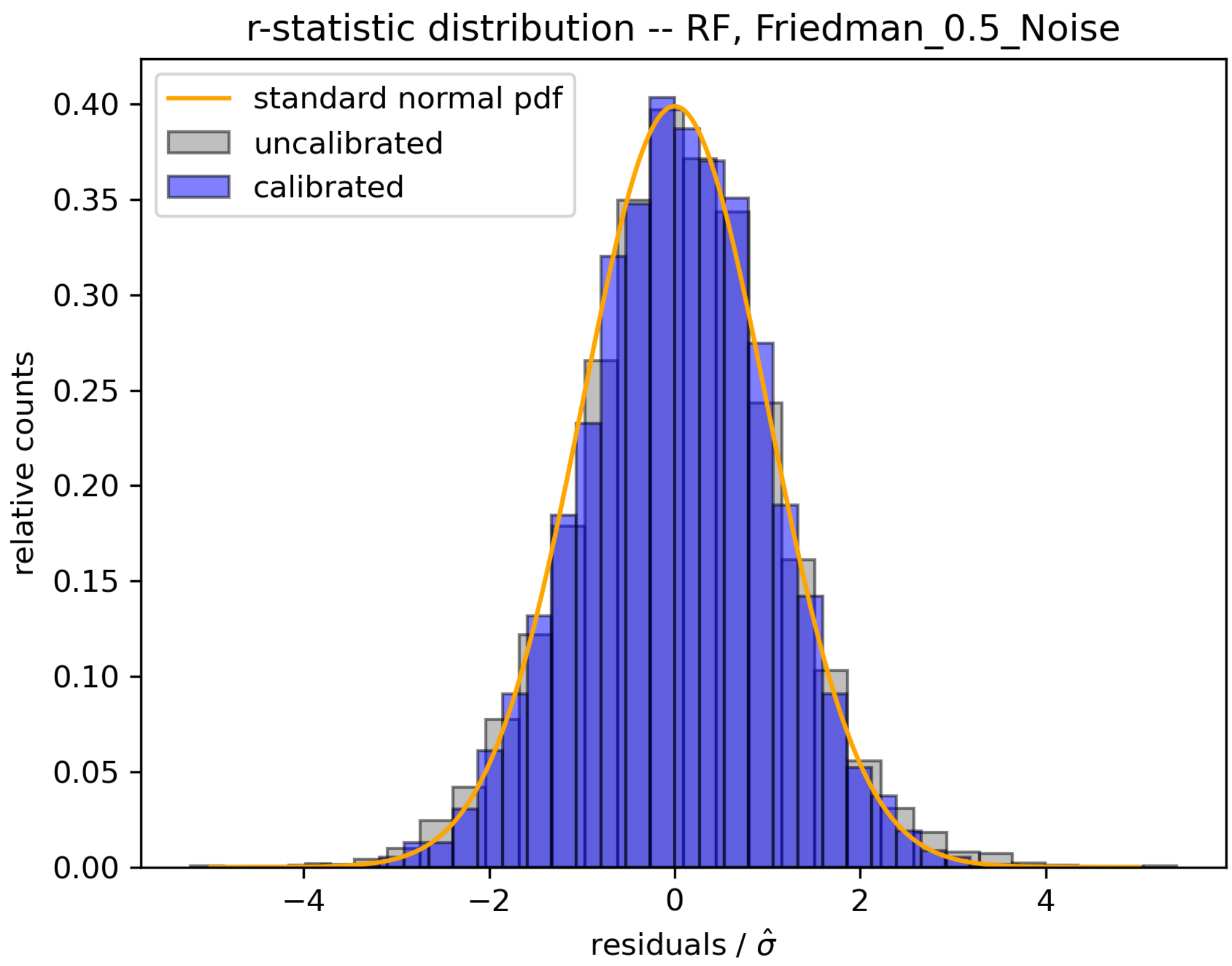

*Figure 31: Friedman with 0.5 sigma gaussian noise r-statistic*

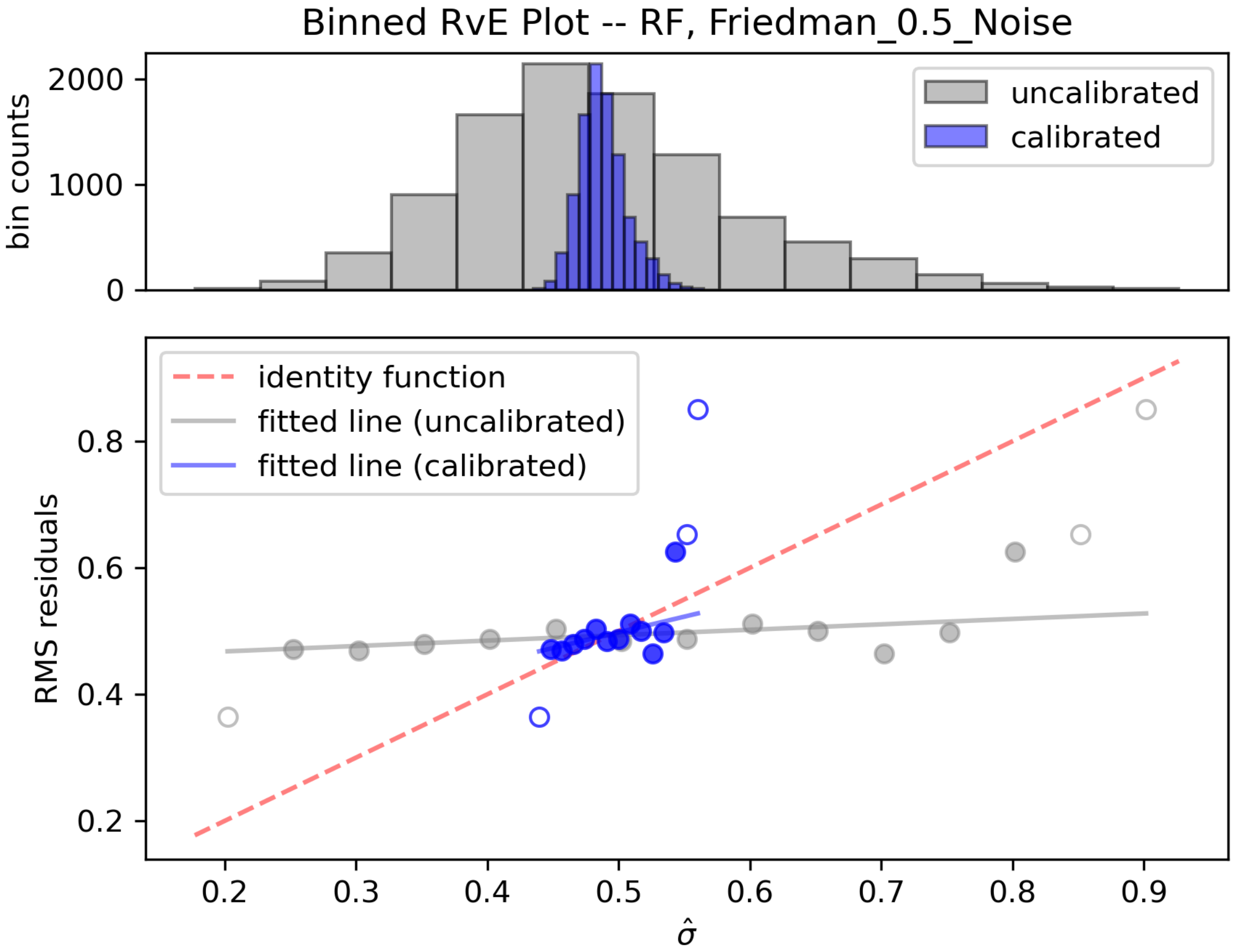

*Figure 32: Friedman with 0.5 sigma gaussian noise RvE*

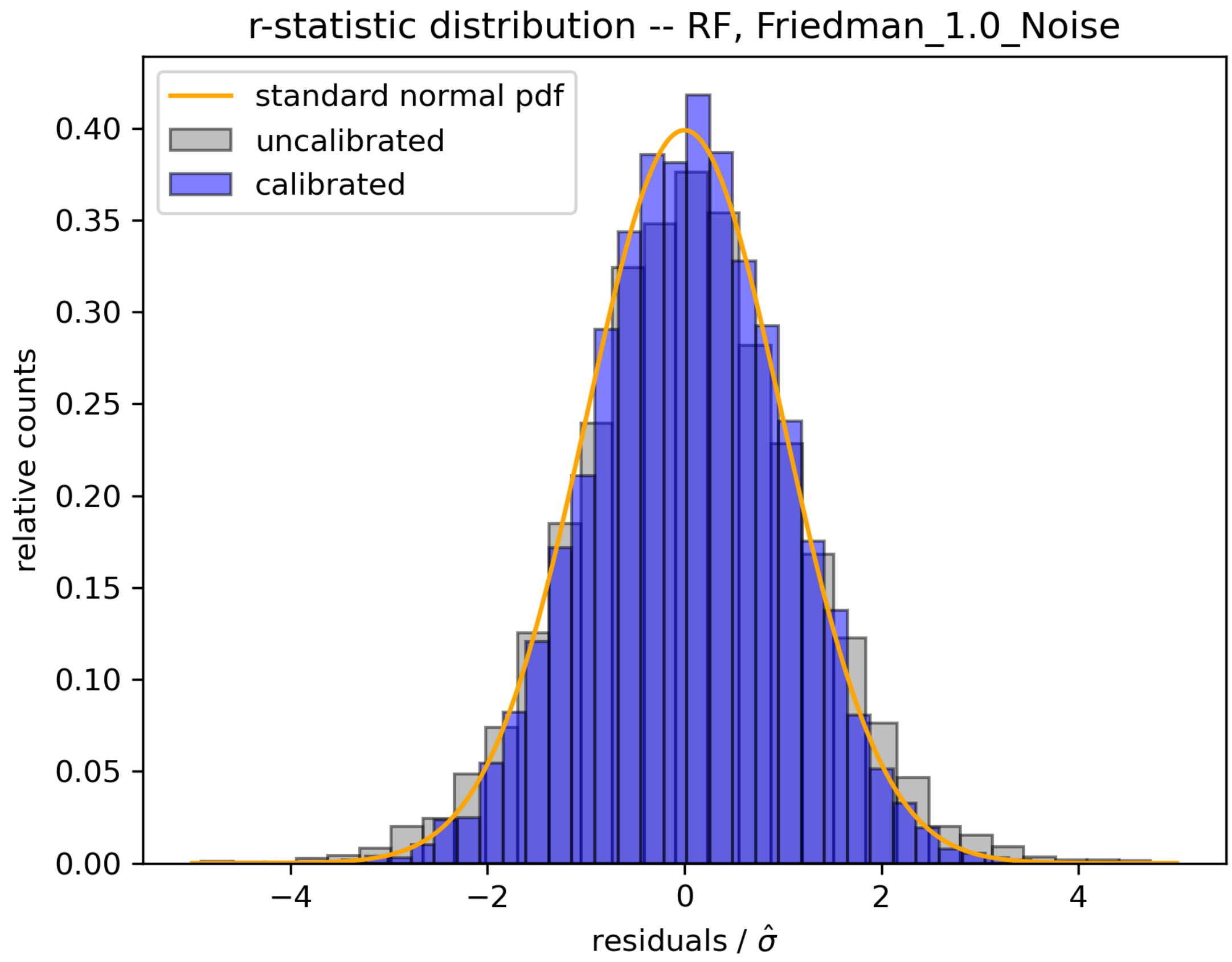

*Figure 33: Friedman with 1.0 sigma gaussian noise r-statistic*

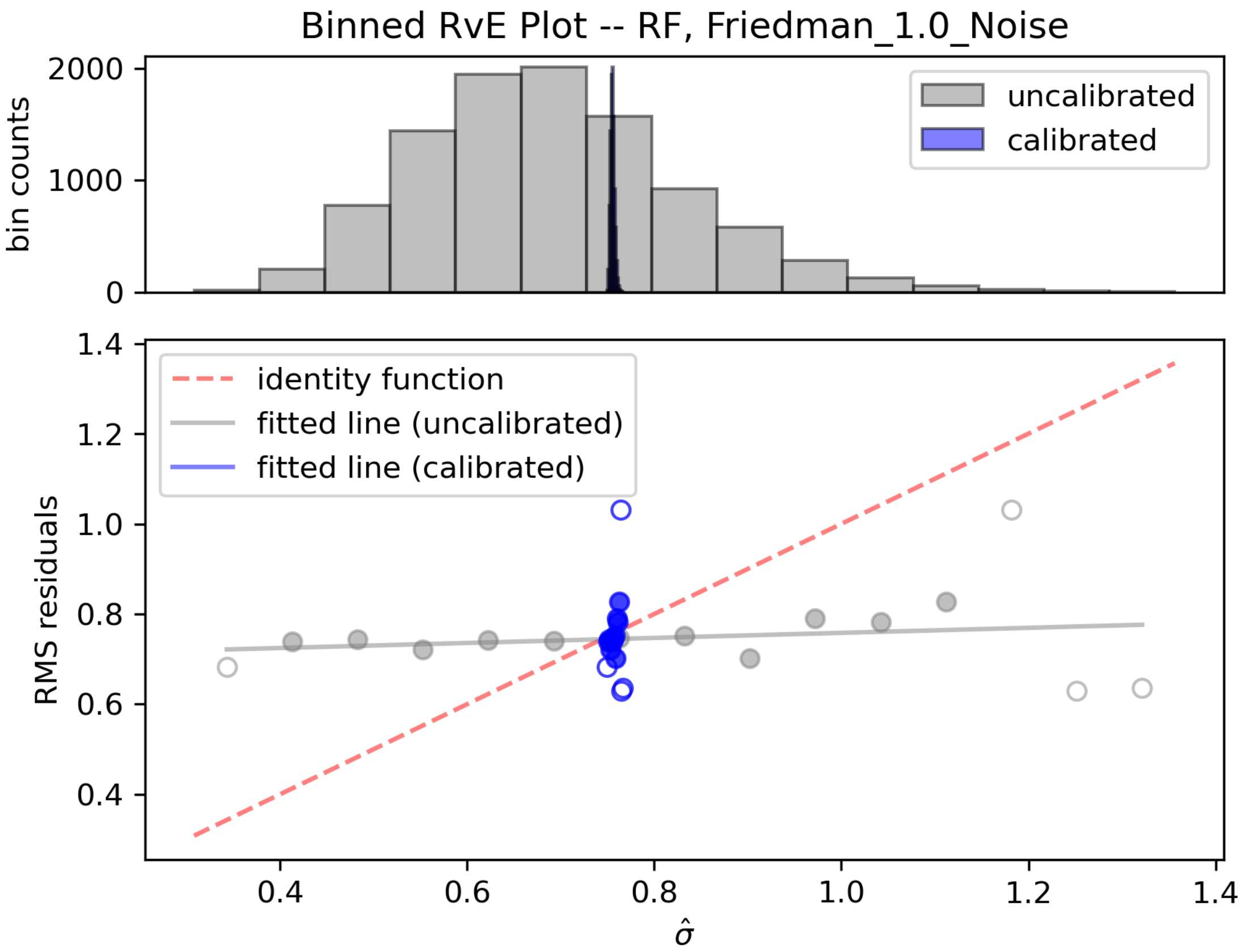

*Figure 34: Friedman with 1.0 sigma gaussian noise RvE*

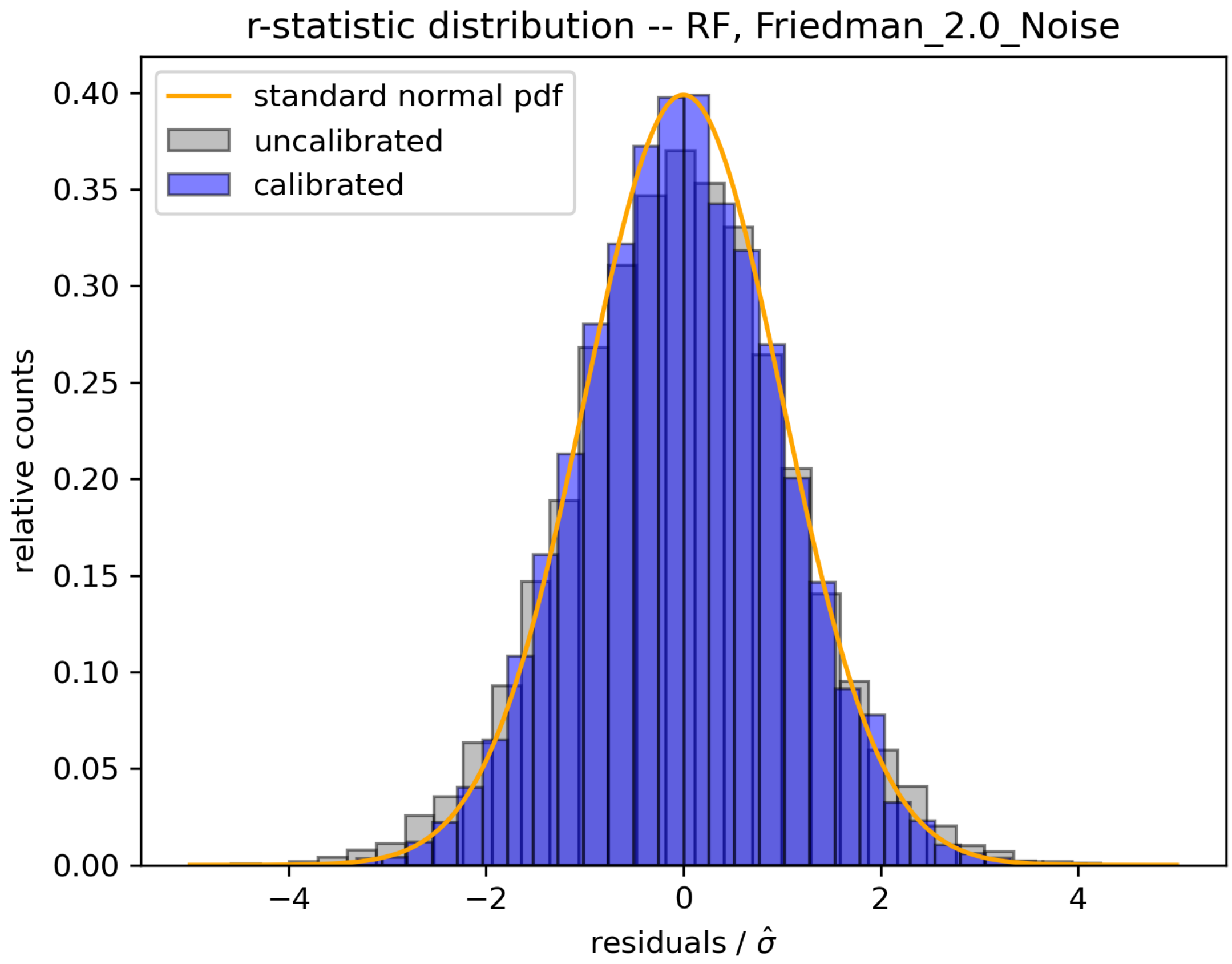

*Figure 35: Friedman with 2.0 sigma gaussian noise r-statistic*

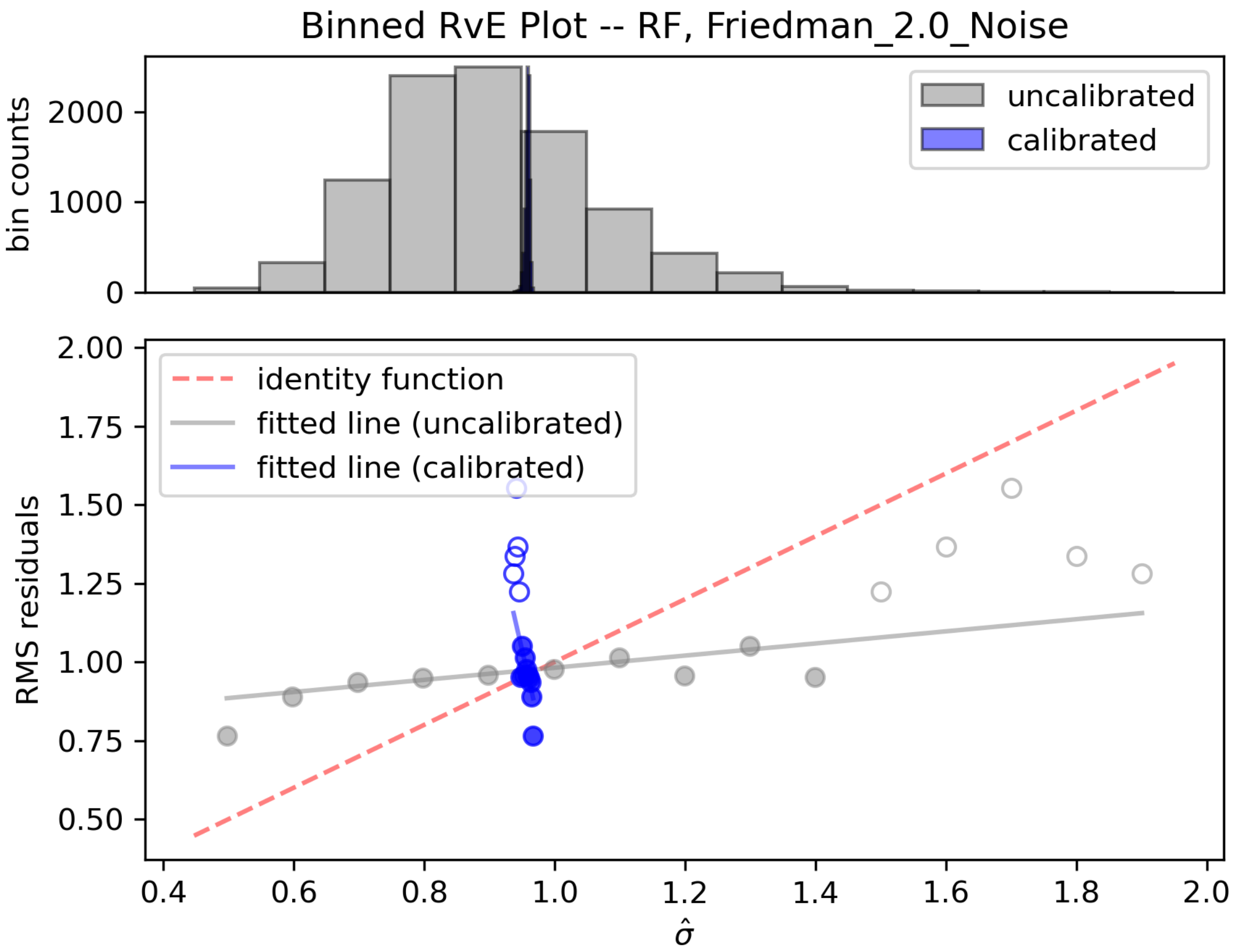

*Figure 36: Friedman with 2.0 sigma gaussian noise RvE*

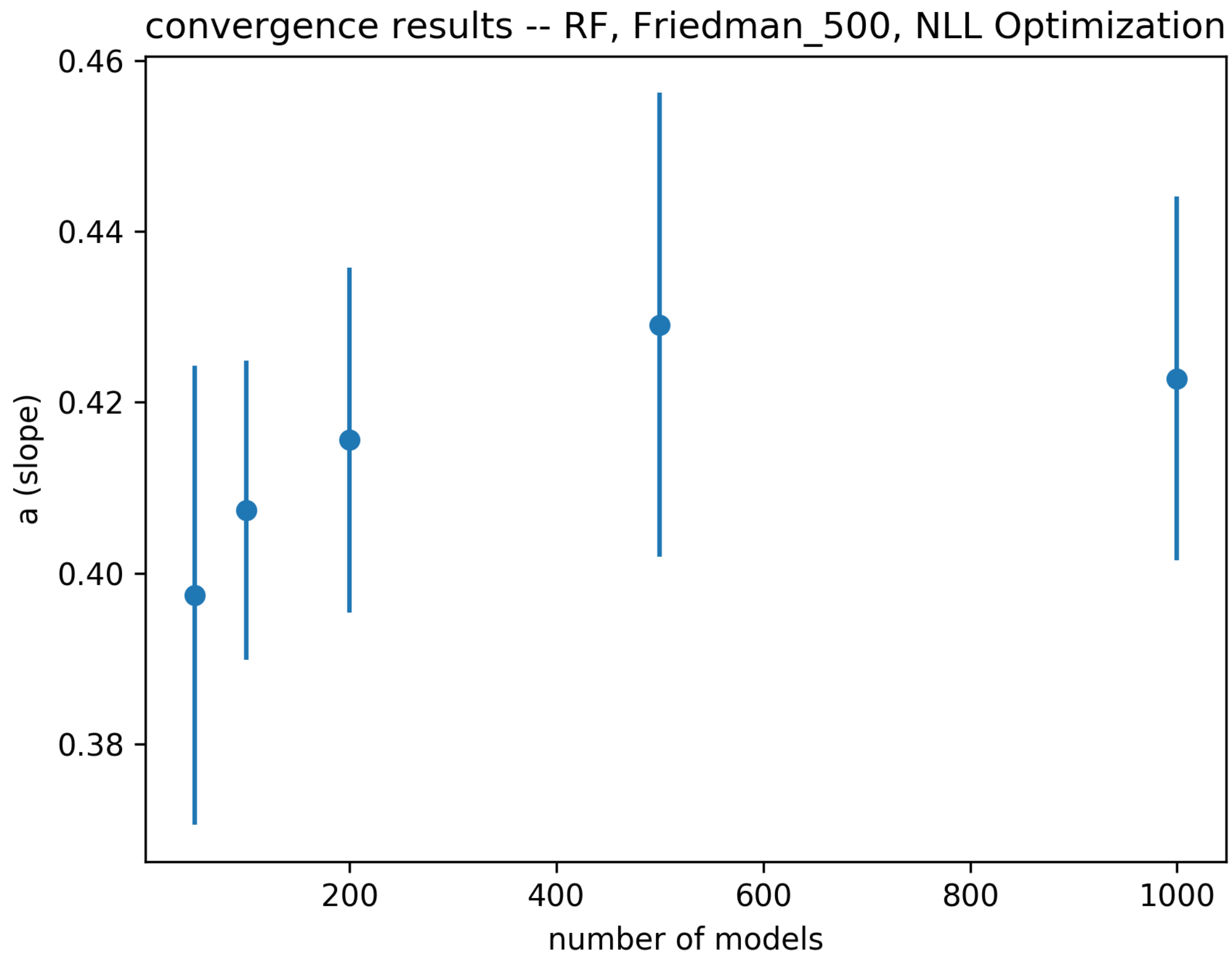

*Figure 37: Convergence of a for Friedman with random forest*

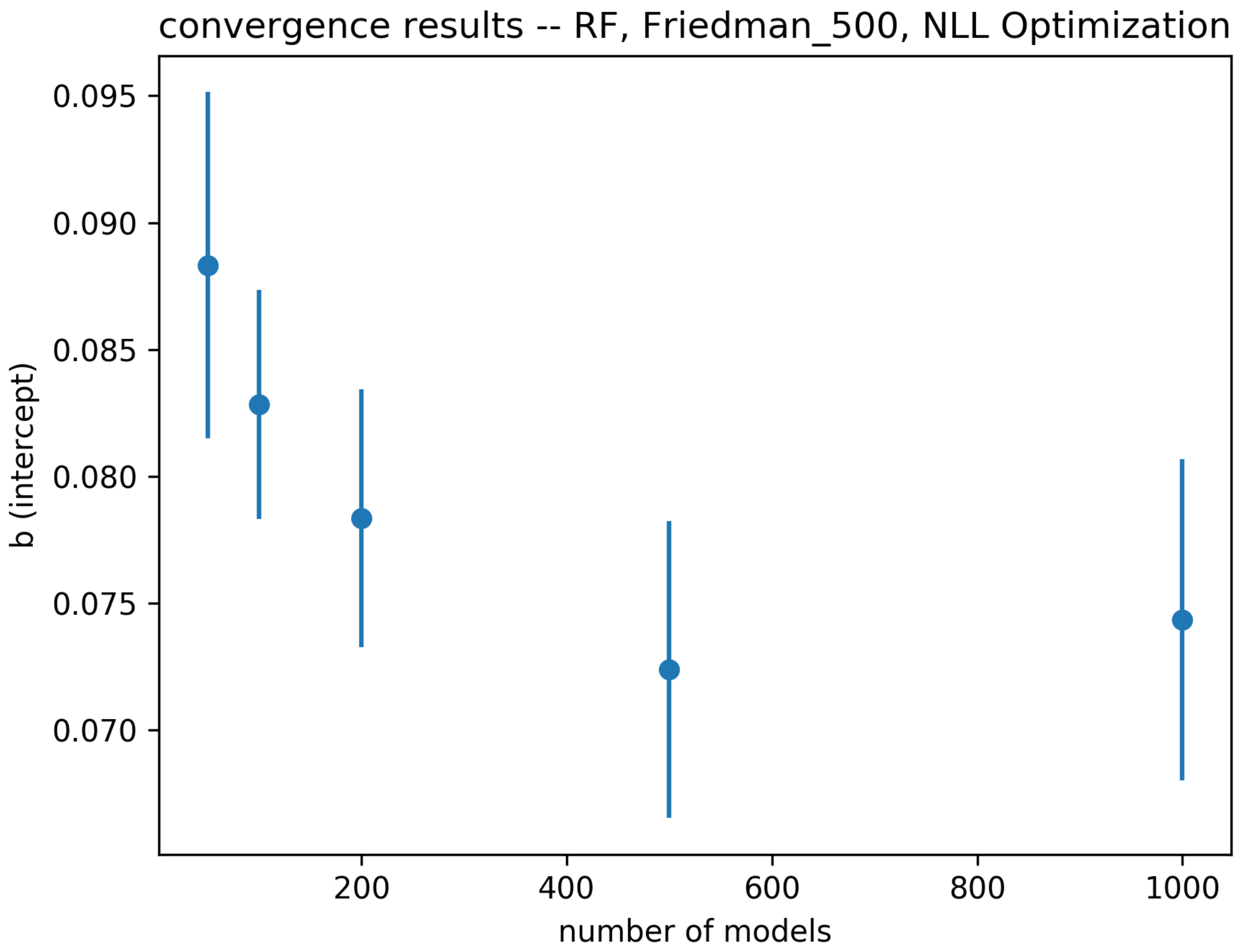

*Figure 38: Convergence of b for Friedman with random forest*

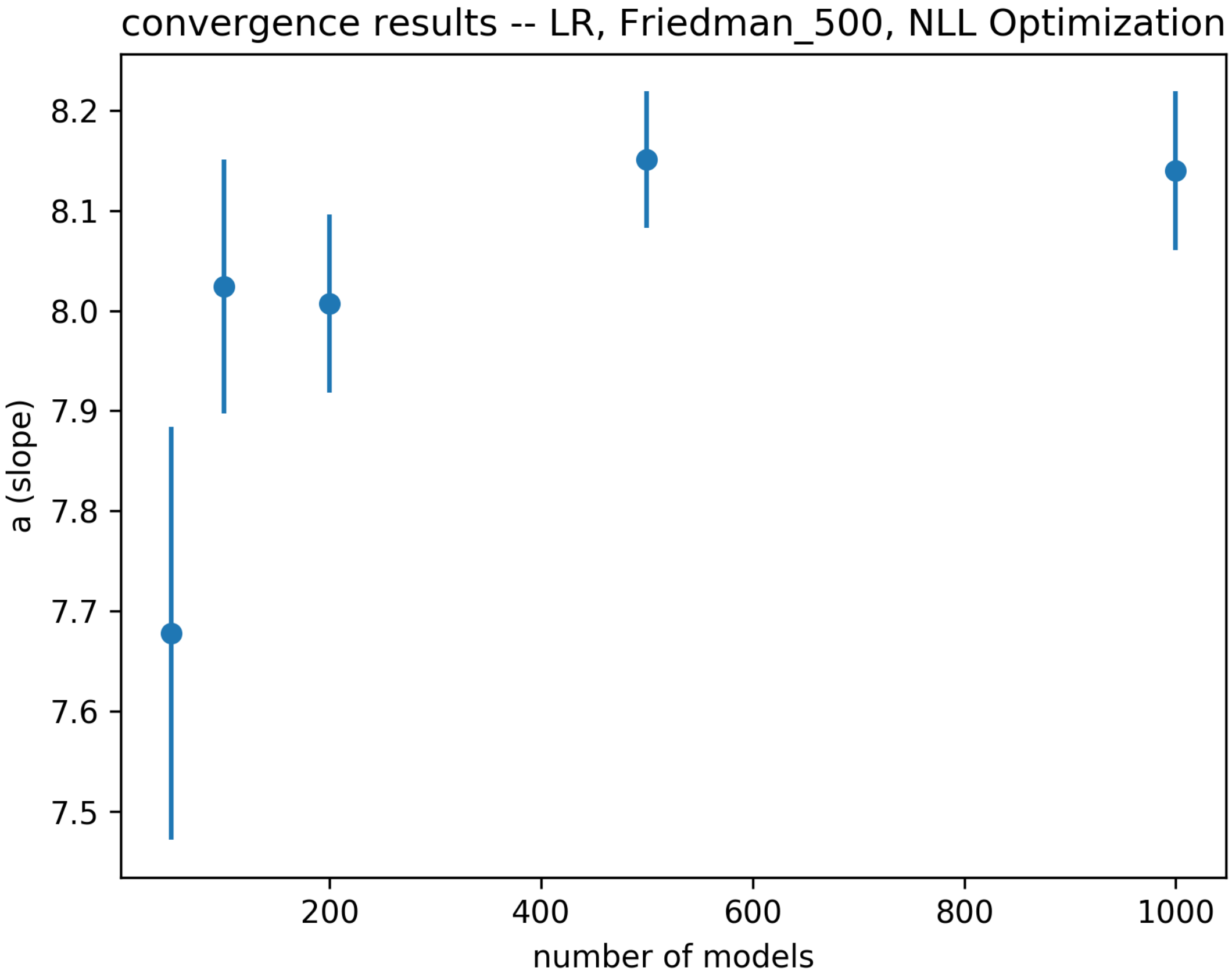

*Figure 39: Convergence of a for Friedman with linear ridge regression*

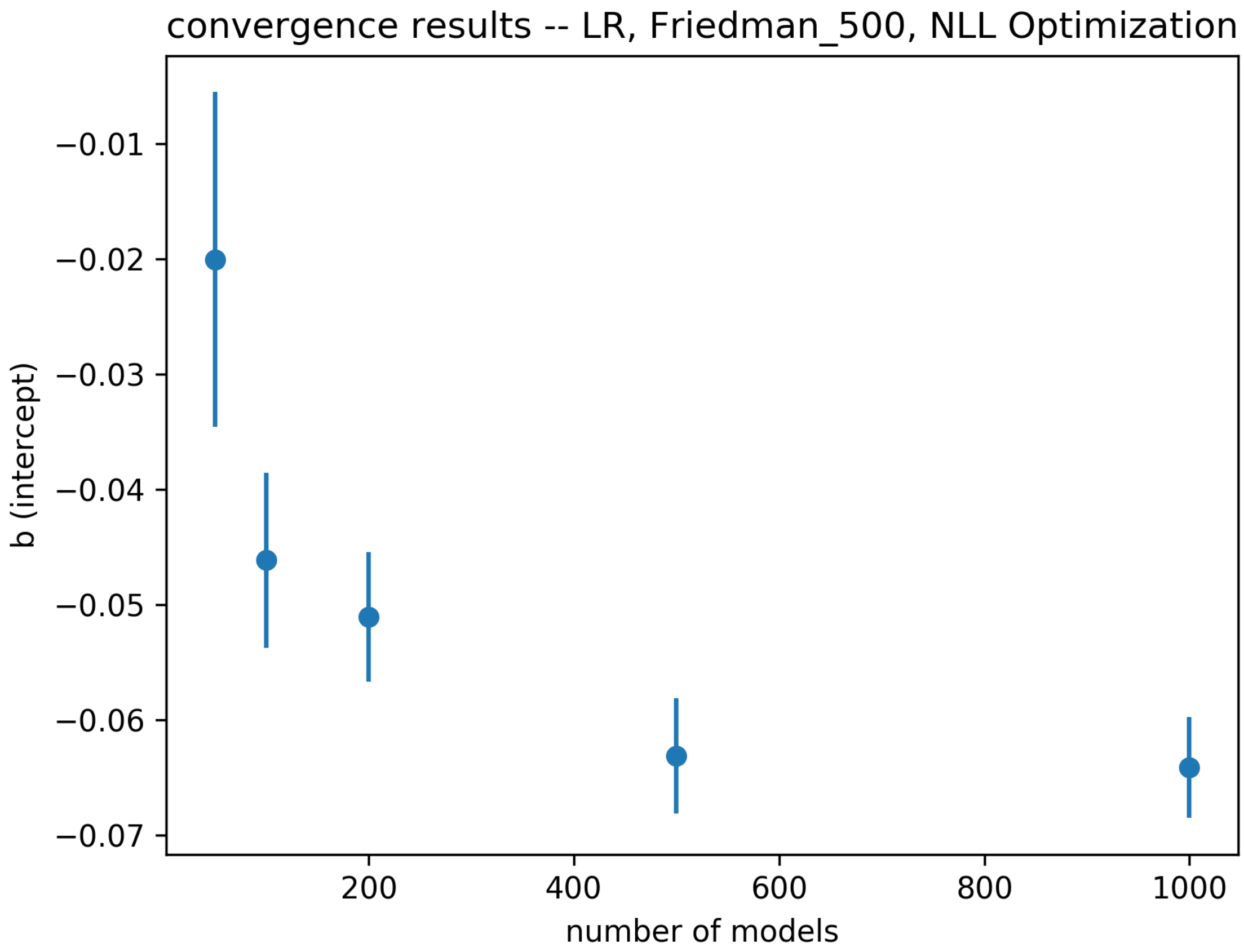

*Figure 40: Convergence of b for Friedman with linear ridge regression*

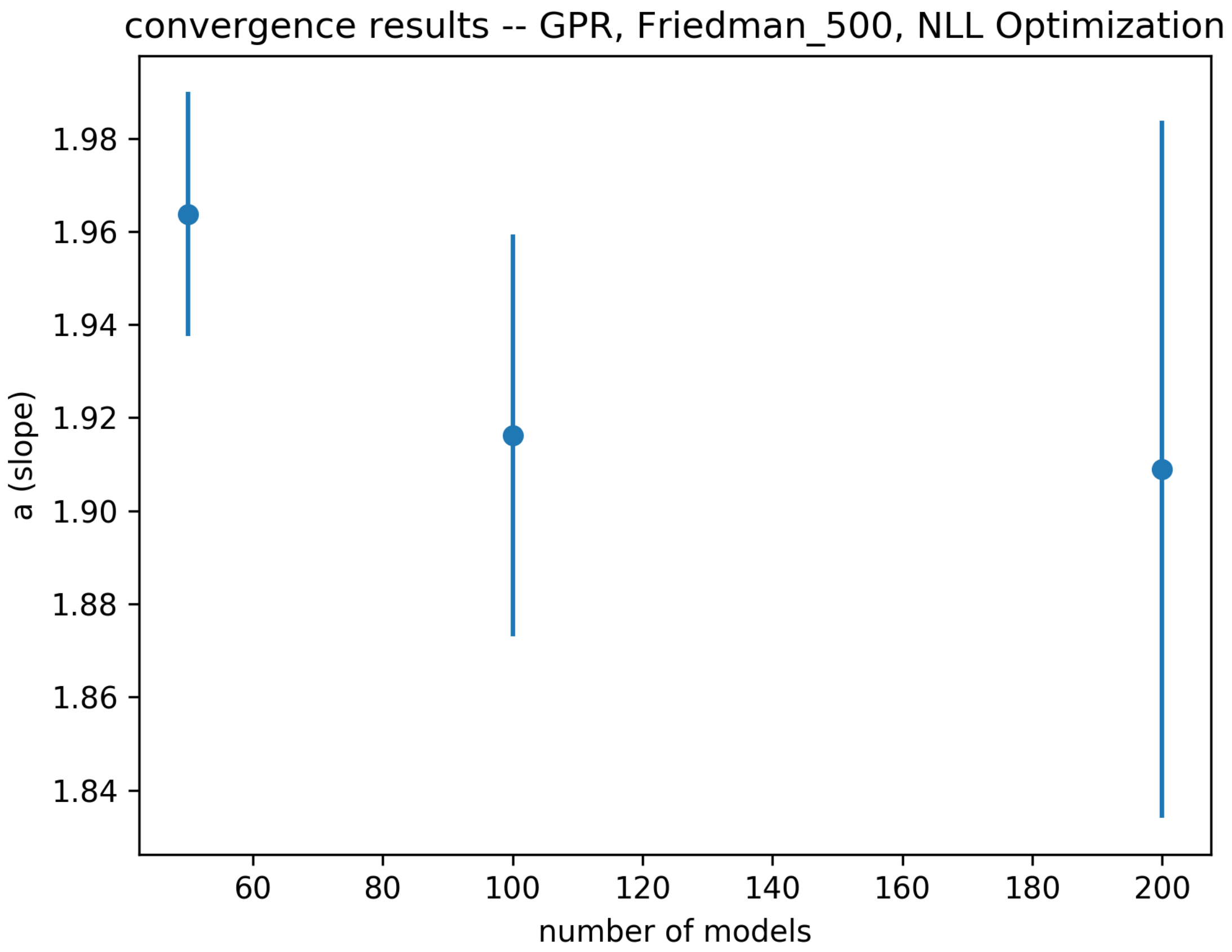

*Figure 41: Convergence of a for Friedman with GPR*

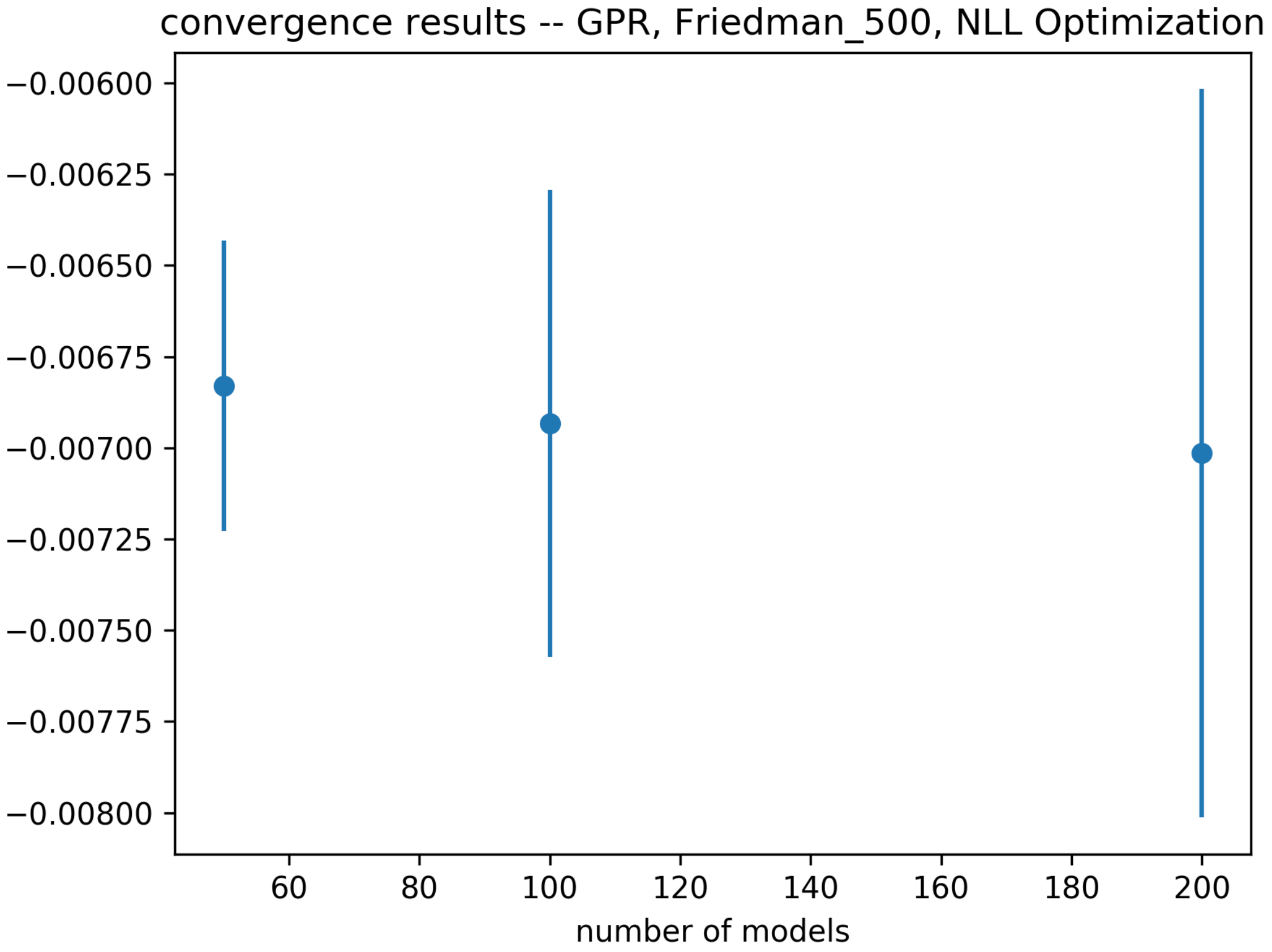

*Figure 42: Convergence of b for Friedman with GPR*

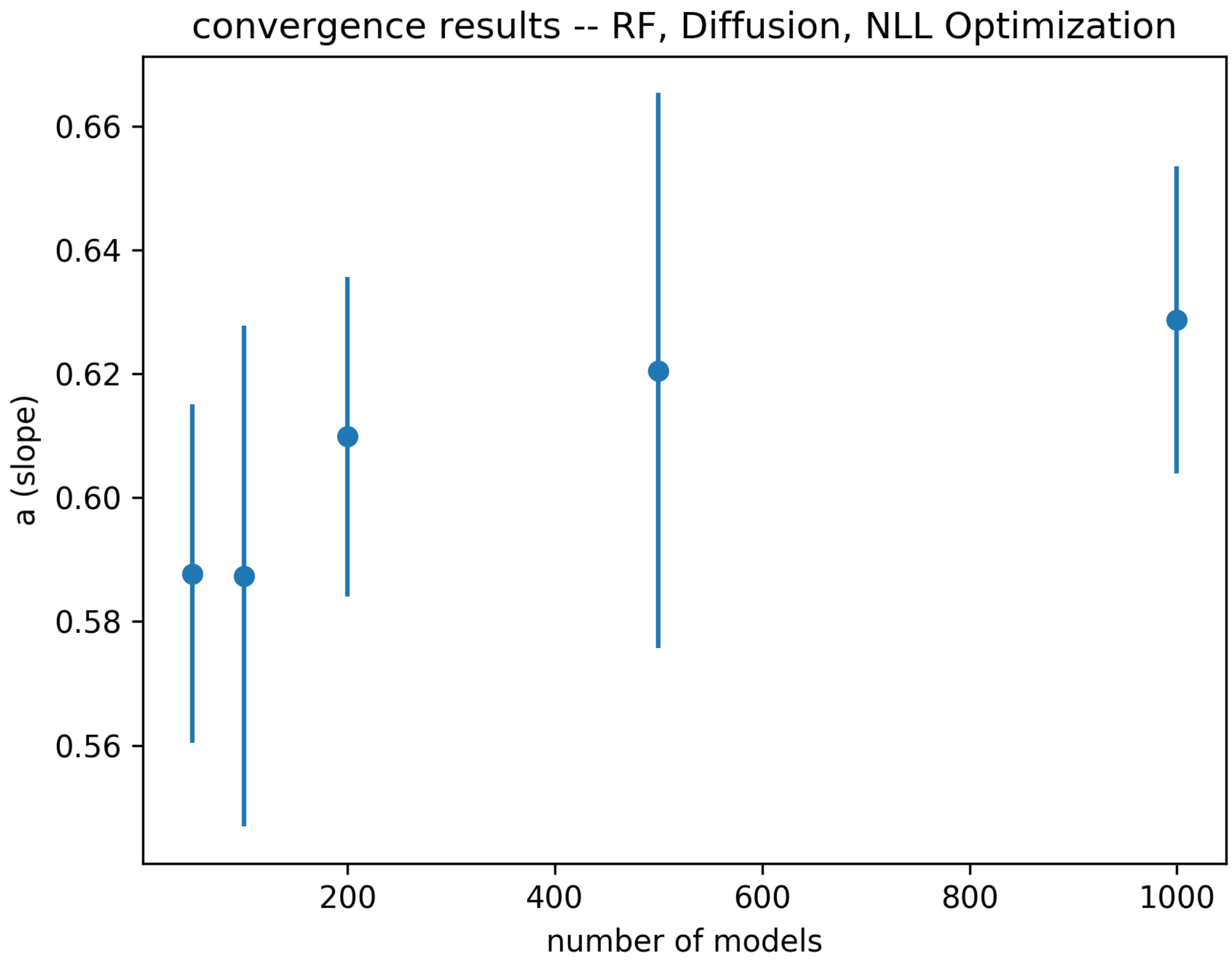

*Figure 43: Convergence of a for Diffusion with random forest*

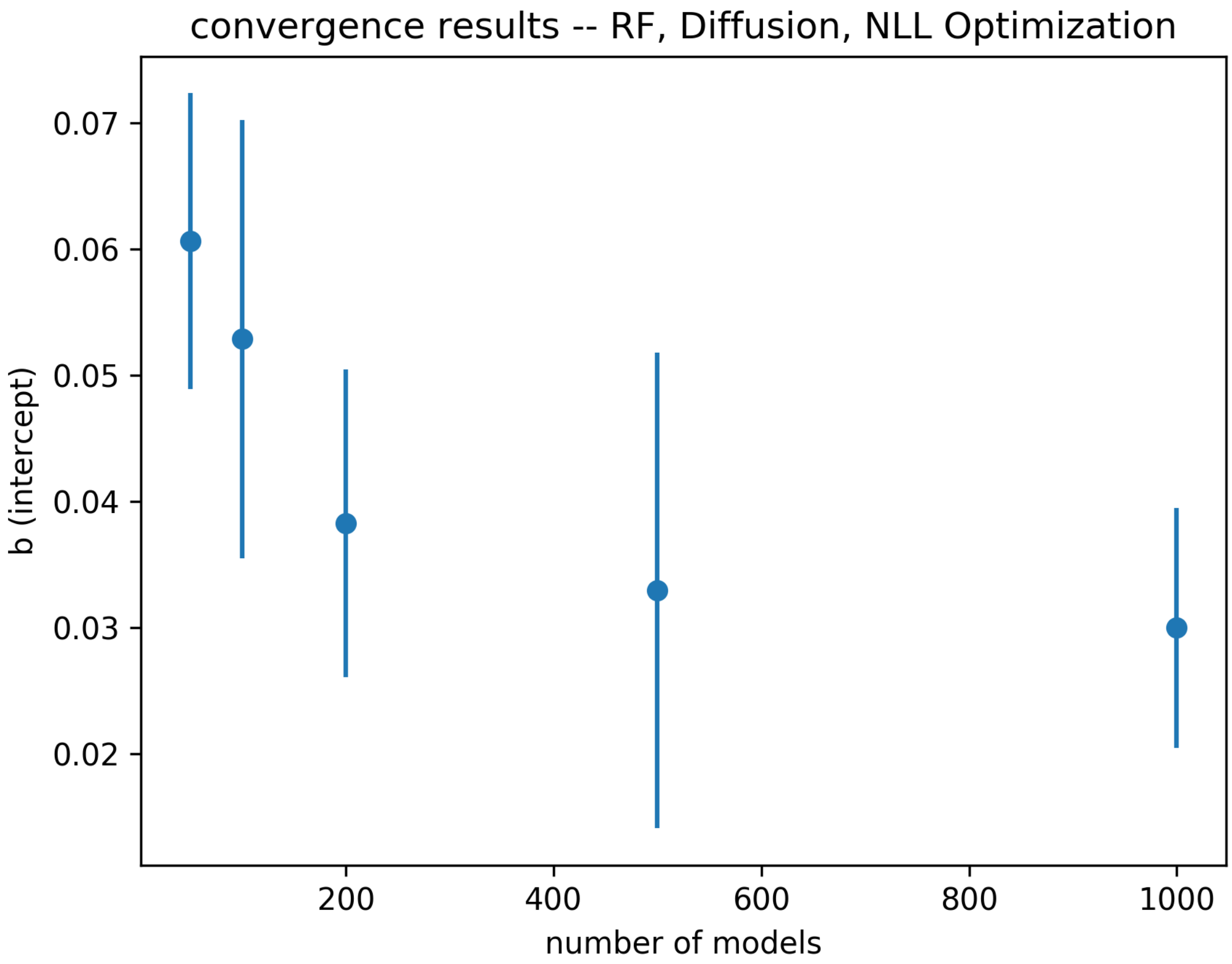

*Figure 44: Convergence of b for Diffusion with random forest*

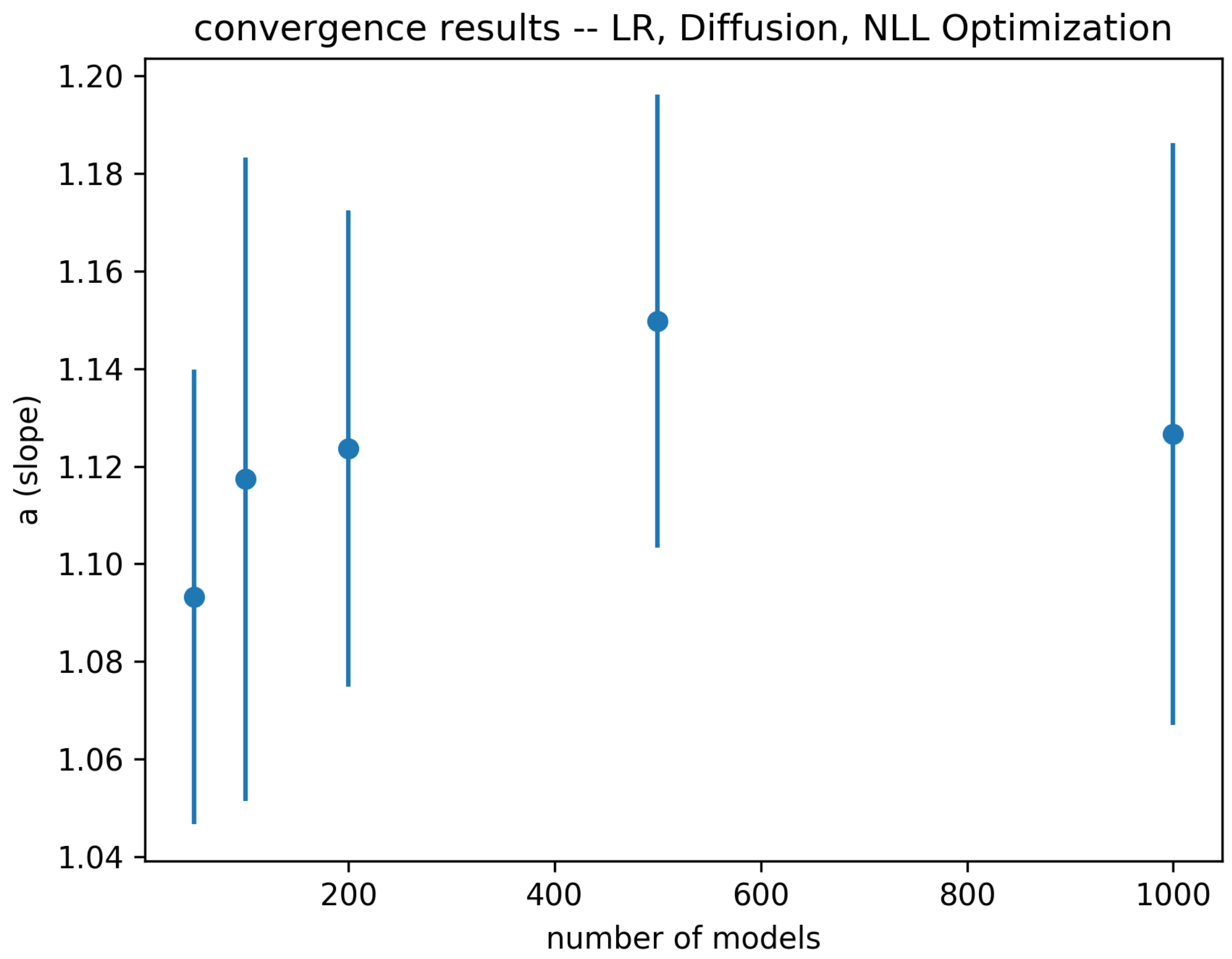

*Figure 45: Convergence of a for Diffusion with linear ridge regression*

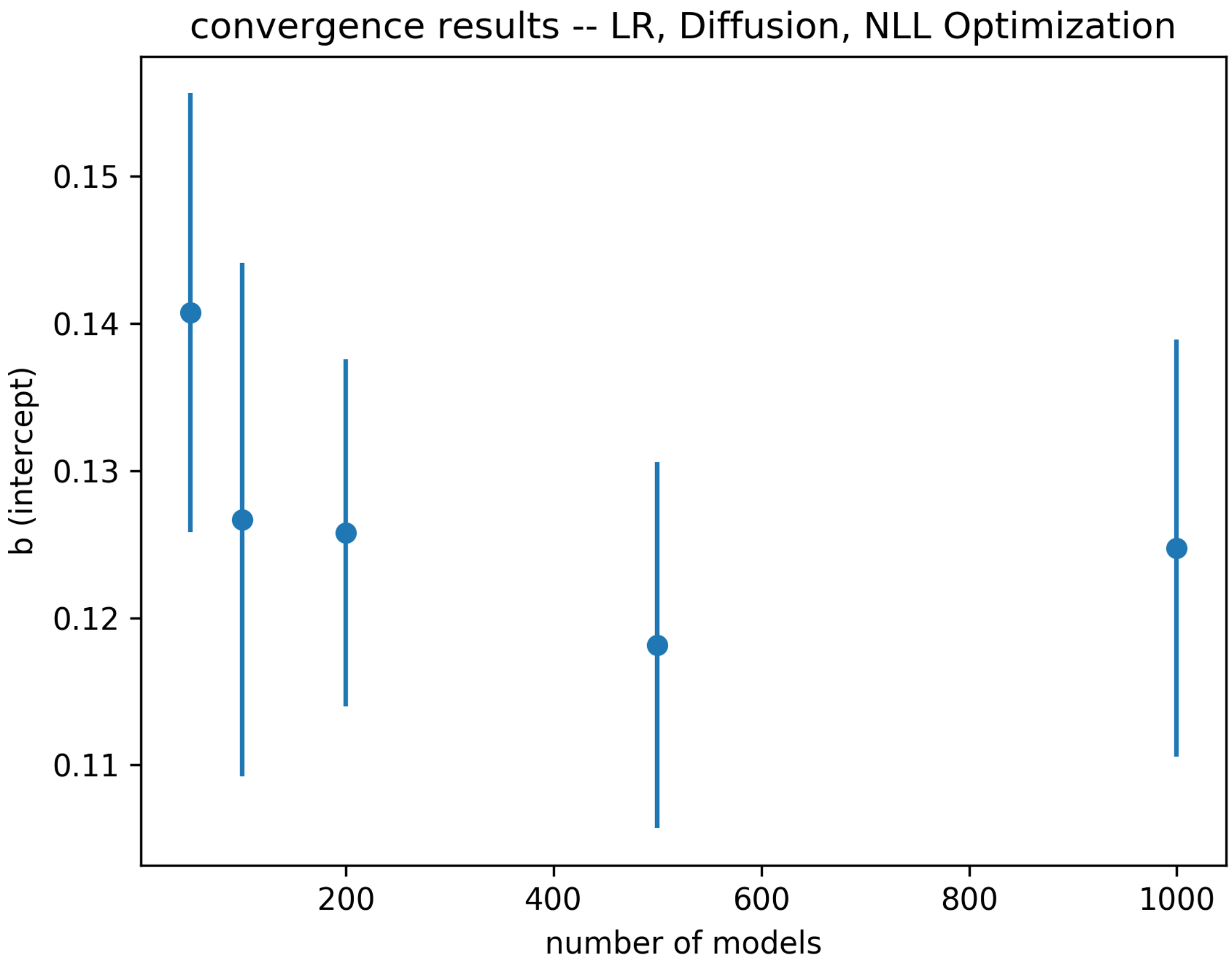

*Figure 46: Convergence of b for Diffusion with linear ridge regression*

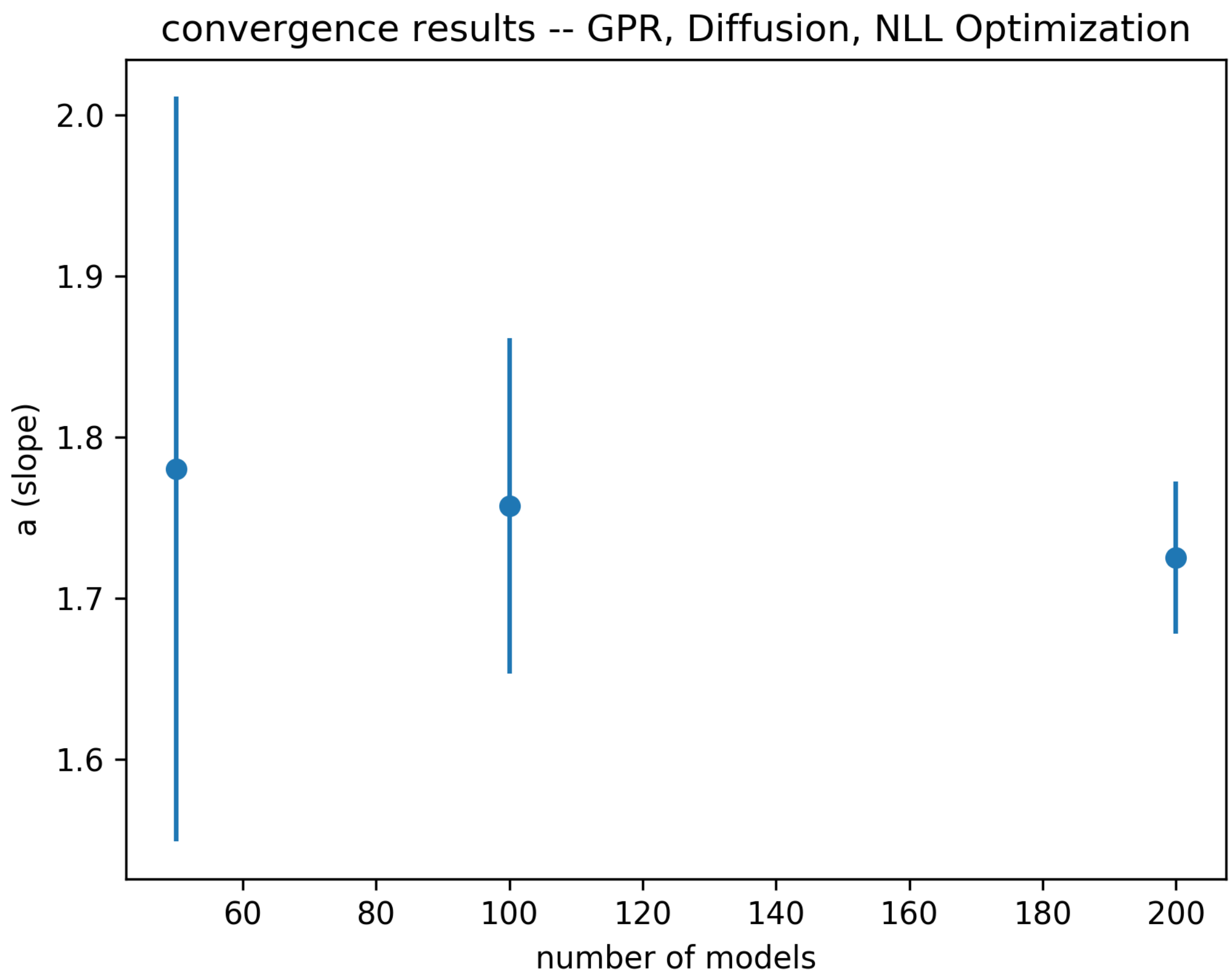

*Figure 47: Convergence of a for Diffusion with GPR*

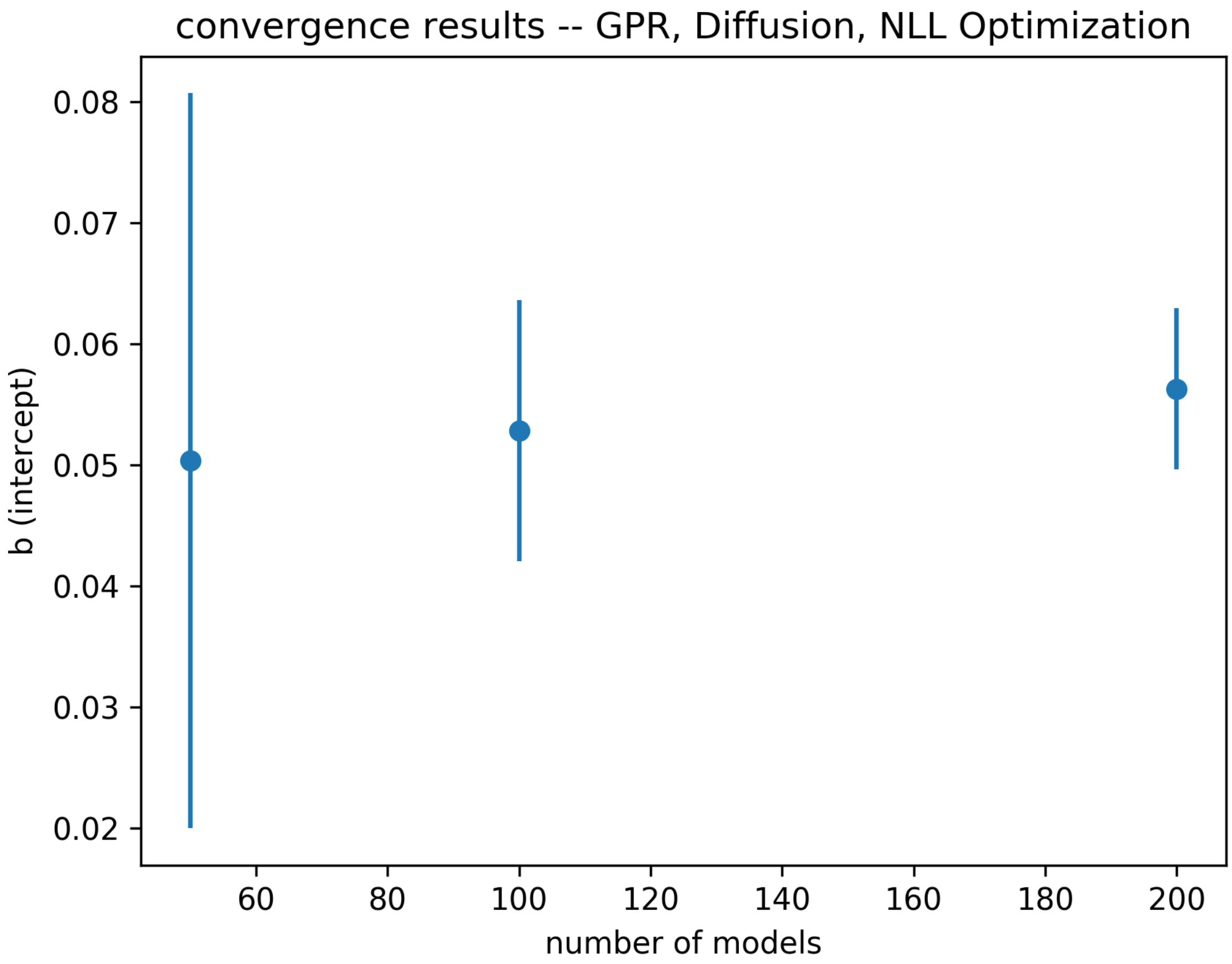

*Figure 48: Convergence of b for Diffusion with GPR*

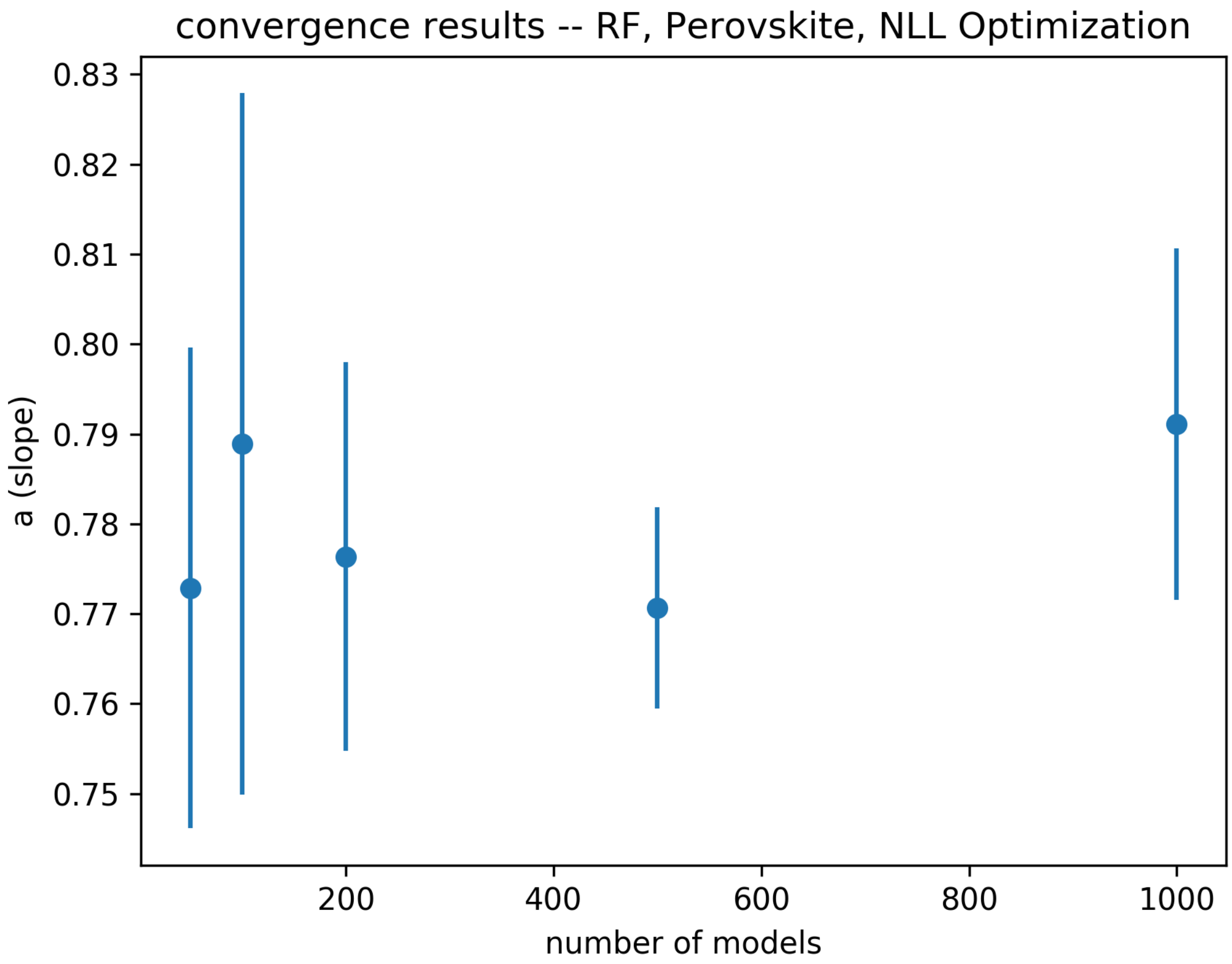

*Figure 49: Convergence of a for Perovskite with random forest*

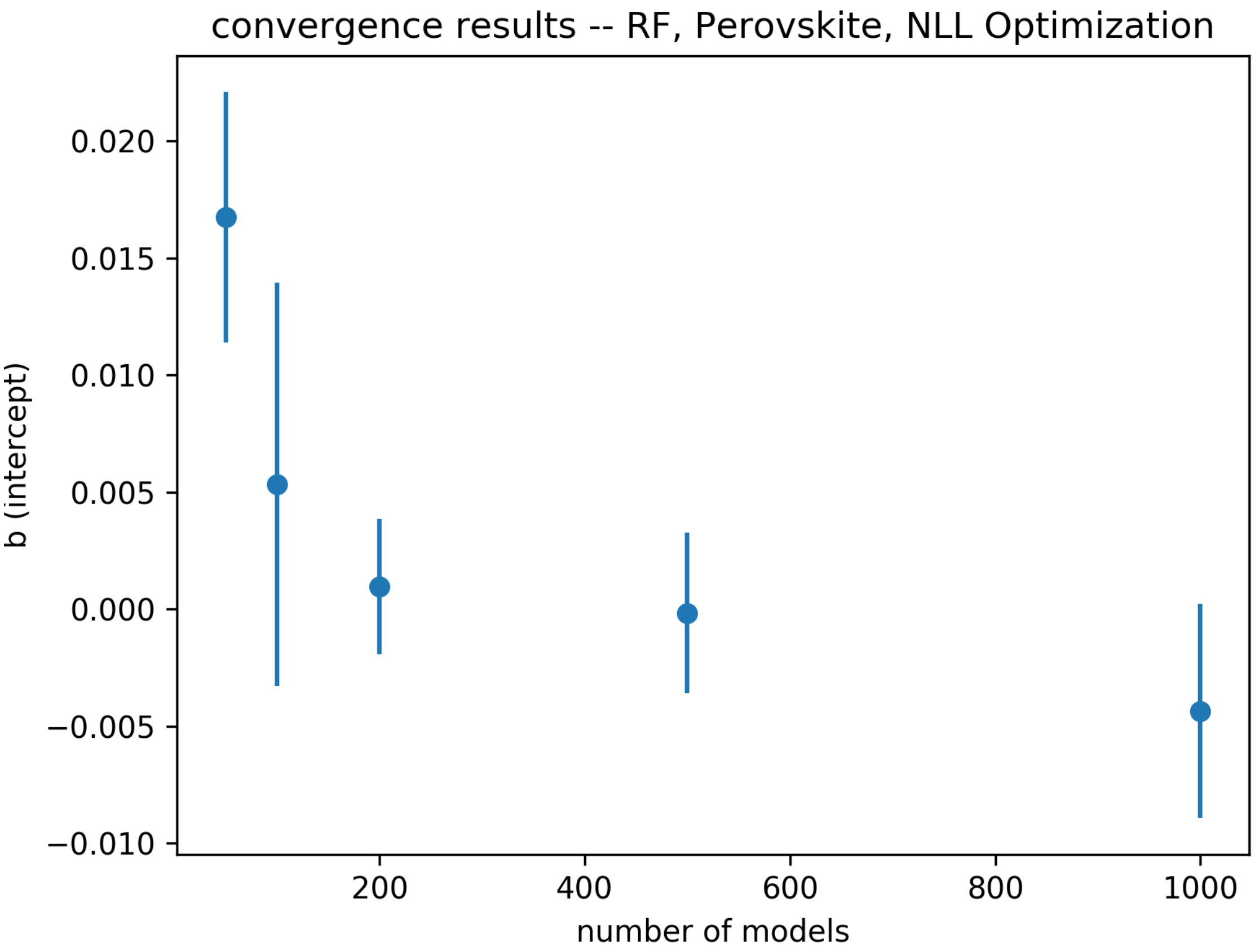

*Figure 50: Convergence of b for Perovskite with random forest*

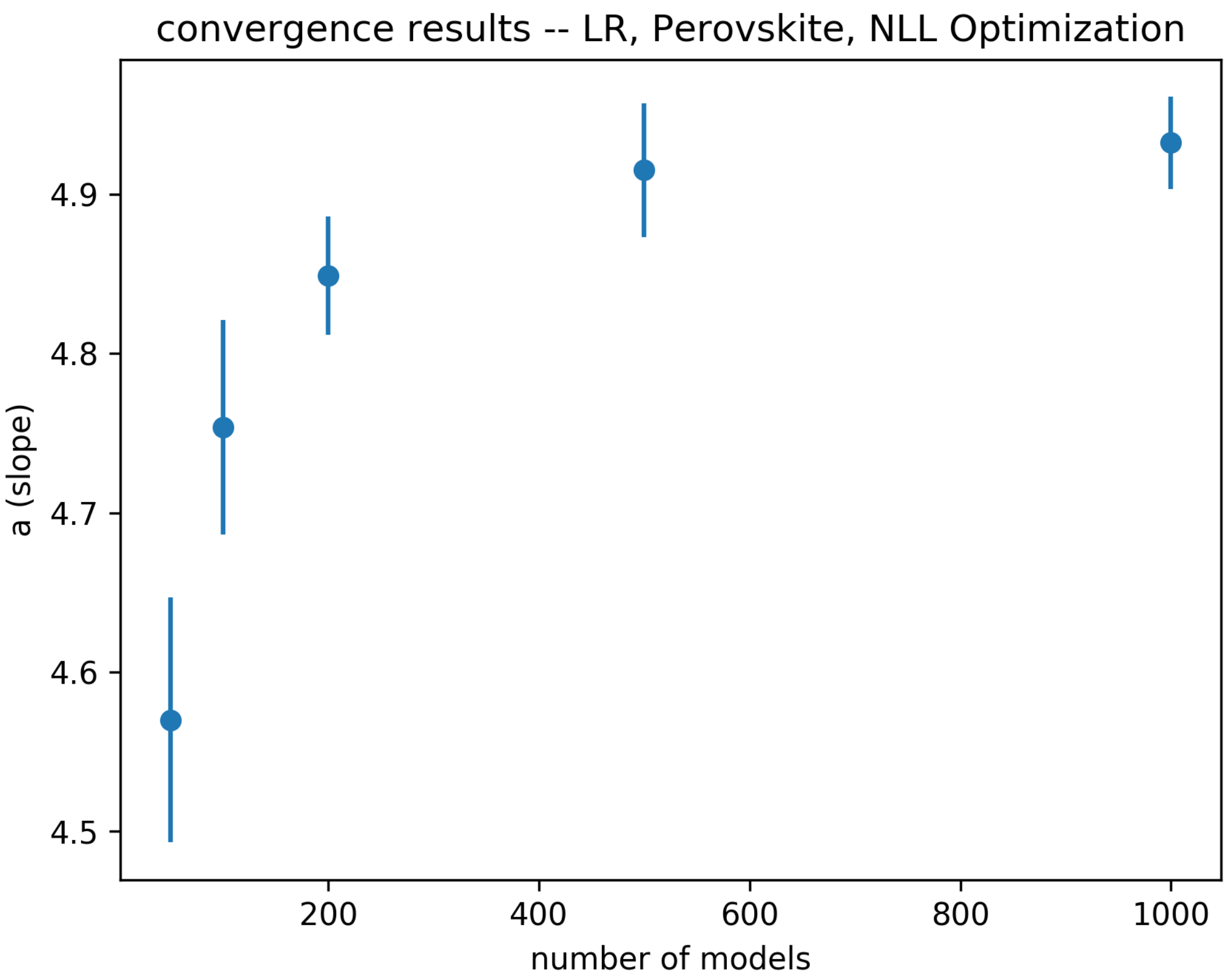

*Figure 51: Convergence of a for Perovskite with linear ridge regression*

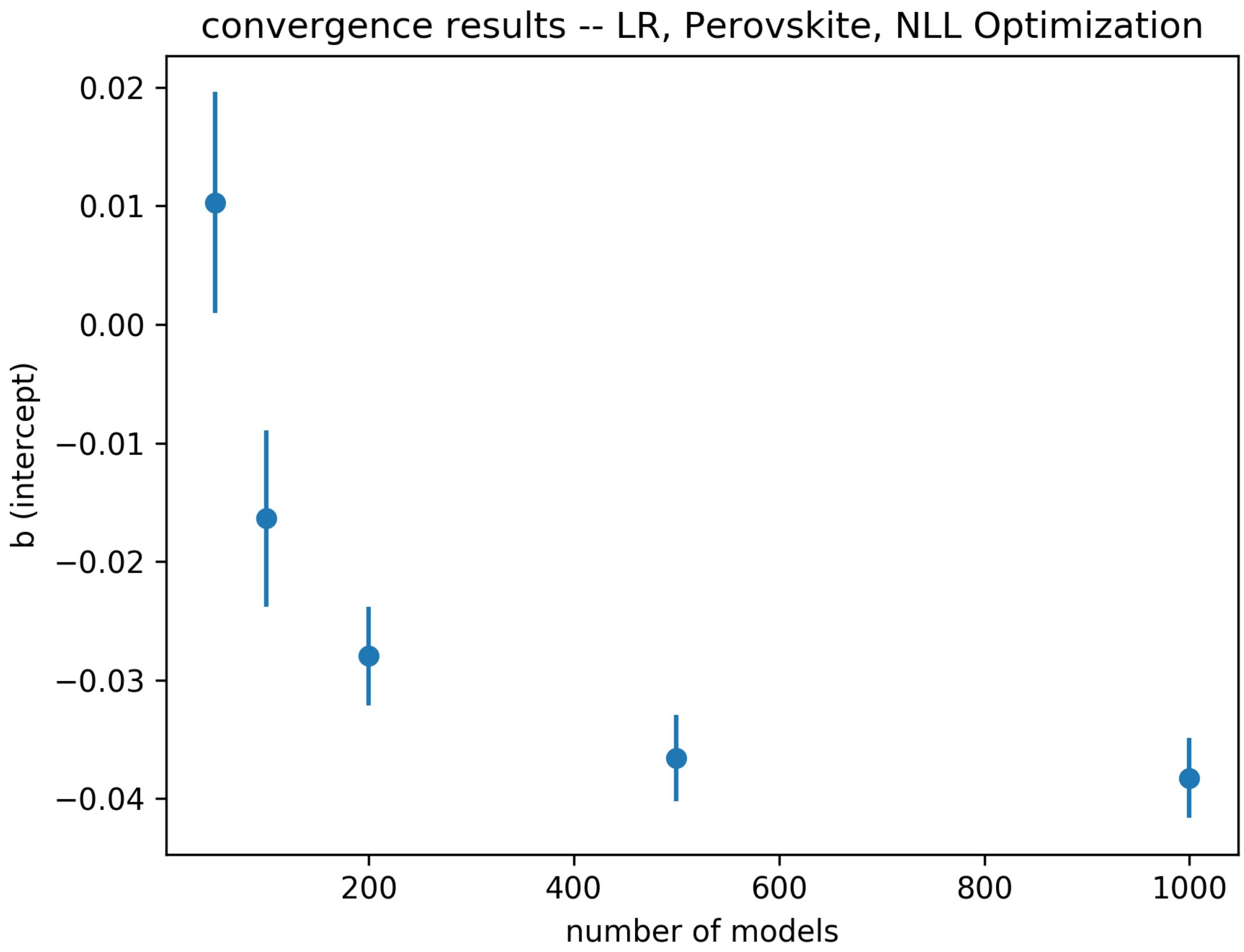

*Figure 52: Convergence of b for Perovskite with linear ridge regression*

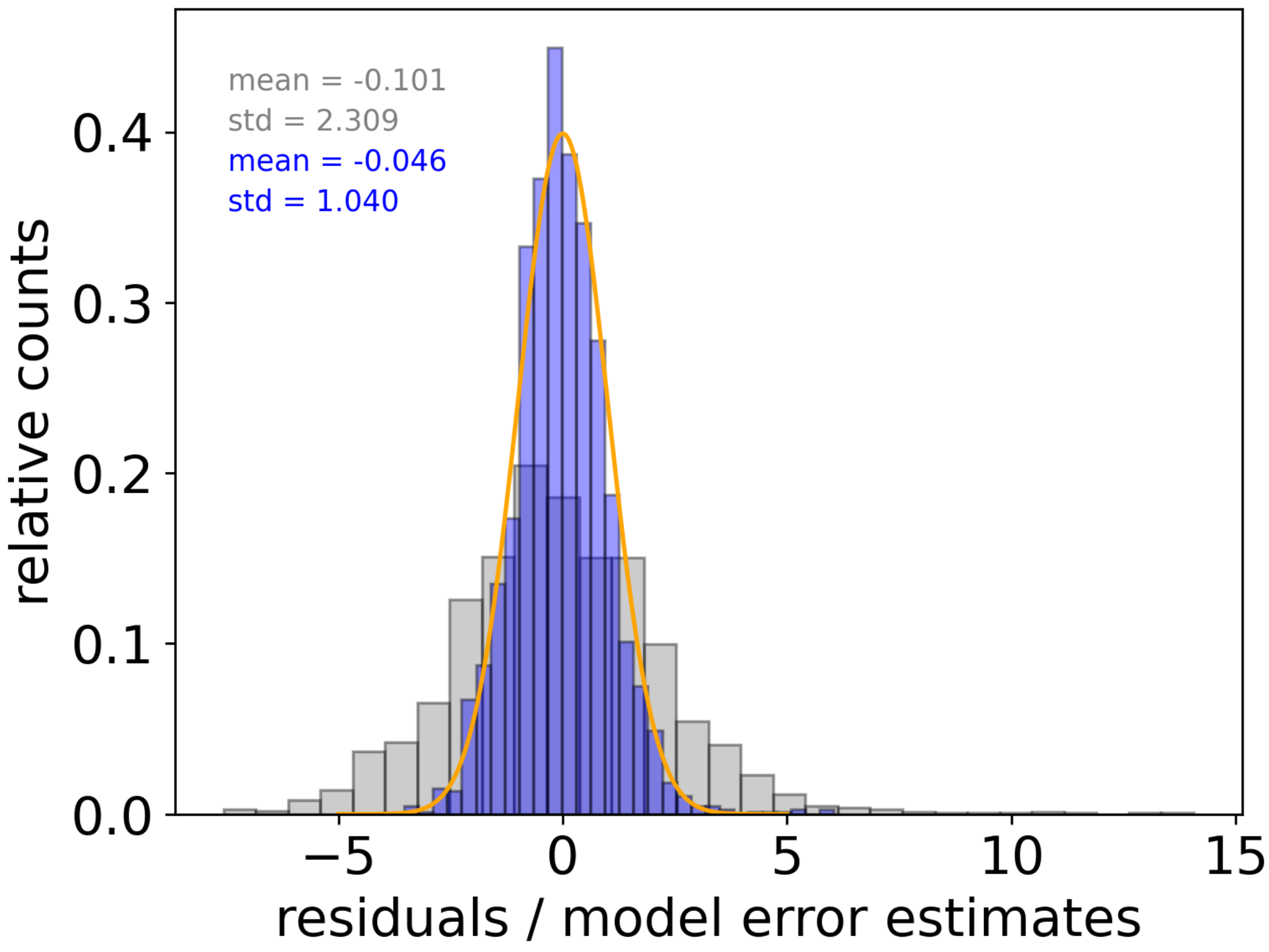

*Figure 53: Diffusion Neural Network (ensemble of 25) r-statistic*

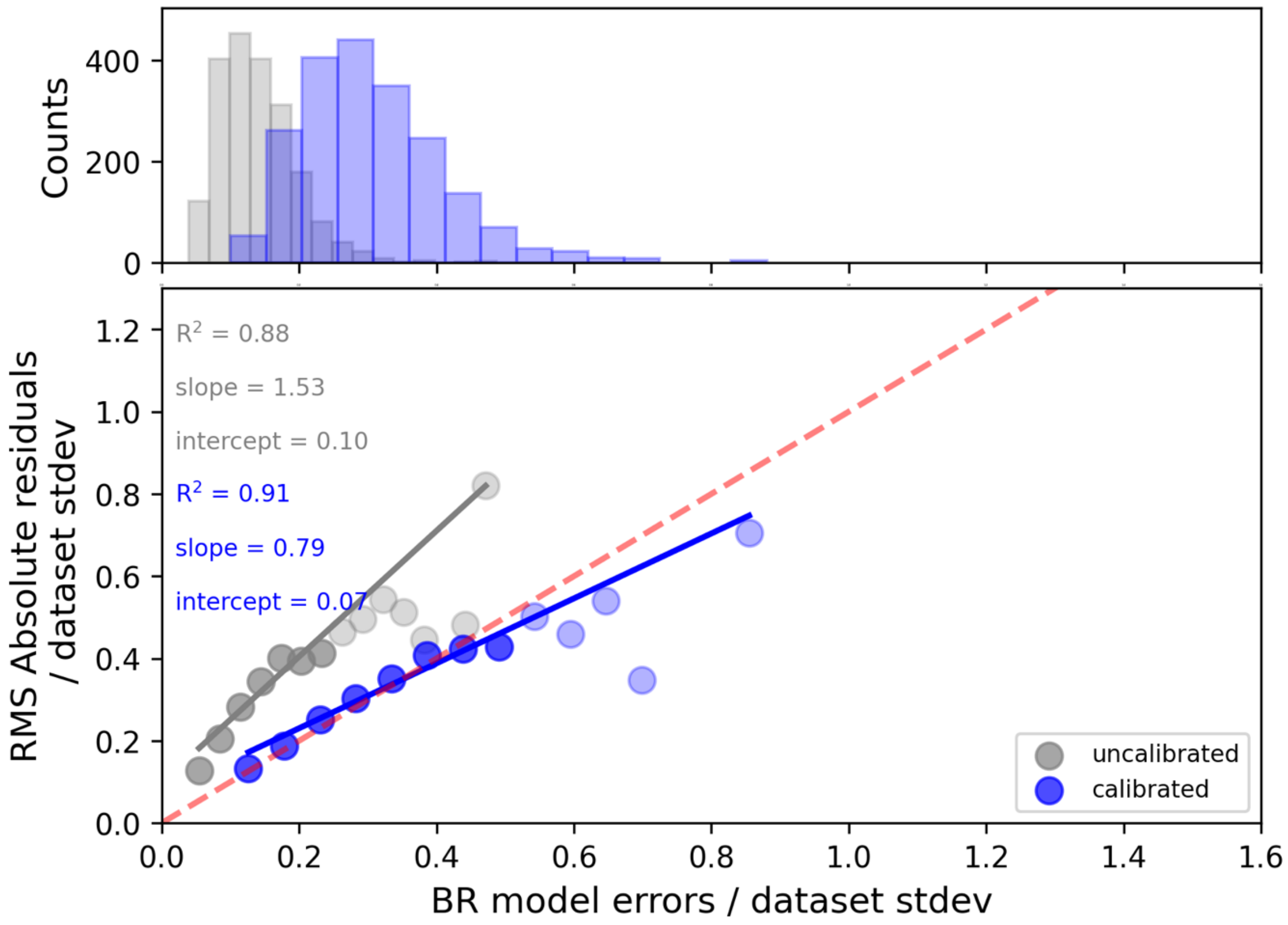

*Figure 54: Diffusion Neural Network (ensemble of 25) RvE*

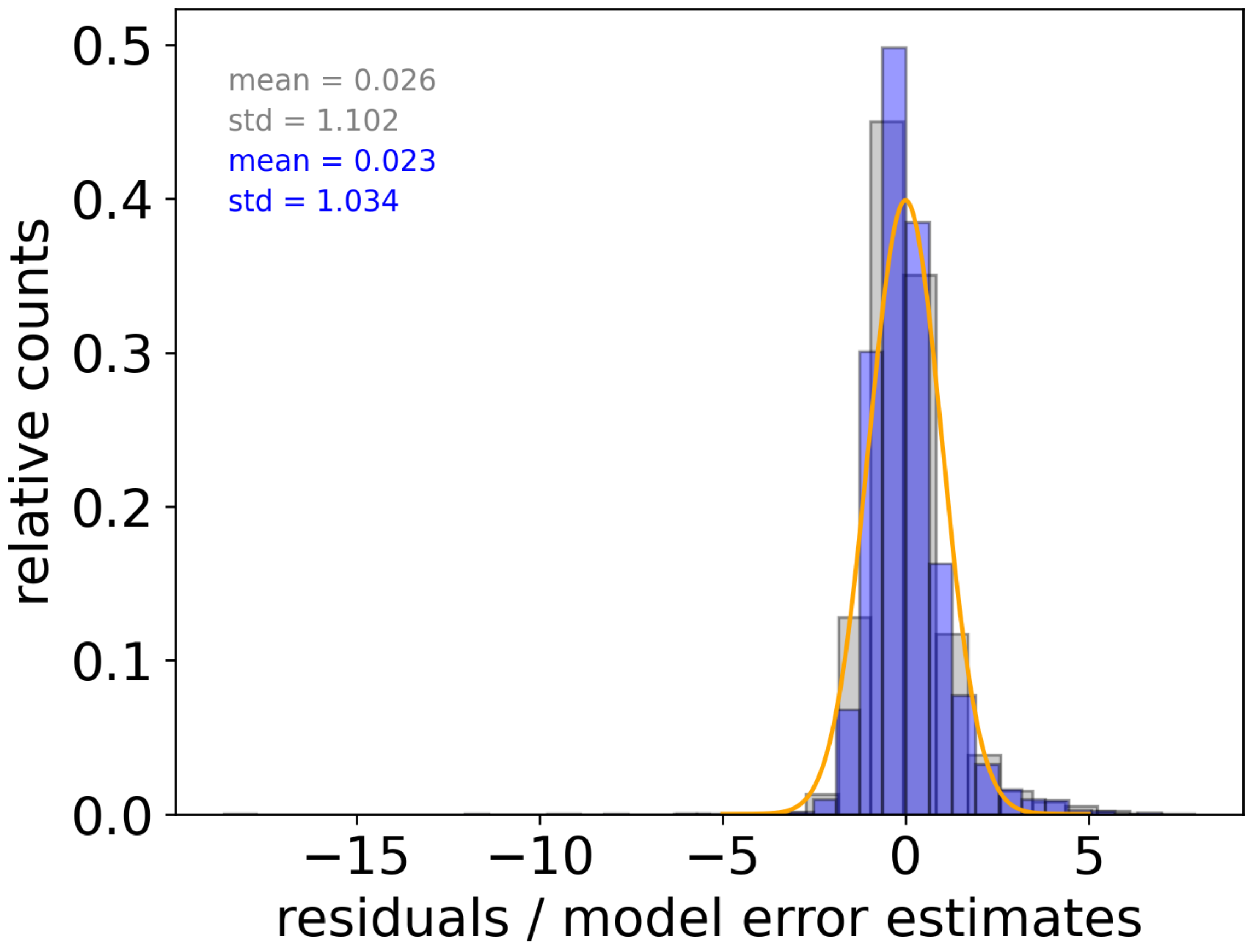

*Figure 55: Bulk modulus random forest r-statistic*

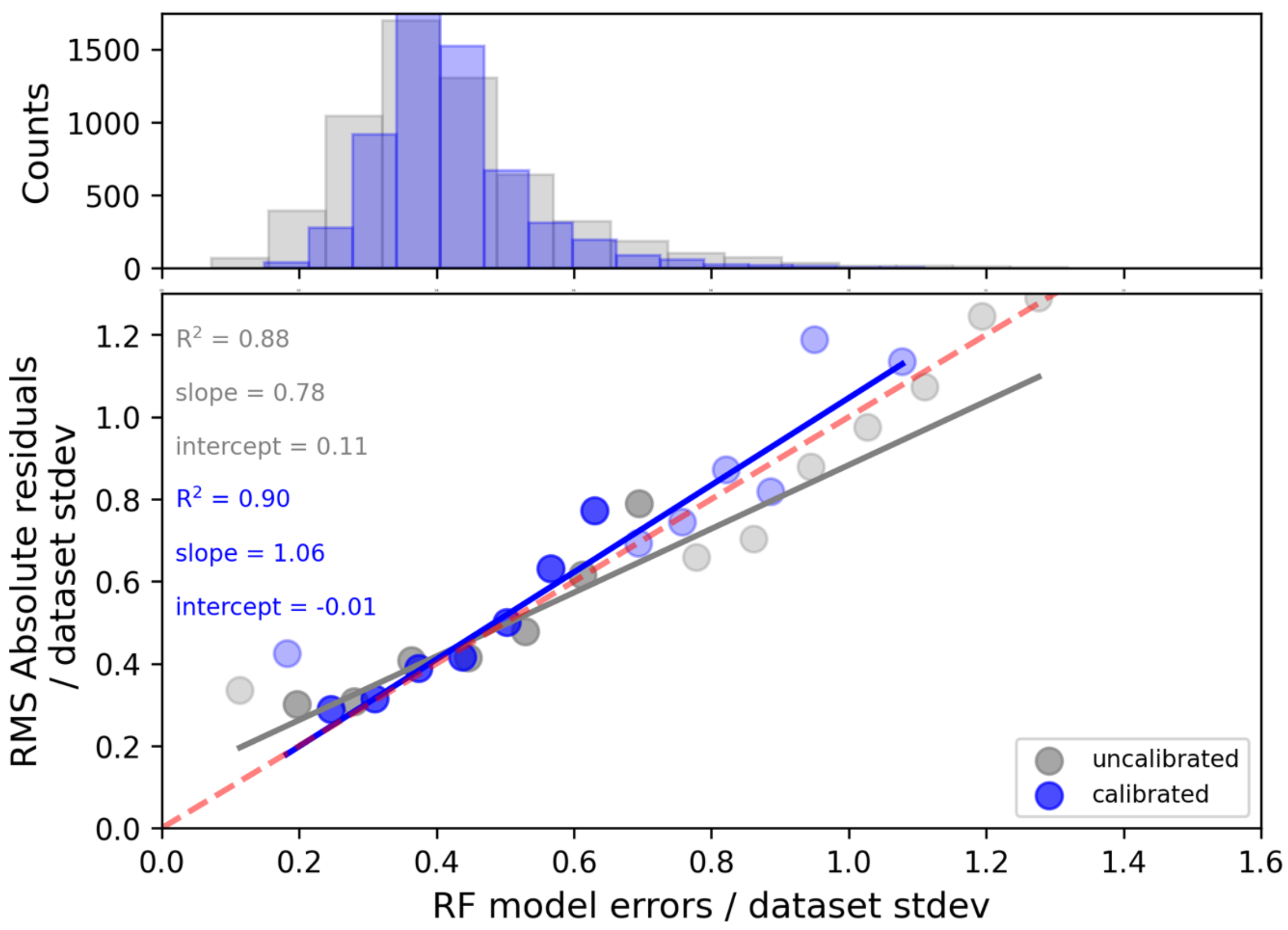

*Figure 56: Bulk modulus random forest RvE*

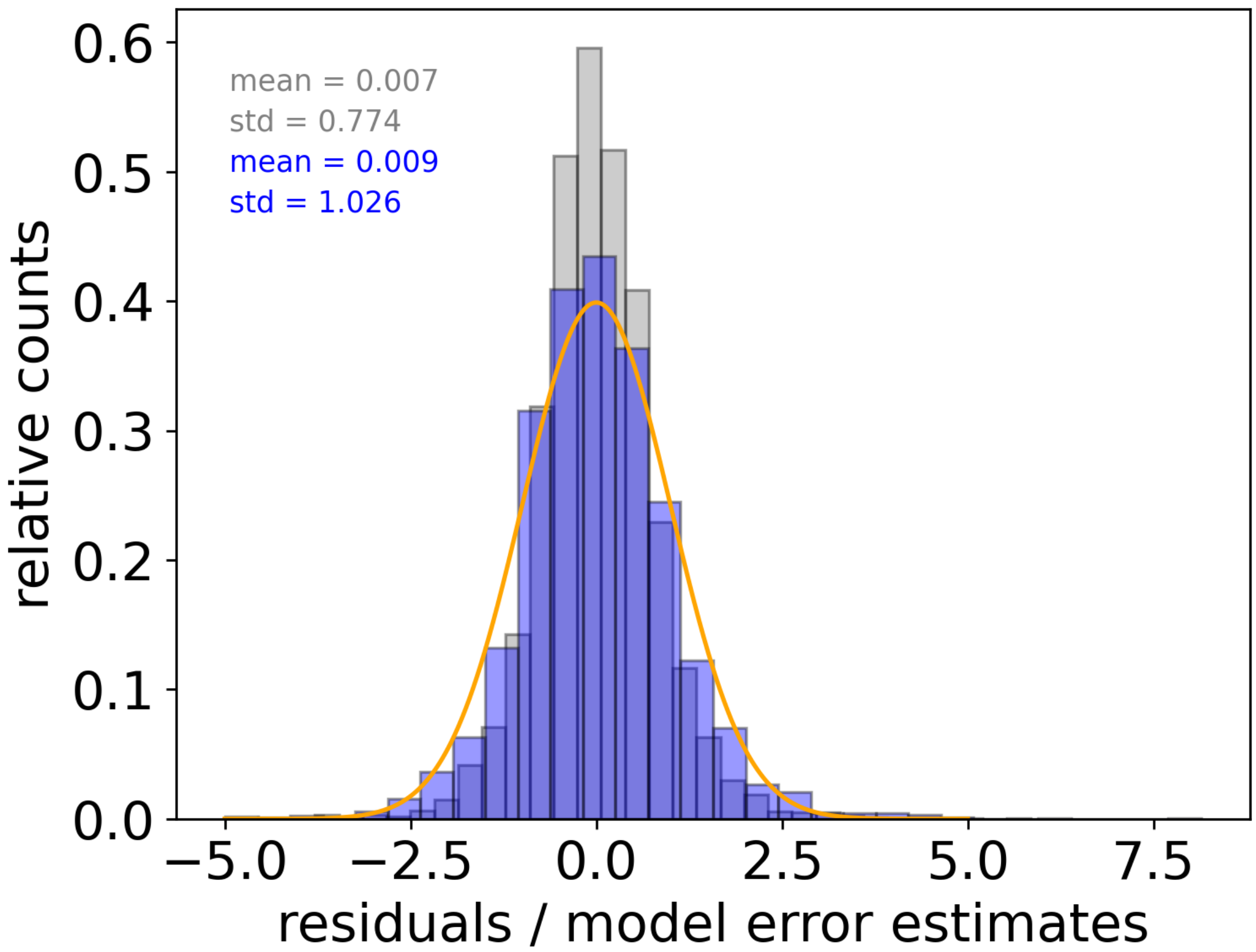

*Figure 57: Double perovskite gap random forest r-statistic*

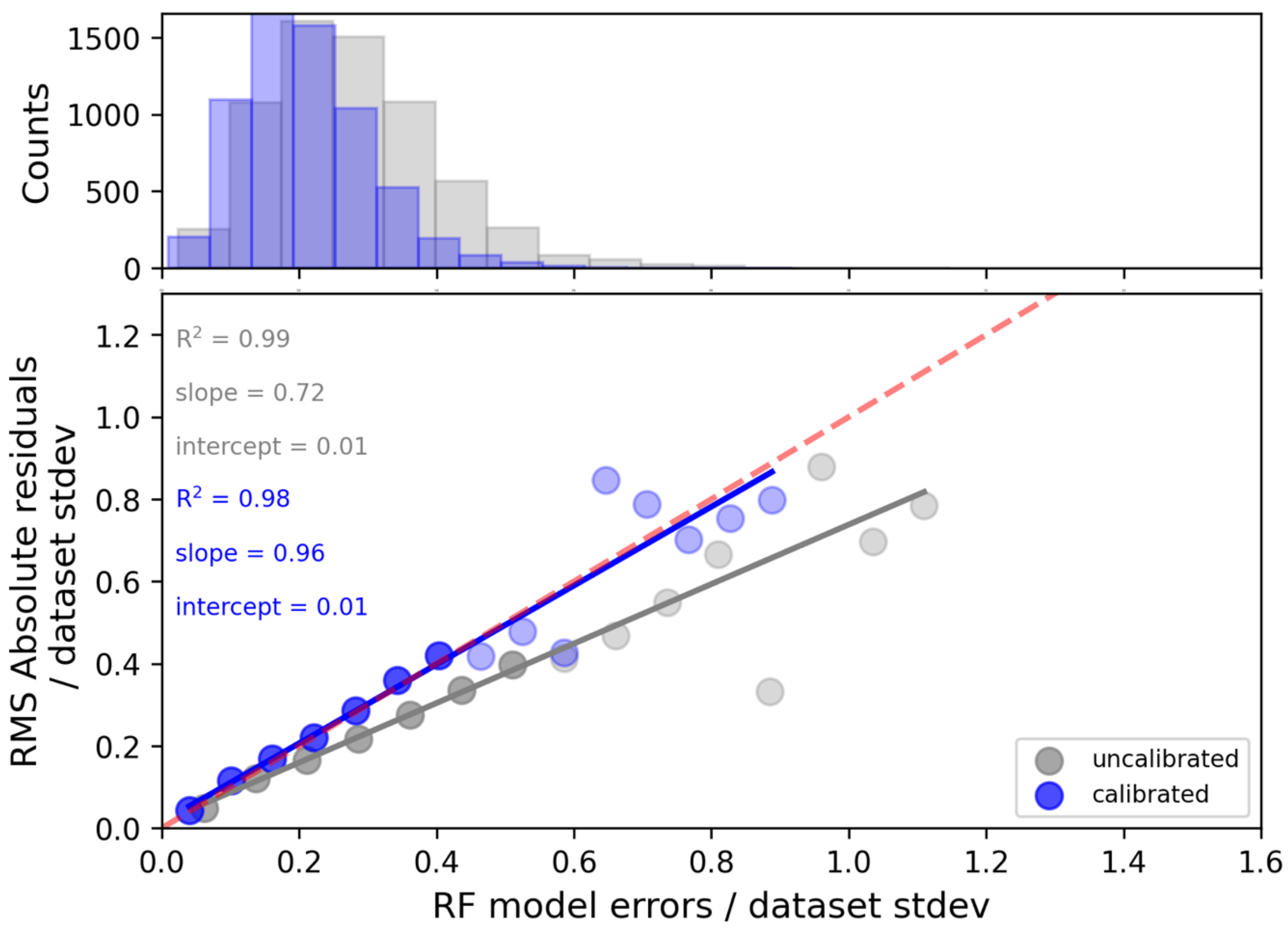

*Figure 58: Double perovskite gap random forest RvE*

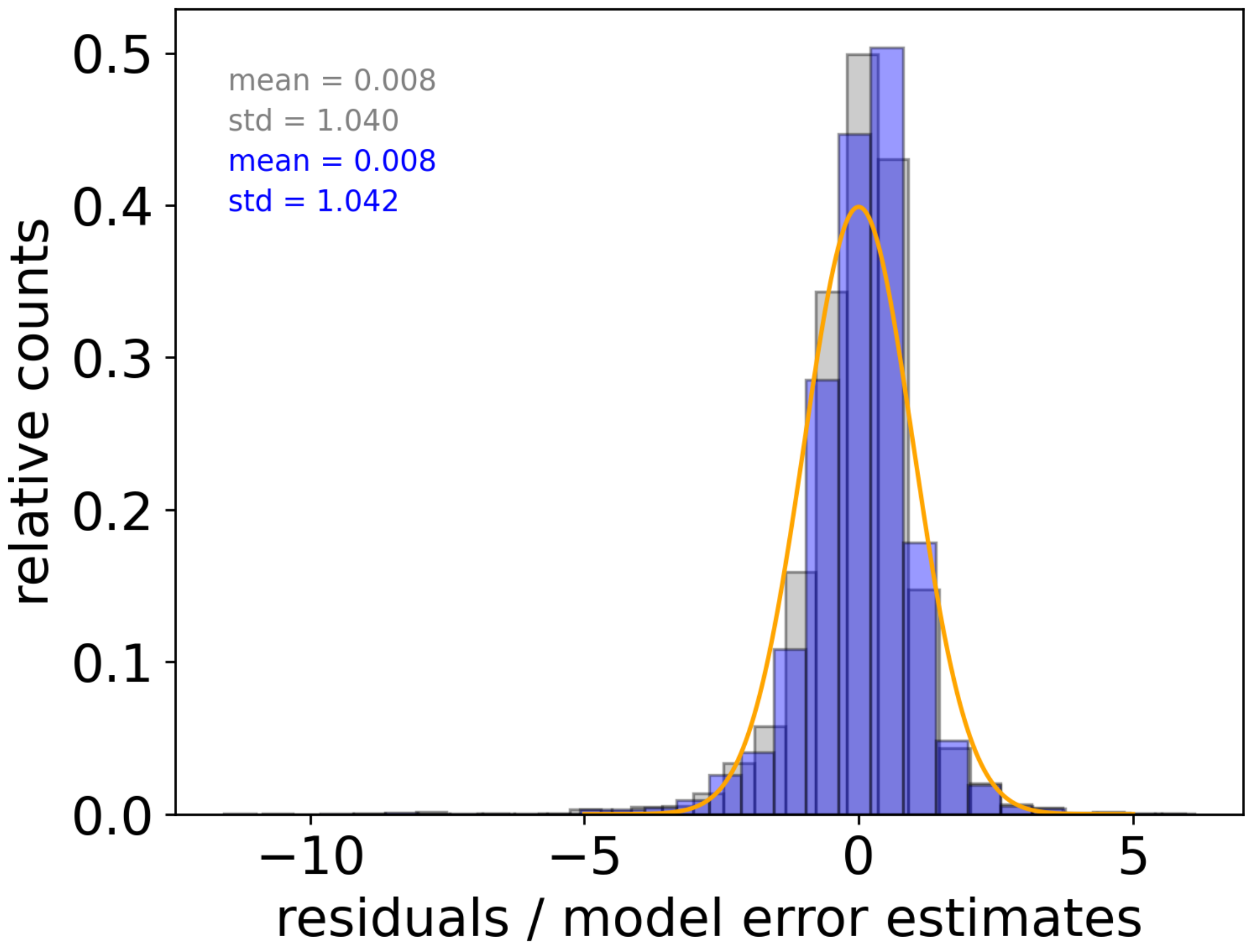

*Figure 59: Heusler random forest r-statistic*

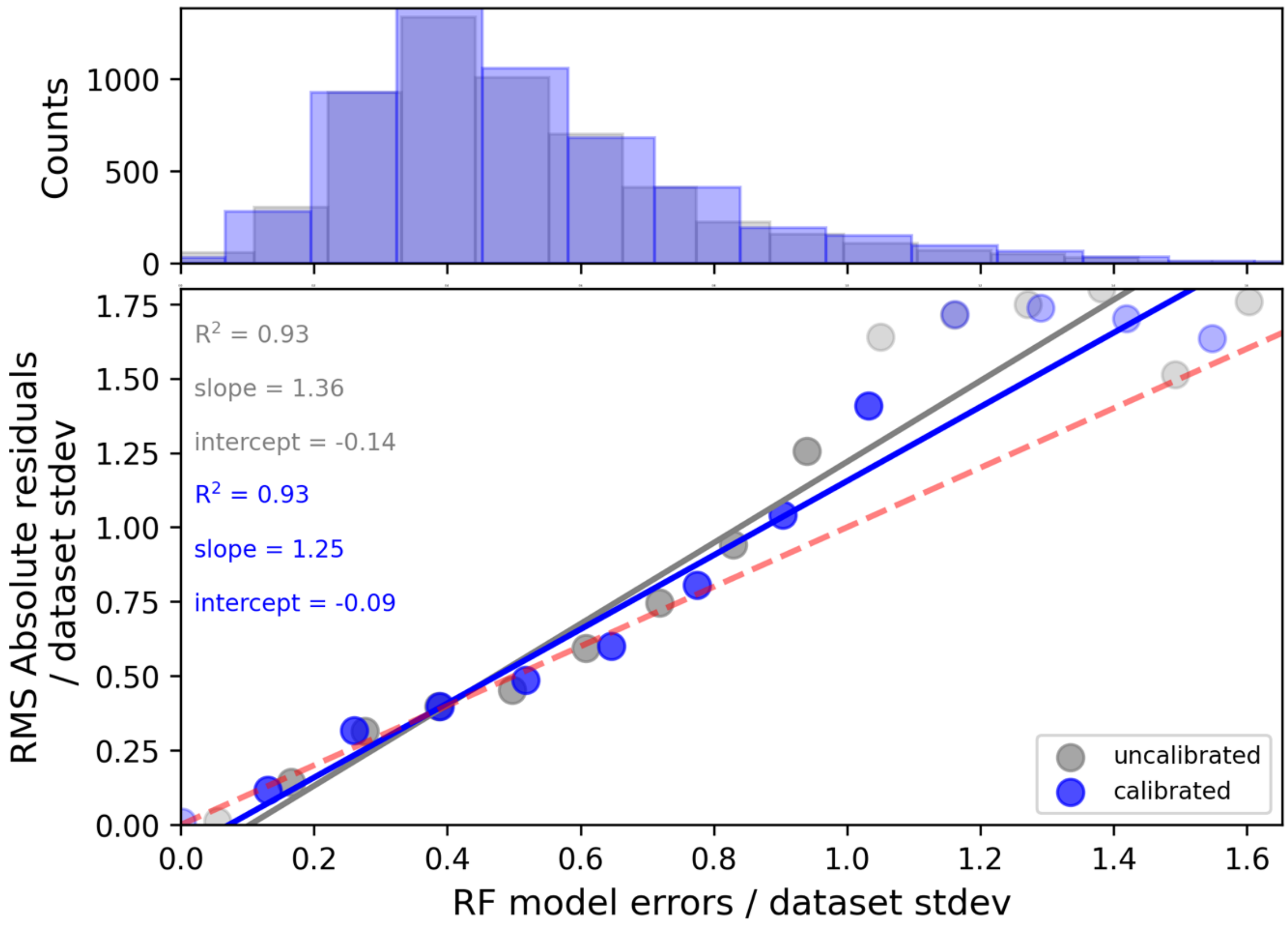

*Figure 60: Heusler random forest RvE*

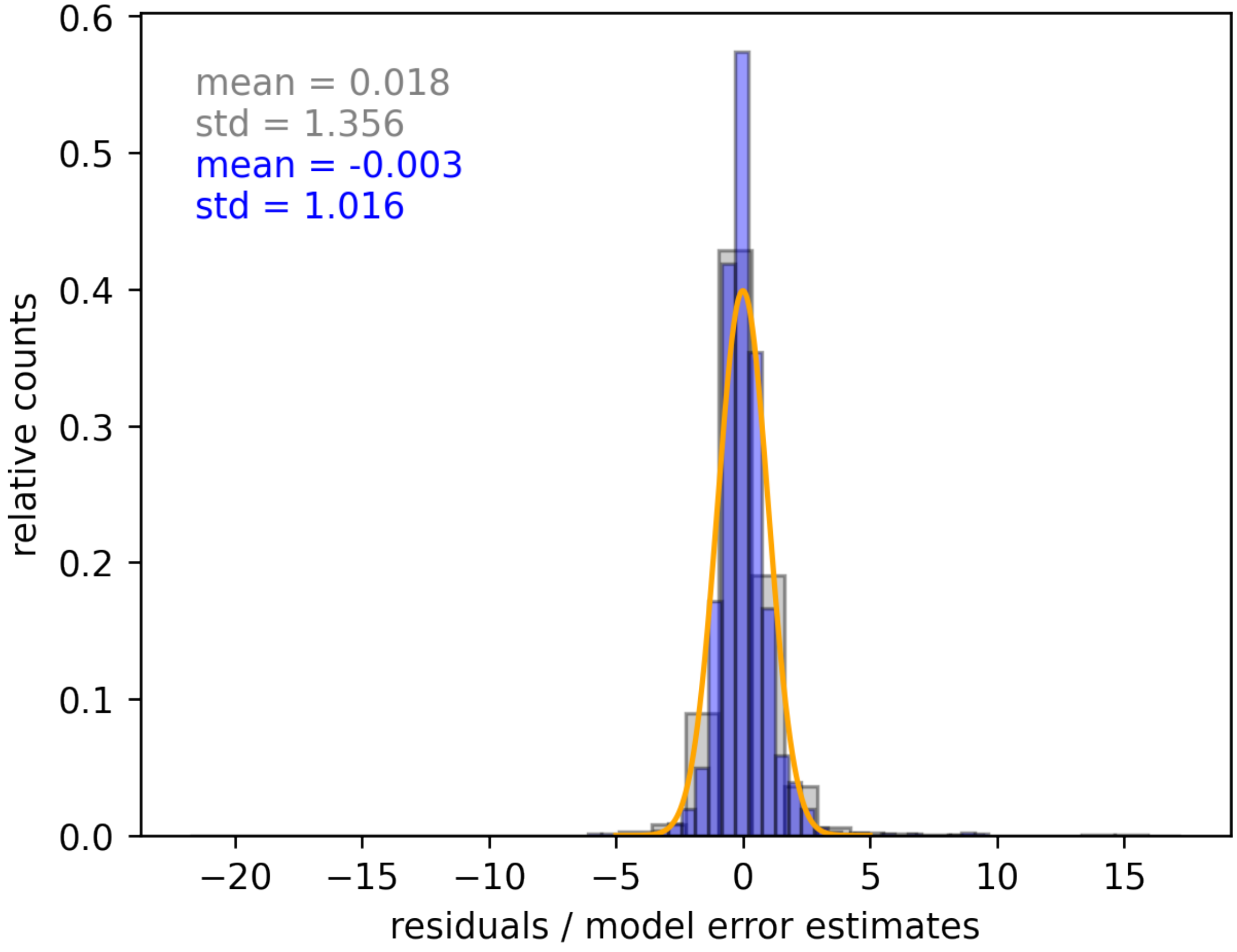

*Figure 61: Piezo random forest r-statistic*

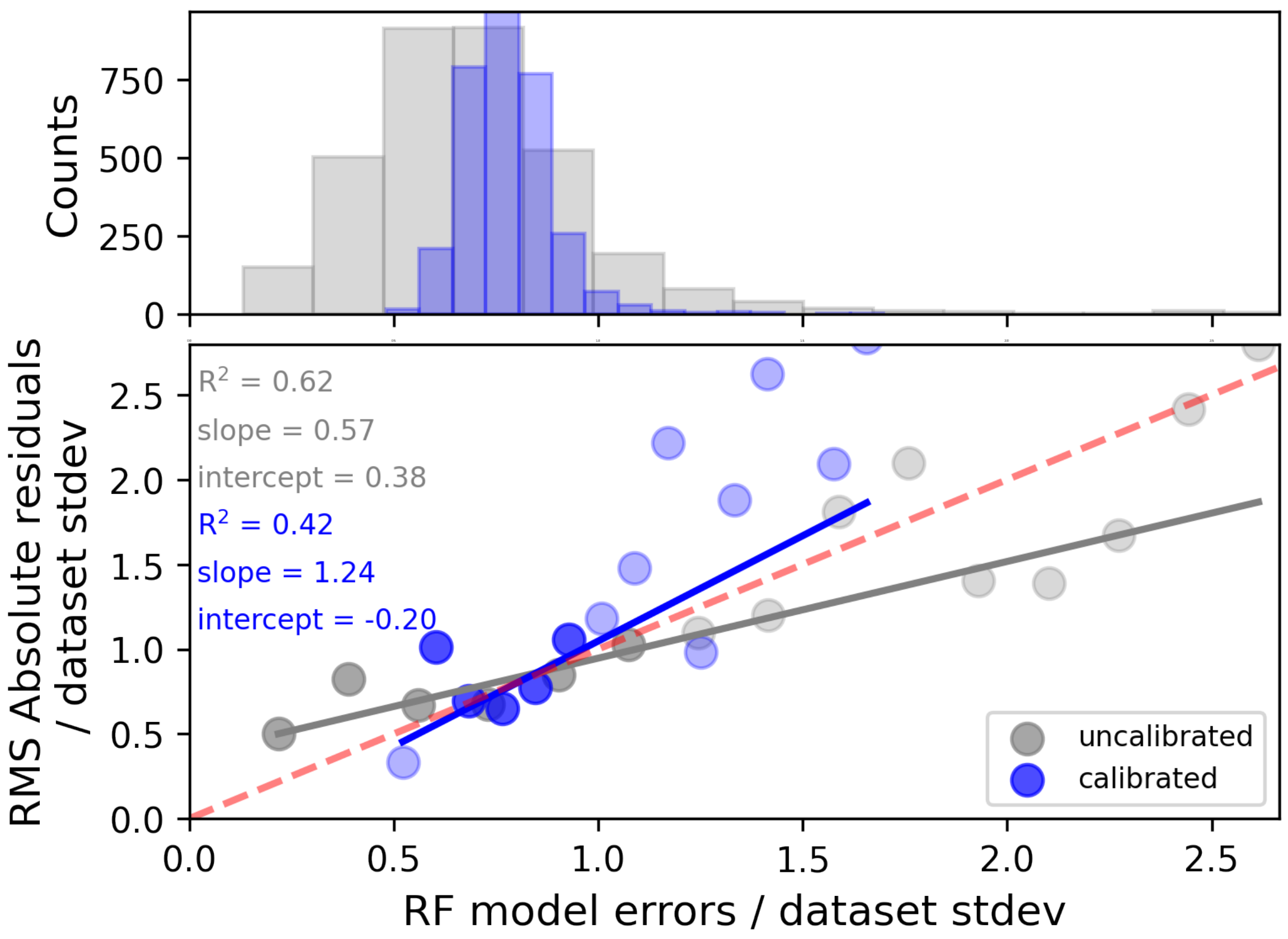

*Figure 62: Piezo random forest RvE*

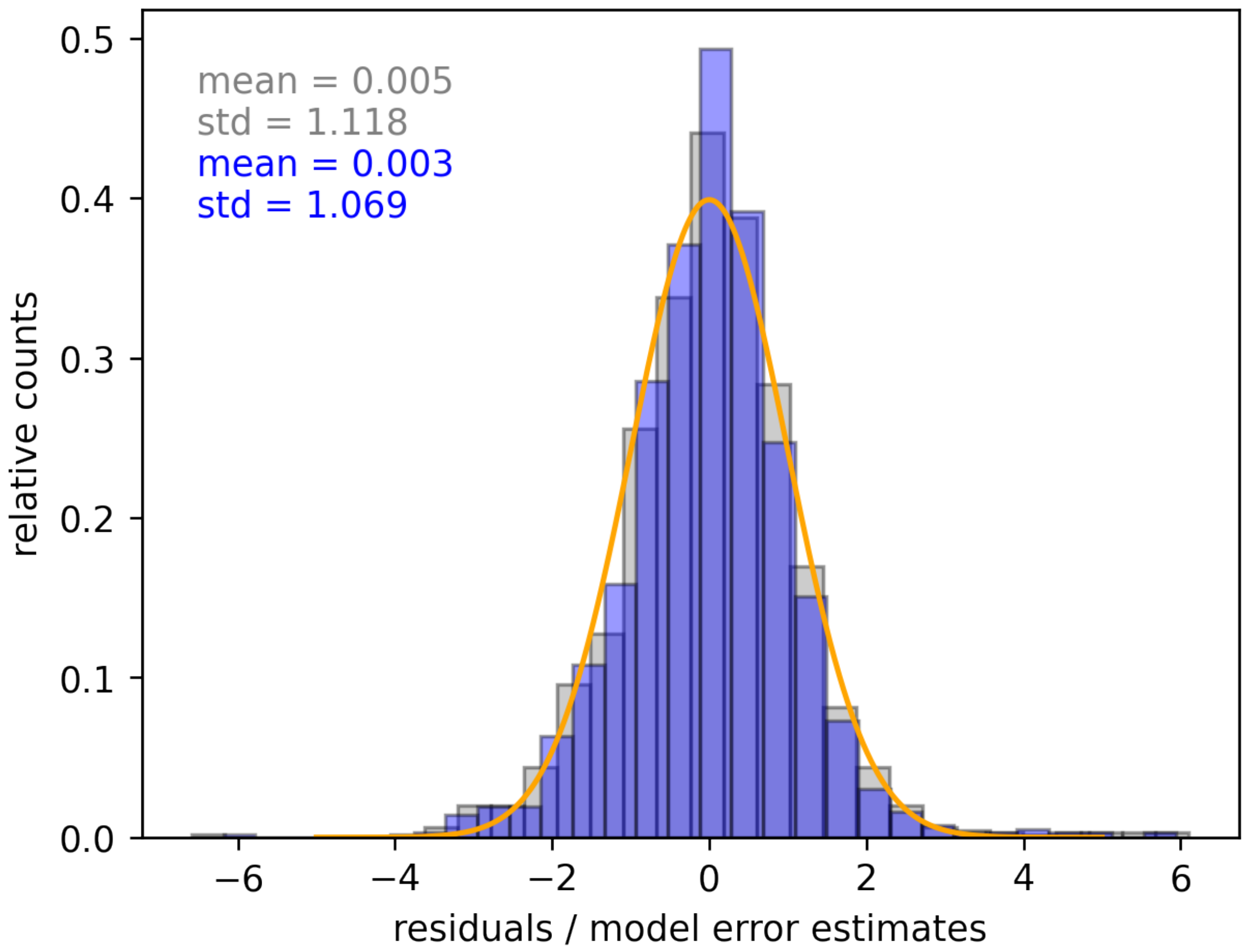

*Figure 63: Steel strength random forest r-statistic*

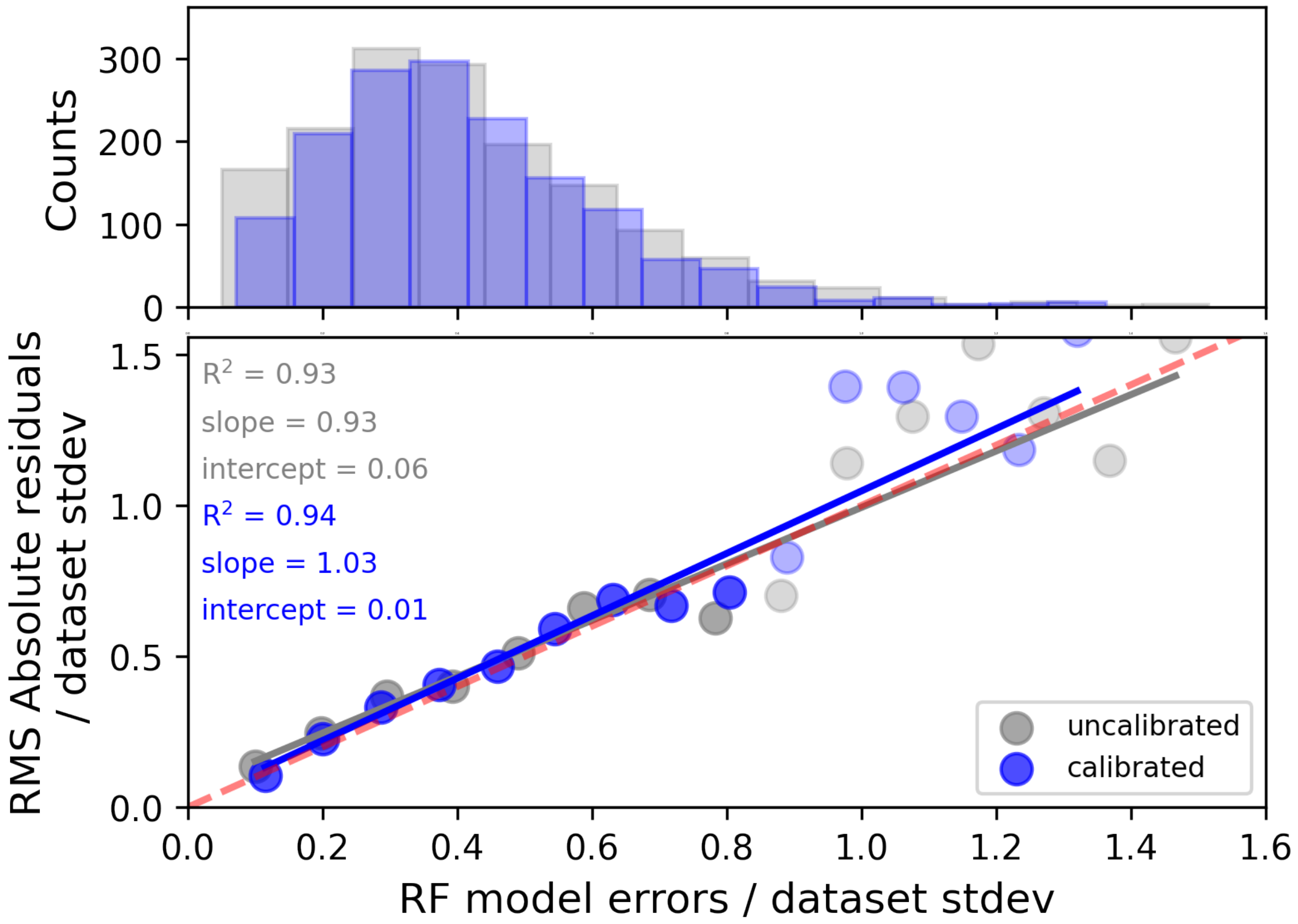

*Figure 64: Steel strength random forest RvE*

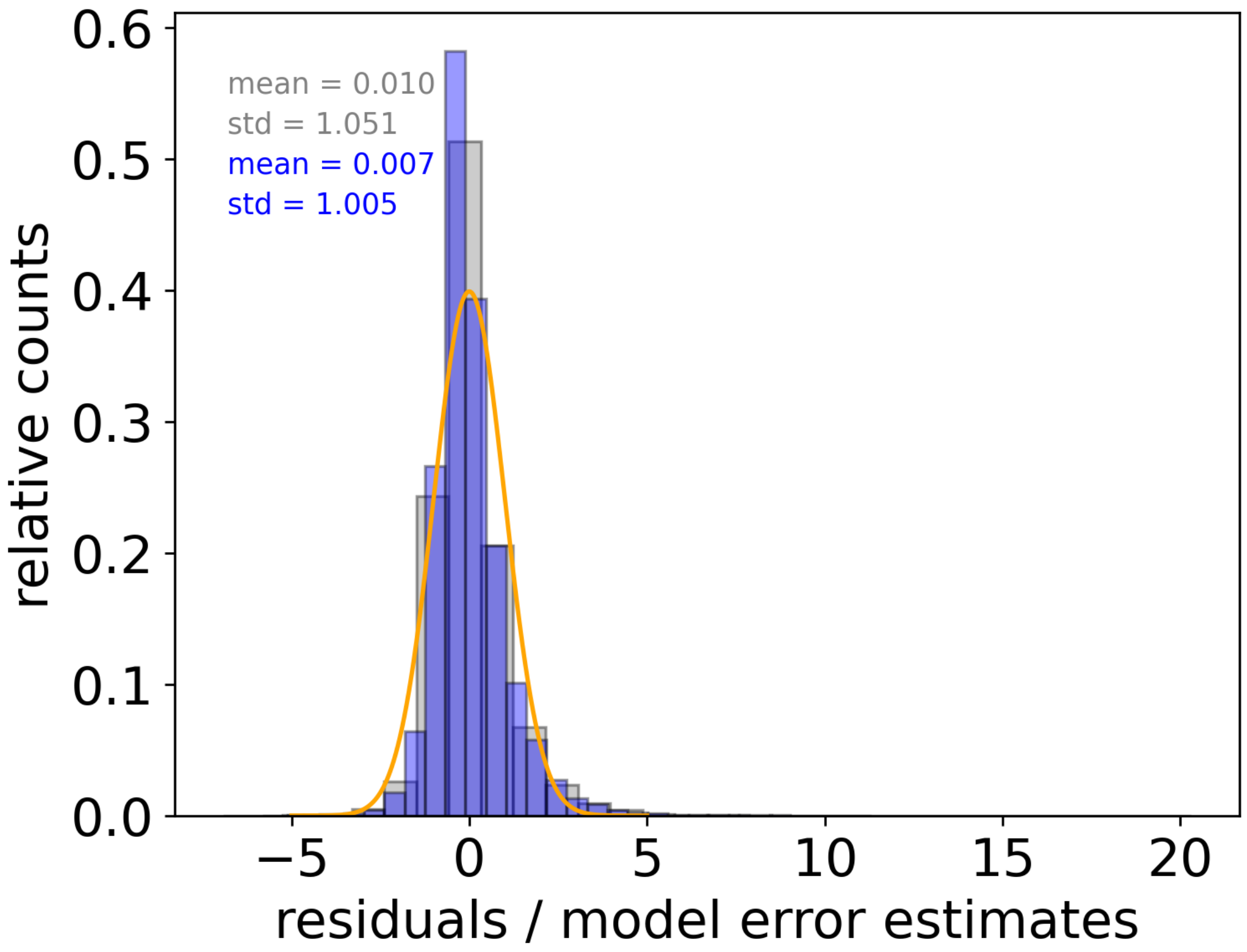

*Figure 65: Superconductor random forest r-statistic*

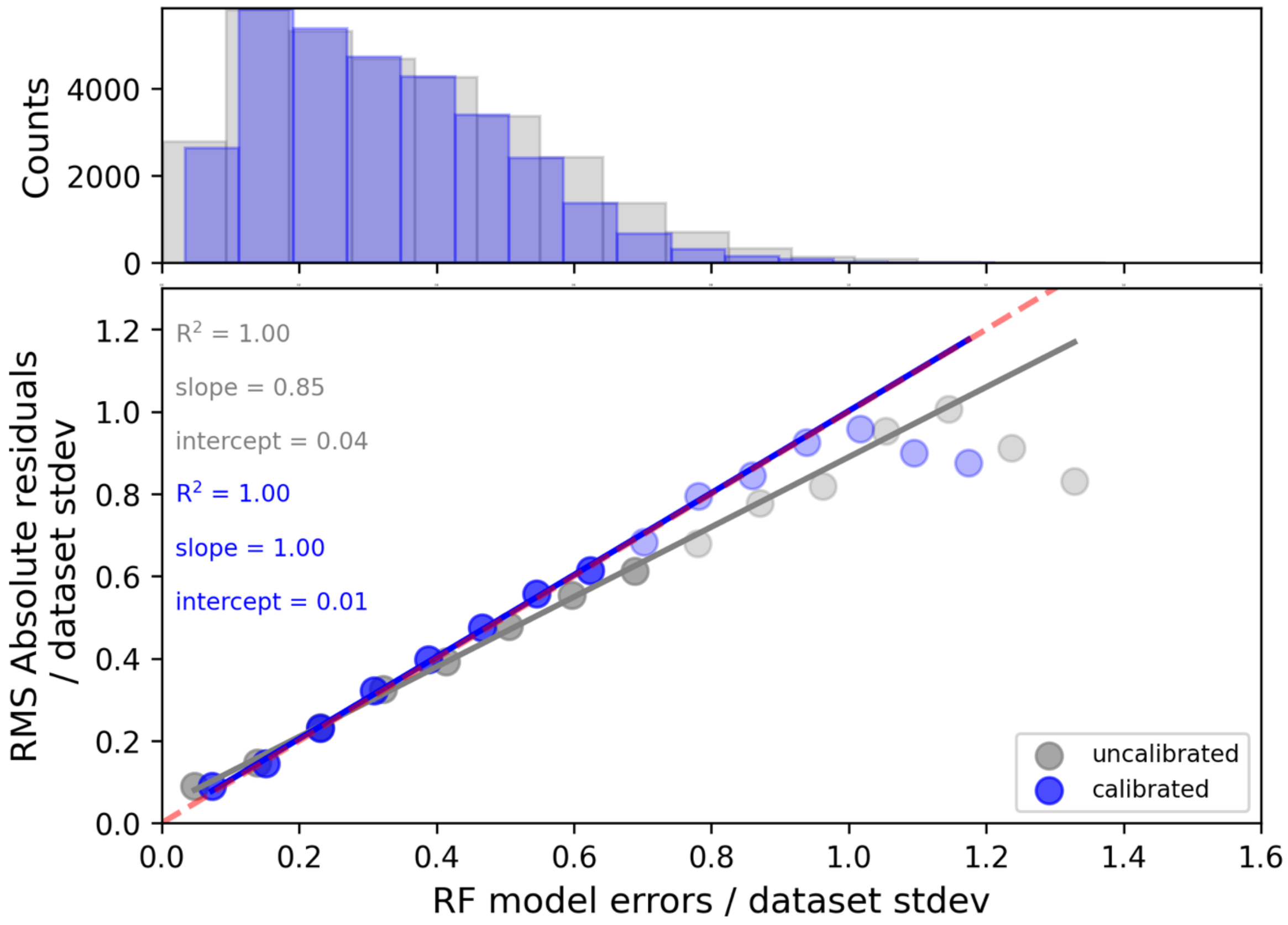

*Figure 66: Superconductor random forest RvE*

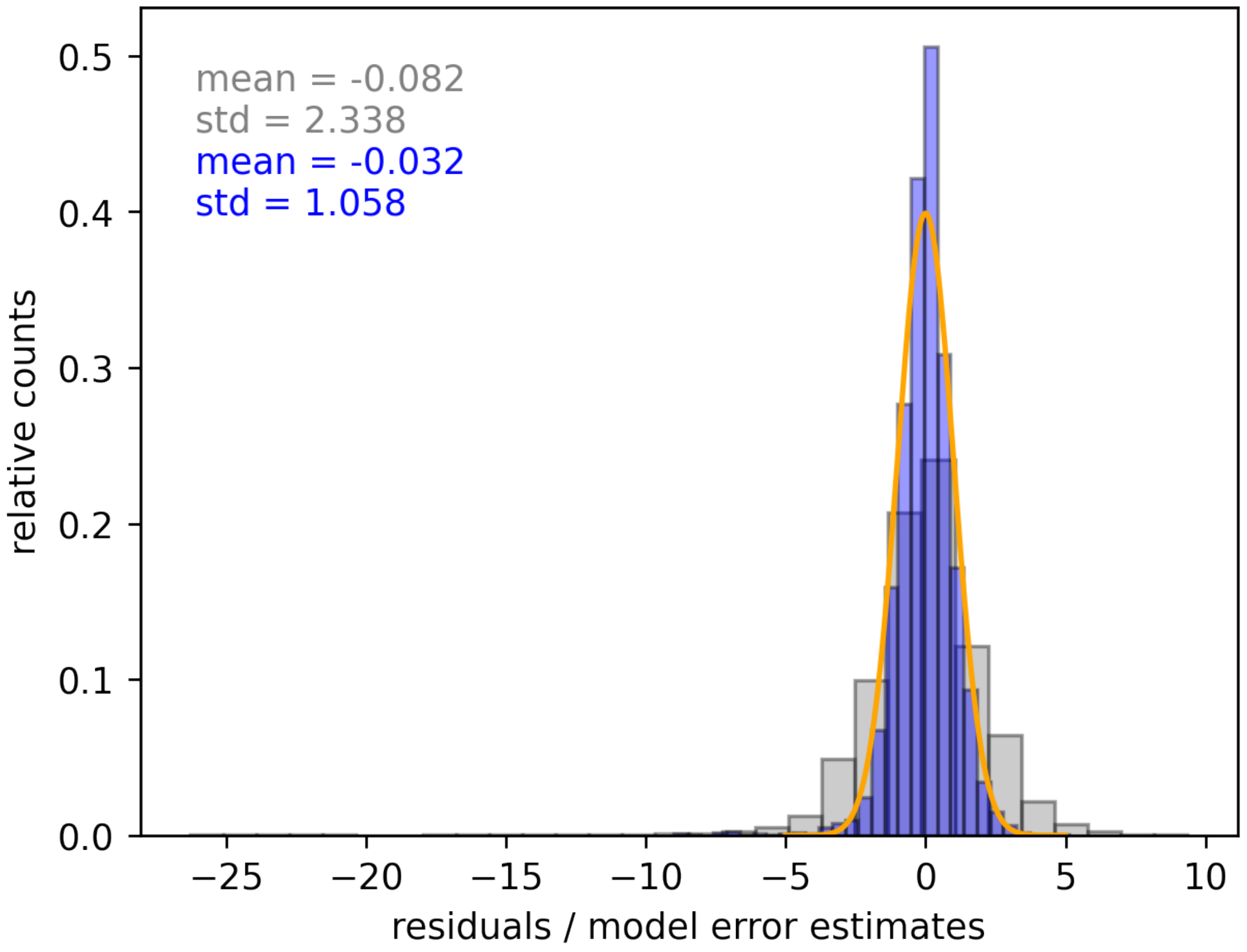

*Figure 67: Thermal conductivity random forest r-statistic*

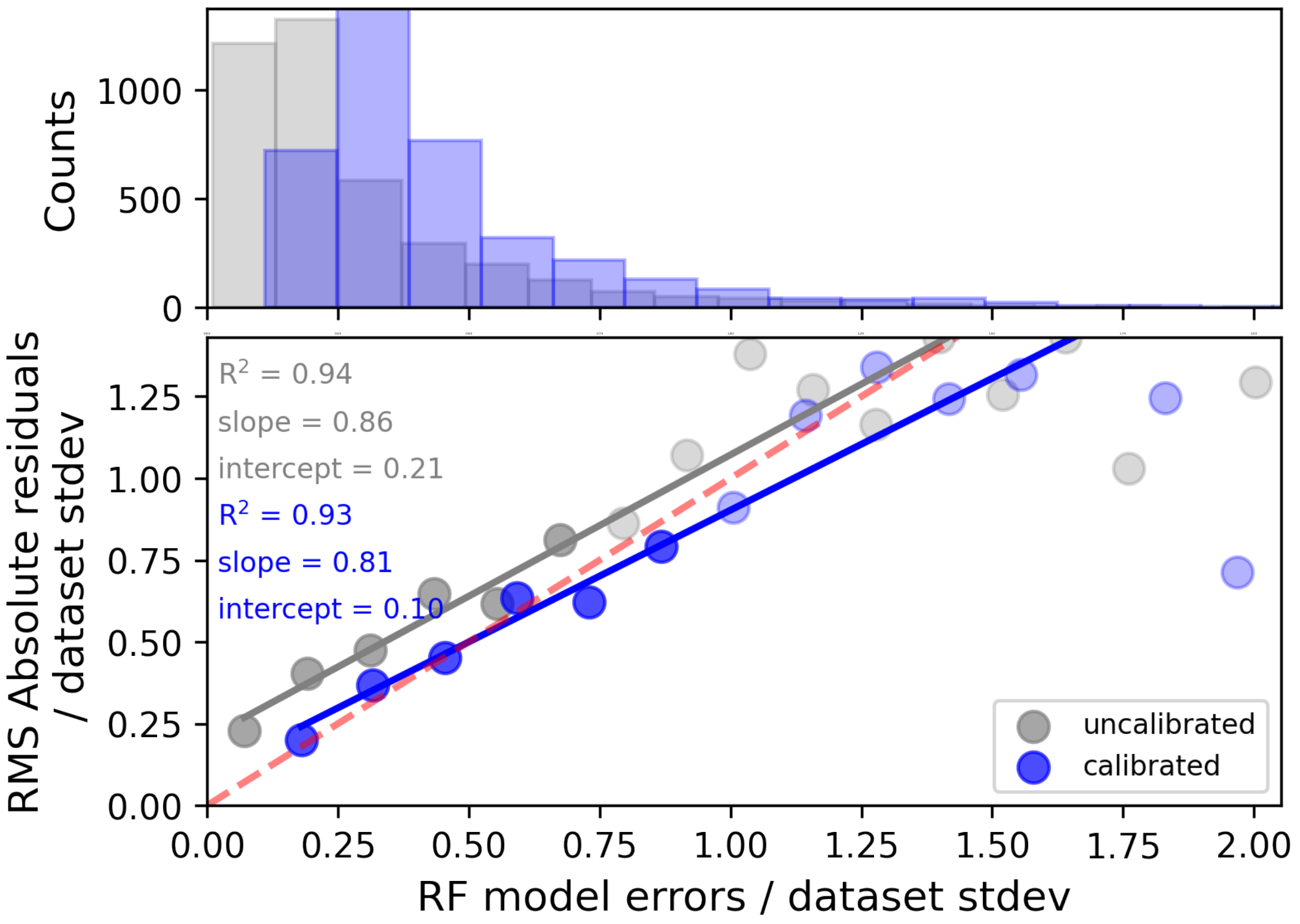

*Figure 68: Thermal conductivity random forest RvE*

In Figures 69-72, we apply our uncertainty quantification and recalibration method to the diffusion data with carefully chosen test sets that are expected to represent test data that is outside the domain of the trained model. Predictions on out of domain data are expected to result in poor model predicative ability and poor ability to accurately predict the model error bars. We use chemical intuition and domain knowledge to construct four different training-test set splits where the test set consists of materials we expect should be outside the training set. The approach is generally to create the test set by leaving out a specific type of chemistry for test. Test set 1 (Figure 69): this is a “leave out non-transition metal host” test where the left out data are Al, Ca, and Mg hosts, which were chosen because they are the only host materials in the dataset that are not transition metals. Test set 2 (Figure 70): this is a “leave out 3d transition metal host” test where the left out data are Fe, Cu and Ni hosts, which were chosen because they are the only host materials in the dataset that are transition metals belonging to the 3d row of the periodic table. Test set 3 (Figure 71): this is a “leave out 4d/5d transition metal host” test where the left out data are Ag, Pd, Ir, Au and Pt hosts, which were chosen because they are the only host materials in the dataset that are transition metals belonging to the 4d or 5d row of the periodic table. Test set 4 (Figure 72): this is a “leave out refractory metal host” test where the left out data are Mo, W and Zr hosts, which were chosen because they are the only host materials in the dataset that are refractory elements. In each case we show the RvE and parity plots, which demonstrate that the calibration does not work and that the model is highly inaccurate, respectively, both results consistent with the test data being outside the domain of the trained model.

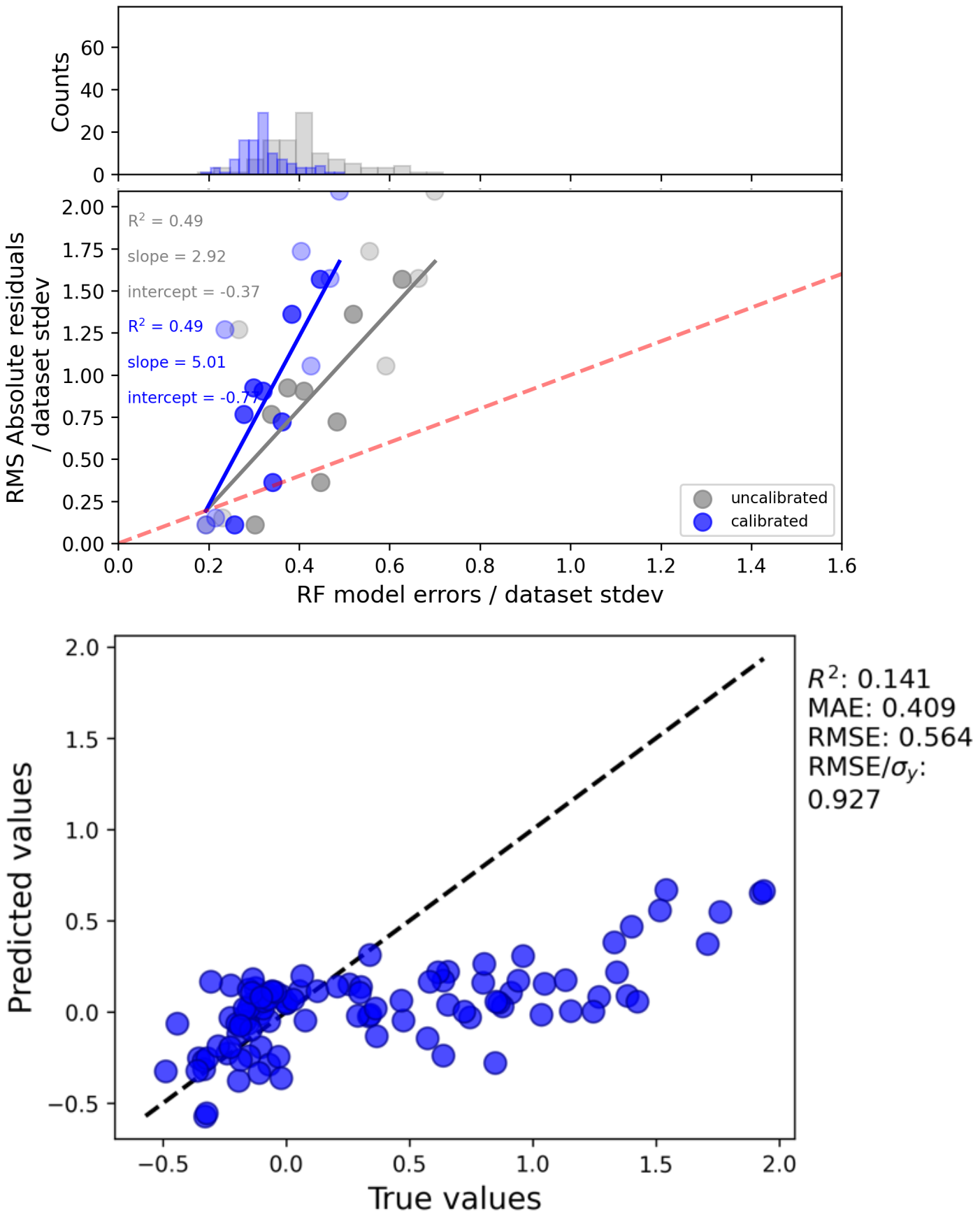

*Figure 69: Diffusion data random forest RvE (left-hand side) and parity plot (right-hand side) for a "leave out non-transition metal host" domain test where the Al, Ca and Mg hosts were left out and the remaining host data (except Pb) was used for training. Recalibration factors were obtained from 5 iterations of 5-fold CV (25 splits) on the training data*

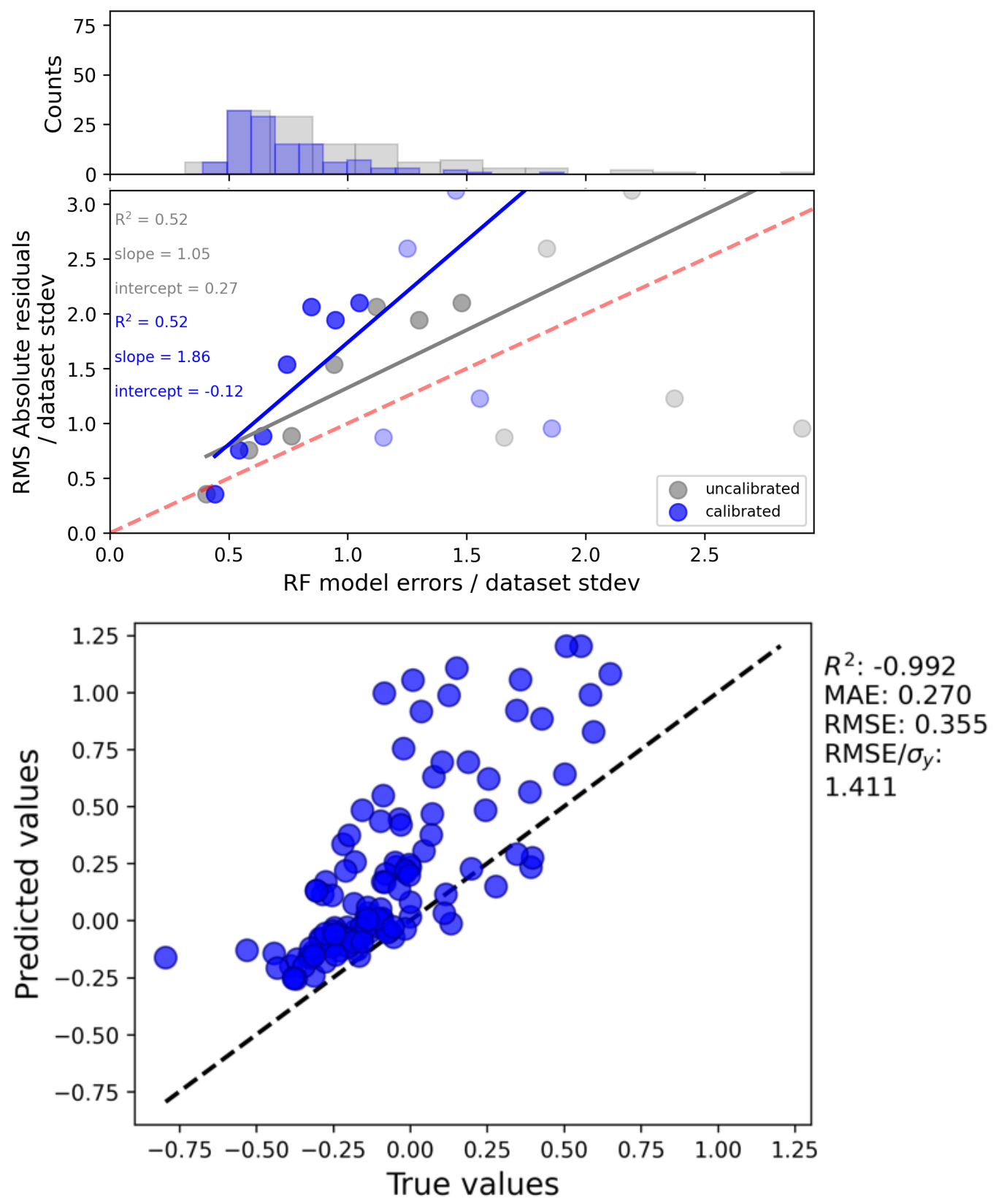

*Figure 70: Diffusion data random forest RvE (left-hand side) and parity plot (right-hand side) for a "leave out 3d transition metal host" domain test where the Fe, Cu and Ni hosts were left out and the remaining host data (except Pb) was used for training. Recalibration factors were obtained from 5 iterations of 5-fold CV (25 splits) on the training data.*

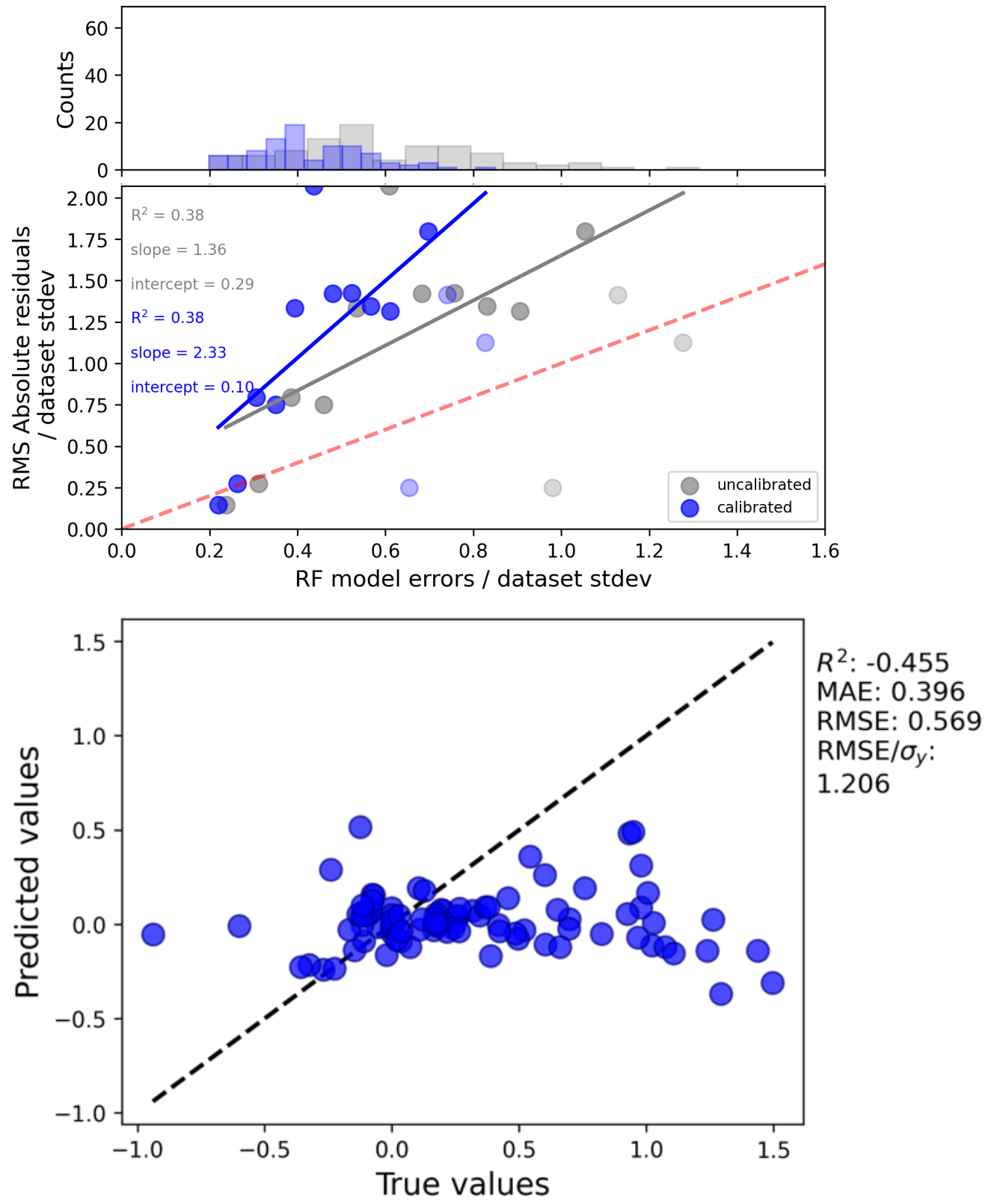

*Figure 71: Diffusion data random forest RvE (left-hand side) and parity plot (right-hand side) for a "leave out 4d/5d transition metal host" domain test where the Ag, Pd, Ir, Pt and Au hosts were left out and the remaining host data (except Pb) was used for training. Recalibration factors were obtained from 5 iterations of 5-fold CV (25 splits) on the training data.*

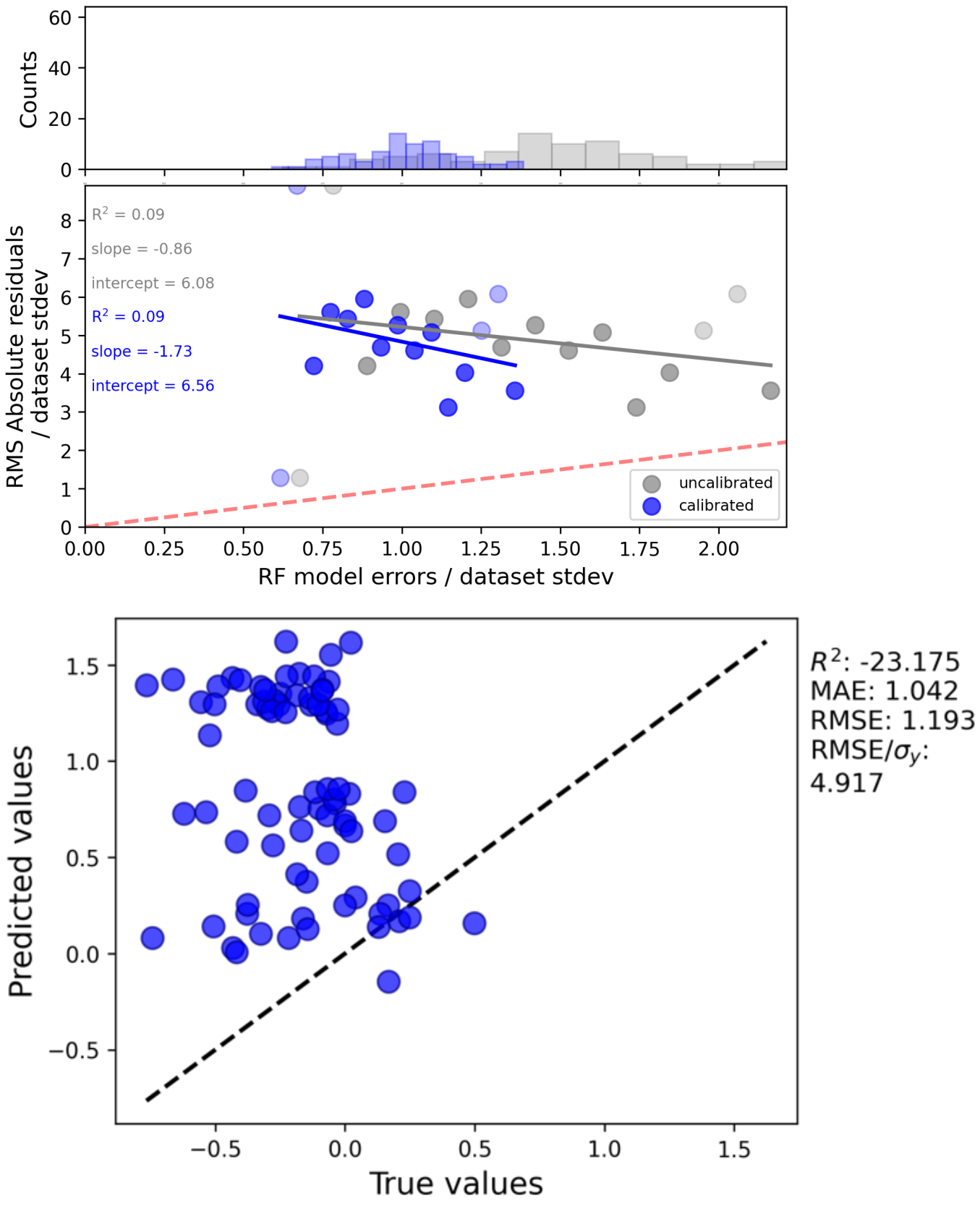

*Figure 72: Diffusion data random forest RvE (left-hand side) and parity plot (right-hand side) for a "leave out refractory metal host" domain test where the Mo, W and Zr hosts were left out and the remaining host data (except Pb) was used for training. Recalibration factors were obtained from 5 iterations of 5-fold CV (25 splits) on the training data.*